\documentclass[pre,twocolumn,amsmath,amssymb,aps]{revtex4-2}
\usepackage{dcolumn}
\usepackage{amsmath, amsthm, amssymb}
\usepackage{algorithm}
\usepackage{algpseudocode}
\usepackage{bm}
\usepackage{graphicx}
\usepackage{tikz}
\usepackage{caption}
\usepackage{subcaption}
\usepackage[export]{adjustbox}         
\newcommand{\vect}[1]{\boldsymbol{#1}}

\providecommand{\abs}[1]{\lvert#1\rvert}
\providecommand{\norm}[1]{\lVert#1\rVert}
\providecommand{\bracket}[1]{\ensuremath{\left(#1\right)}}
\newenvironment{frcseries}{\fontfamily{frc}\selectfont}{}

\usepackage{etoolbox}

\makeatletter
\AtBeginDocument{

}
\makeatother
\renewcommand\thesubfigure{(\alph{subfigure})}

\makeatletter
\patchcmd{\subcaption@place}
  {\box}
  {\rlap{\raisebox{0.9\ht\@tempboxa}{\hspace{0.5em}\subcaplabelfont\thesubfigure}} \box}
  {}{}
\makeatother
\graphicspath{{./figures/}}
\usepackage{epsf,epsfig}
\usepackage{hyperref}
\hypersetup{bookmarks=false}\usepackage{color}
\usepackage{enumerate}
\usepackage{lineno}

\begin{document}
\title{Training cell stress patterns in 3D cellular packings}
\author{Shabeeb Ameen$^1$}
\email{mameen@syr.edu}
\author{Tao Zhang$^2$}
\email{zhangtao.scholar@sjtu.edu.cn}
\author{J. M. Schwarz$^{1,3}$}
\email{jmschw02@syr.edu}

\affiliation{$^1$Department of Physics and BioInspired Institute, Syracuse, Syracuse University, Syracuse, NY 13244, USA} 
\affiliation{$^2$School of Chemistry and Chemical Engineering, Shanghai Jiao Tong University, Shanghai 200240, China}
\affiliation{$^3$Indian Creek Farm, Ithaca, NY, 14850, USA}
\date{\today}

 \begin{abstract}

 The task of learning patterns is typically associated with systems that update parameters on fixed architectures, such as neural networks, where learning proceeds through continuous optimization. Here, we demonstrate that pattern learning can also emerge in reconfigurable cellular tissues, where both mechanical parameters and network topology evolve. Using a three-dimensional vertex model, we show that multicellular packings can be trained to realize prescribed cellular stress patterns through a contrastive learning algorithm that updates hidden-cell shape indices.
We find that learning is intrinsically collective, requiring coordinated, system-wide parameter adjustments even for local targets, and that learnability is governed by an interplay between mechanical state, capacity, and training protocol. In particular, the rigidity of the tissue controls an effective exploration–exploitation tradeoff: fluid-like regimes enhance exploration through cellular rearrangements, while rigid regimes constrain dynamics and favor exploitation of existing configurations. These rearrangements introduce discontinuous learning dynamics, enabling the system to transition between distinct local minima in the cost function landscape. The ratio of target cells to total number of cells in the packing, or constraint load, sets an effective capacity, analogous to over- and underparameterized regimes in conventional learning systems. As this constraint load increases, learning becomes slower, more heterogeneous, and increasingly dependent on rare rearrangements that allow escape from geometrically constrained states. Finally, training cells in sequence, in contrast to parallel protocols, provides an alternative route that can be more robust but generally takes longer to train for the constraint loads studied. These results suggest a learning phase diagram governed by constraint load, cell packing rigidity, and training protocol. By enabling the training of localized internal states, this work positions tissues not only as adaptive materials, but as nonconventional AI platforms. 
\end{abstract}

\maketitle

\section{Introduction}
\label{sec_introduction}

If all biological systems are learning systems, what then constitutes a biological system? Organisms, tissues, cells, proteins? It has long been established that organisms with brains can learn numerous tasks ranging from association formation to classification to mathematical calculations. Single-celled organisms devoid of brains, however, can also learn to anticipate signals, learn to habituate harmless signals, among other tasks~\cite{boussard2021adaptive,alim2026decision,rajan2023single,gershman2021reconsidering,rajan2025receptor}. So if multi-cellular and single-celled organisms can learn, why not a collection of proteins? Intriguingly, researchers have argued that protein learning is possible by considering the complex phase diagram of a multicomponent protein mixture driven their phase separation capability~\cite{zentner2025information}. Any mechanism, such as changes in relative concentrations, by which modification of the phase boundaries can move could lead to learning classification. Going to even  smaller scales, experiments have demonstrated that DNA systems with engineered interactions can learn patterns~\cite{evans2024pattern}. 

All biological systems are also physical systems and so there must a direct link between learning and physics that goes far beyond the brain. For some, the direct link is to consider conventional Artificial Intelligence platforms within a statistical mechanics framework~\cite{Bahri2020, Gabrie2023}. For others, it is to train physical systems, such as flow networks, to perform tasks using physical processes. In these systems, learning occurs not through abstract weight updates in a convolutional neural network but through the updating of physical parameters as prescribed by a physical learning algorithm~\cite{wright2022deep,stern2023learning,anisetti2023emergent,momeni2025training}. A growing family of contrastive-style physical learning algorithms has been developed in this context, including equilibrium propagation, coupled learning, multi-mechanism learning, frequency propagation, {\it in situ} backpropagation, and kernel propagation~\cite{scellier2017equilibrium,Anisetti2023,Anisetti2024,li2024training,falk2025temporal}. Despite differences in implementation, these approaches share a common structure: the system is alternately allowed to relax freely and then gently nudged toward a desired target state via gradient descent in the cost function landscape. These contrastive learning approaches have primarily focused on networks with a fixed architecture, which makes the contrast more feasible. For instance, Hexner \cite{Hexner_stress_patterns} demonstrated that disordered spring networks can be trained to realize prescribed internal stress patterns using local learning rules. In this system, the rest lengths of the springs act as tunable mechanical weights, and contrastive updates adjust these so that, after training, the network spontaneously produces the desired stress distribution. 

As the architecture of many physical systems, including biological ones, is not fixed, it behooves one to ask how does learning happen with rearrangements, be it the brain with its neuronal plasticity~\cite{hebb1949organization}, the cytoskeleton with its binding and unbinding of filaments~\cite{gopinathan2007branching,fletcher2010cell} or even an extremely minimal systems of a collection of interacting particles? Recent work demonstrating that a particle packing can learn associations using local cyclic driving begins to help realize the expressivity of such systems \cite{guo2026learningassociationsreconfigurableparticle}. Such a result provides fundamental bounds on the expressivity of more complex systems, such as living ones. 

 Our biological system of choice is a three-dimensional cellular packing. From a biomechanical perspective, a confluent cell packing can be viewed as a multi-body generalization of a spring network: vertices transmit forces, cell shapes encode mechanical energy, and target shape indices play the role of trainable parameters. This analogy naturally raises a central question: can we train a cellular tissue to realize a specified internal stress pattern, just as one trains a disordered spring network to a precise cellular stress pattern? Cellular packings, however, go beyond conventional spring networks in a crucial way in that they can {\it reconfigure}. Cells can exchange neighbors, and therefore, alter contact the topology of the multibody spring network. This structural plasticity introduces additional degrees of freedom absent in fixed-topology elastic networks and makes the learning landscape richer but also more complex. In this sense, cellular tissues share important parallels with reconfigurable particle systems, where contact networks evolve under driving and memory can be stored in rearrangements rather than solely in parameter values. Learning in a tissue therefore involves not only tuning effective mechanical parameters, but also navigating a space of possible cellular rearrangements, coupling stress propagation, topology change, and mechanical adaptation into a unified learning process. More specifically,  cellular rearrangements act as discrete transitions between mechanically distinct configurations, enabling the system to explore otherwise inaccessible regions of the cost function landscape. In the exploration versus exploitation sense, cellular rearrangements can perhaps be viewed as exploratory events~\cite{berger2014exploration}. 

The ability to train stress patterns in living tissues is not merely a theoretical curiosity - rather, it has profound biological and biomedical implications. Indeed, tissues are multiscale adaptive systems in which mechanics, cytoskeletal organization, nuclear structure, and chromatin state are tightly coupled across scales~\cite{Zhang2026}. Learning at the cellular and tissue levels may underlie phenomena such as habituation to repeated mechanical cues, anticipation of cyclic environments, and the formation of mechanochemical associations between distant regions of a tissue \cite{nasrollahi_past_2017}. Conversely, diseases such as fibrosis and cancer can be reframed as forms of {\it mislearning}, where maladaptive stress patterns and mechanochemical feedback loops become self-reinforcing. Developing algorithms that allow us to train, probe, and ultimately retrain stress patterns in tissues therefore provides a path toward decoding how tissues learn in health and how they might be {\it retrained} in disease.

\section{Energetics, Stresses, and Minimization Procedure}

\subsection{Cell packing energetics}
The cellular packing is represented by a 3D vertex model with periodic boundary conditions \cite{Honda2004,Okuda2013,Okuda2015,Tao_vertex_model}. Specifically, each cell is described as a polyhedron and shares faces with neighboring cells, i.e. confluent. The collection of cells is specified by positions of the vertices ${\vect{r_\mu}}$, where $\mu$ denotes the spatial indices. The mapping of vertices to edges, the mapping of edges to polygonal cell surface facets, and finally the mapping of the polygonal cell surface facets to the cells provide the link from the vertices to the cells. The energy of the $jth$ cell is then given by

\begin{align}
\label{eq:energy}
E_j = K_V\bracket{V(\{\vect{r_\mu,j}\}) - V_0}^2 
+ K_S\bracket{S(\{\vect{r_\mu,j}\}) - S_{0j}}^2,
\end{align} 
where $V(\{\vect{r_\mu,j}\})$, the volume of the polyhedron and $S(\{\vect{r_\mu,j}\})$, the surface area of the polyhedron, are both  functions of $\{\vect{r_\mu}\}$, the set of position vectors of polyhedron vertices; $V_0$ are the $S_0$ are the target volume and target surface area, respectively; and $K_V$ and $K_S$ are volume and surface stiffnesses, respectively. The total energy of the cell packing is the sum of the individual cell energies. Physically, the volume term represents the bulk elasticity of the cell with $V_0$ denoting a target volume, while the area term represents the isotropic contractility of the acto-myosin cortex. The larger the $S_{0}$, the less
 isotropically contractile the cell is, and vice versa. The contractility is isotropic in the sense that all faces are equivalent.

\subsection{Cellular Stress}
As we will be training for cell stress patterns, it behooves us to compute cellular stresses. Luckily, given the cellular resolution of the total energy in Eq.~\ref{eq:energy}, we have recently derived a cell stress tensor $\vect{\sigma}$ as (~\cite{Ameen2026_stress_tensor}) 

\begin{align}\label{stress_tensor_final}
    \vect{\sigma} 
    =& 
    -2k_V(V-V_0)\mathbb{I}
    - \frac{2k_S(S-S_0)}{V}
    S \mathbb{I}
    \nonumber\\
    &
    + \frac{2k_S(S-S_0)}{V}
    \sum_\mathcal{F} 
    \frac{1}{\norm{\vect{S}_\mathcal{F}}}
    \bracket{
    \vect{S}_\mathcal{F} 
    \otimes 
    \vect{S}_\mathcal{F}
    }, 
\end{align}
where $\vect{S}_\mathcal{F}$ denotes the area vector for a triangular subsection $\mathcal{F}$ on a surface polygon of the cell. Thus, the sum in the third term is taken over a triangulation of the cell surface. 

This cell stress tensor emerges naturally from the Cauchy stress tensor defined microscopically in terms of forces as $\boldsymbol{\sigma} = \frac{1}{V} \sum_\mu \mathbf{r}_\mu \otimes \mathbf{F}_\mu$, which can be interpreted as a virial-like expression capturing how forces on each vertex ($\mathbf{F}_\mu$) are transmitted across a cell. Substituting the force expression reveals that the stress tensor decomposes into a contribution from volume elasticity and surface elasticity, with the latter depending explicitly on the geometry of cell faces. The third term in the above equation arises from the anisotropic contribution of the surface-area elasticity to the cell stress tensor, and can be traced directly to how changes in vertex positions modify the orientation of individual faces rather than just their total area. Specifically, starting from the surface area term in the energy, the force on each vertex involves gradients of the face areas, $\partial \| \mathbf{S}_F \| / \partial \mathbf{r}_\mu$, which depend not only on the magnitude of each face area vector $\mathbf{S}_F$, but also on its direction. When constructing the stress tensor, one must therefore evaluate sums of the form $\sum_{\mu \in F} \mathbf{r}_\mu \otimes \partial \|\mathbf{S}_F\|/\partial \mathbf{r}_\mu$. Using the geometric identity
\begin{align}
\sum_{\mu \in F} \mathbf{r}_\mu \otimes \frac{\partial \|\mathbf{S}_F\|}{\partial \mathbf{r}_\mu}
= \|\mathbf{S}_F\| \mathbb{I} - \frac{1}{\|\mathbf{S}_F\|} \mathbf{S}_F \otimes \mathbf{S}_F,
\end{align}
the surface contribution naturally splits into an isotropic part and a directional correction involving $\mathbf{S}_F \otimes \mathbf{S}_F$. Physically, this term encodes the fact that surface tension acts along the local tangent planes of each face, leading to directional stresses aligned with face normals. In contrast to the volume term (purely isotropic) and the scalar surface-area term (isotropic after summation), this contribution retains information about cell geometry and orientation, thereby capturing how anisotropic cell shapes generate anisotropic stress distributions. 

Several scalar quantities, each carrying some information about the stress of the cell, can be defined from the eigenvalues of this tensor. In particular, if $\sigma_1$, $\sigma_2$, $\sigma_3$ are the eigenvalues of the cellular stress tensor in ascending order, the maximum shear stress $\sigma$ can be defined as 

\begin{align}\label{maximum_shear_stress}
    \sigma_{max} = \frac{1}{2}(\sigma_3-\sigma_1).
\end{align}
It is this cellular scalar quantity that we will elevate to a trainable quantity. 

\subsection{Hybrid Langevin-FIRE minimization}

For computational ease, we work with the following dimensionless version of the total cell packing energy or $e=E/(K_A V_0^{4/3})$: 
\begin{align}
\label{eq:non_dimensionalized_energy}
e= \kappa\sum_{j}(v_{j} - 1)^2+\sum_{j}(s_{j} -s_{0j})^2 
\end{align}
with $\kappa= K_V/(K_A V_0^{4/3})$, $v=V/V_0$ and $s=S/V_0^{2/3}$. Moreover, $s_0 = \frac{S_0}{V_0^{2/3}}$ is the target shape index.
We note in the model defined by Eqn. \ref{eq:non_dimensionalized_energy} that the target shape index for each cell has been made individually addressable. 

For each instance of our system we first generate a non-degenerate Voronoi tessellation of a periodic cube with sides of integer length l; with one randomly distributed seed per unit volume, each system has $l^3$ total cells. This randomly generated configuration is energy optimized using a hybrid overdamped Langevin-FIRE minimization technique discussed below. The resulting minimized configuration is used as an initial configuration for training.

Our vertex model instances are generally initialized on periodic cubes with sides of length $l = 4,5,6$ (in the natural length units of the system). Thus, cellular packings with $N=64,128,256$ cells have been investigated.
The pretrained state of the system is then obtained by initializing the target shape indices of all cells and energy minimizing the corresponding configuration.  We use a  homogeneous assignment of initial target shape indices: $\{s_0^i\}_{i=1}^{l^3} = 5.0$. A number of alternative initial assignments for target shape indices have been considered, including bi-disperse distributions, random normal distributions, and other uniform values, but  generally found to impact negligible variation in resulting performance. We note this assignment corresponds to an initially rigid system. We set the ratio of volume stiffness to surface area stiffness $\kappa = 10$, to model the relative volume incompressibility of real cells. Results for $\kappa = 1$ are presented in the Supplementary Figures.

For a 3D vertex model configuration with $N_v$ vertices, the root mean square dimensionless force $f_{rms}$ is defined as
\begin{align}
    f_{rms} = \sqrt{\frac{1}{3N_V}\sum_{j,\mu} \abs{f_\mu,j}^2}.
\end{align}
Our minimization algorithm aims find a configuration, where $f_{rms}$ below minimization tolerance $\tau_{min} = 10^{-6}$. The minimization algorithm is designed to find any local energy minimized state and it does not discriminate between local minima with different energy values.

FIRE minimization \cite{Bitzek_2006, Bitzek2007} is a useful energy minimization scheme for molecular dynamics. Starting from a non-minimized state, the FIRE algorithm finds a local minimum by changing the state using an equation of motion with modified acceleration. This works so long as energy is differentiable along the path taken in state variables from the initial state to the minimized state. However, if the energy is not differentiable, this can lead to complications. In the vertex model, reconnection events are treated as discrete events with a finite discontinuity in energy. This challenges a straightforward application of the FIRE minimization algorithm on the randomly generated initial condition, since we generally expect a large number of reconnection events for the system to get close to some local minimum in energy.

We resolve this issue by repeating a two-step protocol, which we refer to as {\it hybrid Langevin-FIRE minimization}. In the first step the system undergoes overdamped motion with added Gaussian noise, or Langevin dynamics~\cite{Tao_vertex_model}. A counter is reset at the start of every checking interval  of $n_{OD}$ iterations of the overdamped Brownian motion. The counter stores the number of reconnection events in the $n_{OD}$ iterations. As overdamped motion dissipates energy and the system  approaches a local minimum in energy, there are fewer reconnection events every checking interval. The overdamped stage is exited when no further reconnection events occur in one checking interval.

In the second stage, $i_{max}$ iterations of the usual FIRE minimization algorithm are assigned for the resulting configuration. The second state is exited successfully if $f_{rms}$ of the configuration is reduced below tolerance $\tau_{min}$. If minimization fails after $i_{max}$ iterations of the usual FIRE minimization, it is likely due to insufficient energy dissipation via overdamped motion in the first stage. Accordingly, the two step process is repeated until a energy minimum is found. We note that if a configuration is already approaching a local energy minimum - for example, a perturbed minimized configuration  - we can start from the second stage; that is, start with the usual FIRE minimization algorithm directly to find the corresponding energy minimized configuration. Generally, the hybrid algorithm should yield the same minimized configuration in such a case.


\begin{figure}[htbp]
\centering
\begin{minipage}{0.95\columnwidth}
\hrule
\hrule
\vspace{0.5em}
\textsc{\textbf{Algorithm 1} Hybrid Langevin-FIRE minimization}
\vspace{0.5em}
\hrule
\begin{algorithmic}[1]
\Require $TOL$, $n_{\mathrm{OD}}$, $i_{\max}$
\State Evaluate $f_{\mathrm{rms}}$
\While{$f_{\mathrm{rms}} > TOL$}
  \Repeat
    \State $R_{\mathrm{count}} \gets 0$
    \State Perform $n_{\mathrm{OD}}$ iterations of Langevin dynamics
    \State $R_{\mathrm{count}} \gets N$\Comment{Number of reconnections}
  \Until{$R_{\mathrm{count}} = 0$}
  \For{$i = 1$ to $i_{\max}$}
    \State Perform FIRE minimization iteration
    \State Evaluate $f_{\mathrm{rms}}$
    \If{$f_{\mathrm{rms}} < TOL$}
      \State \textbf{break}
    \EndIf
  \EndFor
\EndWhile
\end{algorithmic}
\vspace{0.5em}
\hrule
\end{minipage}
\end{figure}

\section{Training Algorithm and Measurements}

\subsection{Algorithm}
We now group the cells into three distinct populations:
\begin{enumerate}
    \item Target Cells: For a training task we assign $N_T$ target cells with initial target shape index assignments $\{s_0^T\}$. Maintenance of this assignment of target shape indices shall be referred to as the {\it free state} of the system. In contrast, a {\it clamped state} is achieved by changing $\{s_0^T\}$. The training task seeks to achieve target maximum shear stresses $\{\sigma^T\}$ on the target cells in the free state of the system up to some tolerance.
    \item Hidden Cells: These cells have initial target shape index assignments $\{s_0^H\}$. Training involves changing these assignments using a learning rule. In other words, $\{s_0^H\}$ constitute the learning, or modifiable, parameters of the system.
    \item Passive Cells:  These cells have initial target shape index assignments $\{s_0^P\}$, which will remain unchanged in all stages of the training algorithm. That is, while these cells appear in the total energy of the system, they are not part of the training algorithm. We note that the set of passive cells may be empty.  
    
\end{enumerate}

As we will choose some fraction of cells to train, the loss function must encode the task of training cells to some target maximum shear stress.  Such a loss function can be defined as the average absolute difference between the free state stresses of the target cells $\sigma_{free}^T$ and the target stress values $\sigma^{T}$:

\begin{align}
C = \frac{1}{N_T}\sum_{T}\abs{\sigma^{T} - \sigma_{free}^T}
\end{align}

To achieve such a task, we present below an adaptation of a contrastive learning algorithm appropriate for our cellular packing. Note that we will use $\sigma$ from now on to denote the maximum shear stress for a cell.

\begin{figure}[htbp]
\centering
\begin{minipage}{0.95\columnwidth}
\hrule
\hrule
\vspace{0.5em}
\textsc{\textbf{Algorithm 2} Contrastive Learning Algorithm For Cellular Packings}
\vspace{0.5em}
\hrule
\begin{algorithmic}[1]
\Require  $TOL$, $\{\sigma^{T}\}$, $\{s_0^T\}$,$\{s_0^H\}$,$\{s_0^P\}$, $\gamma$
\Ensure {$C<TOL$}
\State  $C \gets C(\{\sigma^T\},\{\sigma^T_{free}\})$
\While {$C>TOL$}
\State Calculate free state shape indices $\{s_{free}^{H}\}$
\State Clamp target cells using the clamping subroutine
\State Calculate clamped shape indices $\{s_{clamped}^{H}\}$
\State $s_0^{H} \gets s_0^{H} + \gamma(s_{free}^{H}-s_{clamped}^{H})$
\State Unclamp cells (by resetting $\{s_{0}^{T}\}$)
\State Minimize configuration to obtain free state
\State  $C \gets C(\{\sigma^T\},\{\sigma^T_{free}\})$
\EndWhile
\end{algorithmic}

\vspace{0.5em}
\hrule
\end{minipage}
\end{figure}


\begin{figure*}[t]
\centering
    \begin{subfigure}{0.19\linewidth}
        \includegraphics[width=\linewidth]{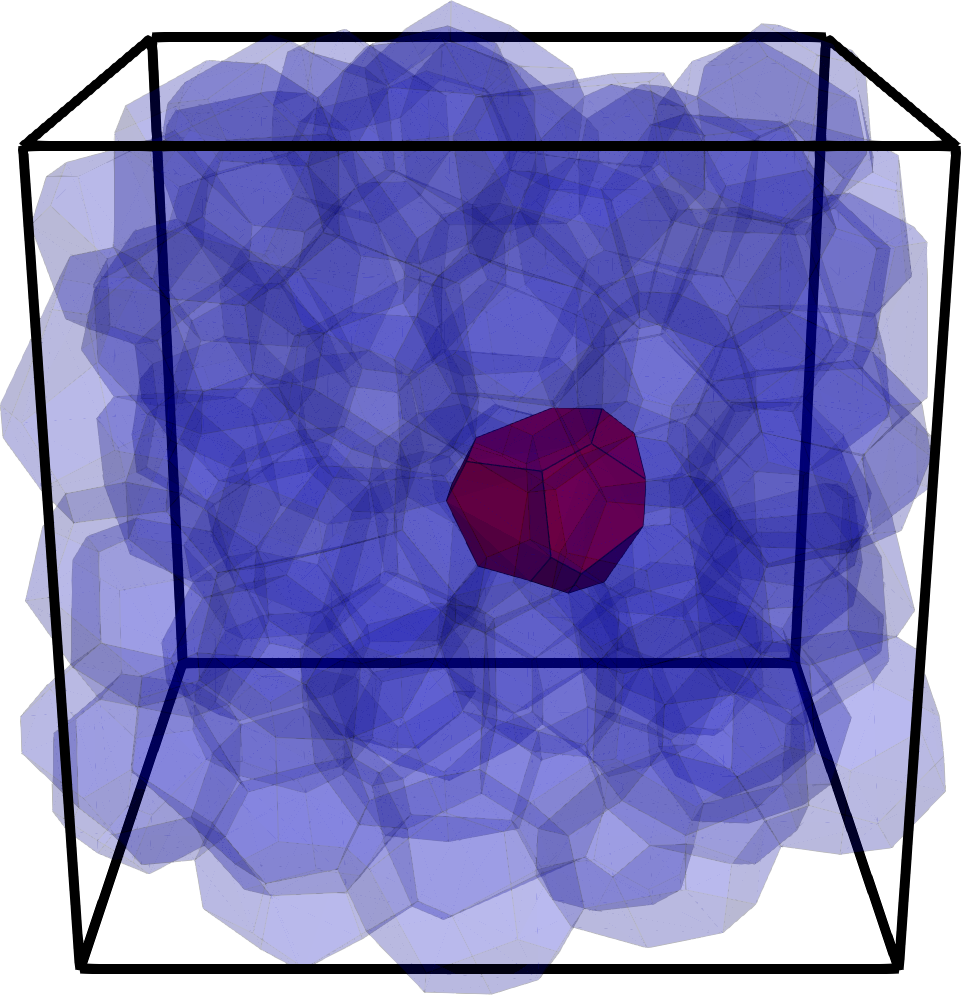}
        \caption{}
        \label{fig:1a}
    \end{subfigure}
    \begin{subfigure}{0.19\linewidth}
        \includegraphics[width=\linewidth]{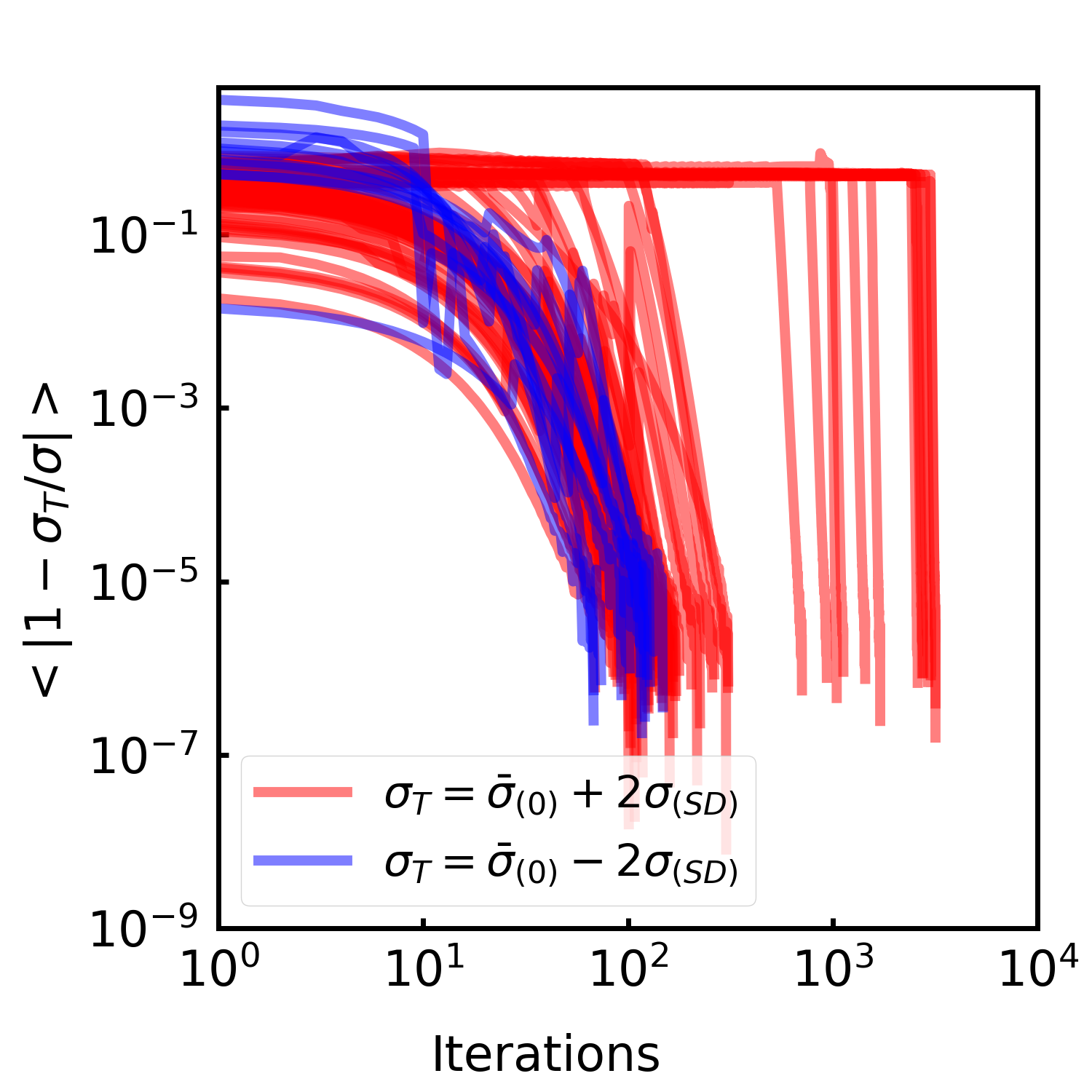}
        \caption{}
        \label{fig:1b}
    \end{subfigure}
    \begin{subfigure}{0.19\linewidth}
        \includegraphics[width=\linewidth]{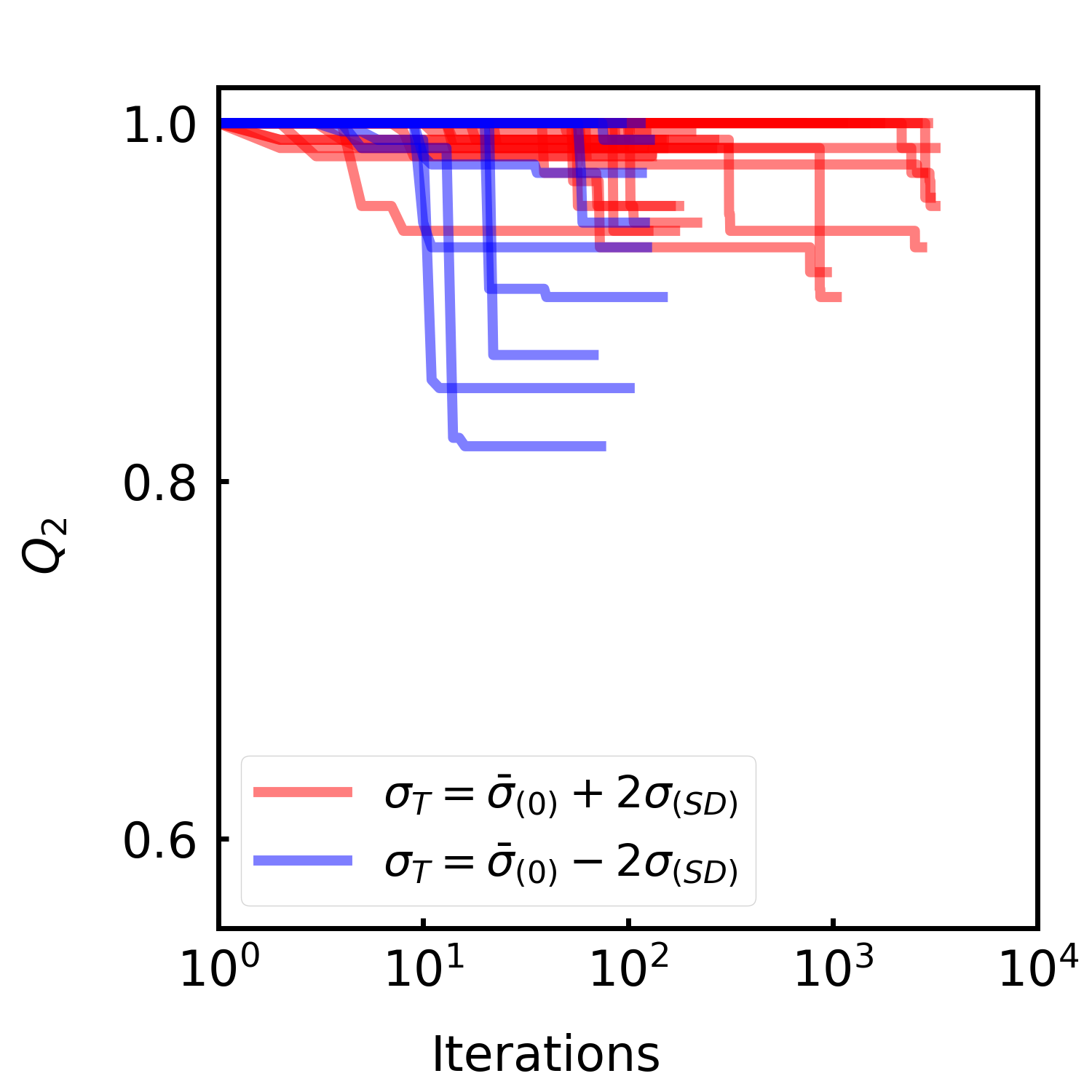}
        \caption{}
        \label{fig:1c}
    \end{subfigure}
    \begin{subfigure}{0.19\linewidth}
        \includegraphics[width=\linewidth]{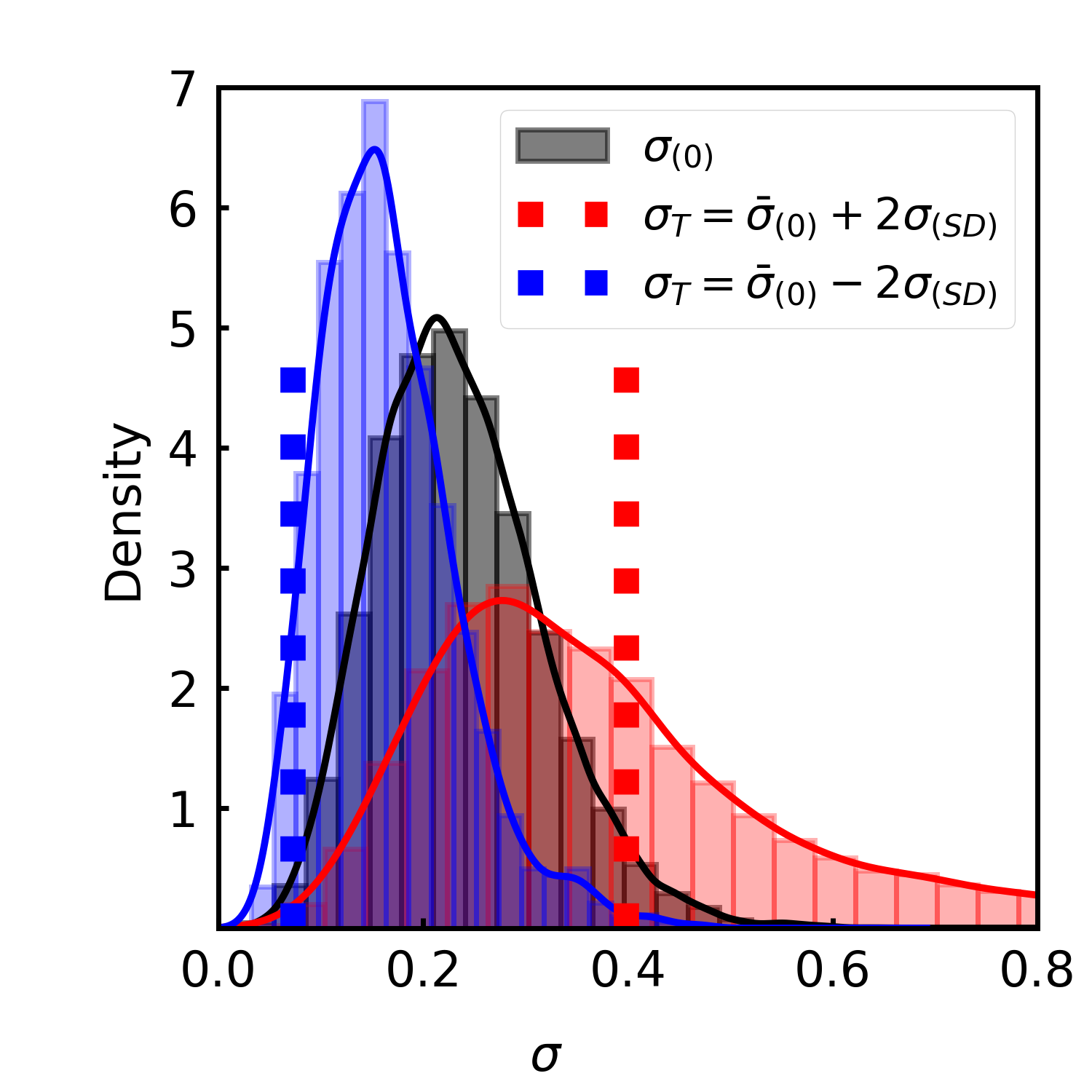}
        \caption{}
        \label{fig:1d}
    \end{subfigure}
    \begin{subfigure}{0.19\linewidth}
        \includegraphics[width=\linewidth]{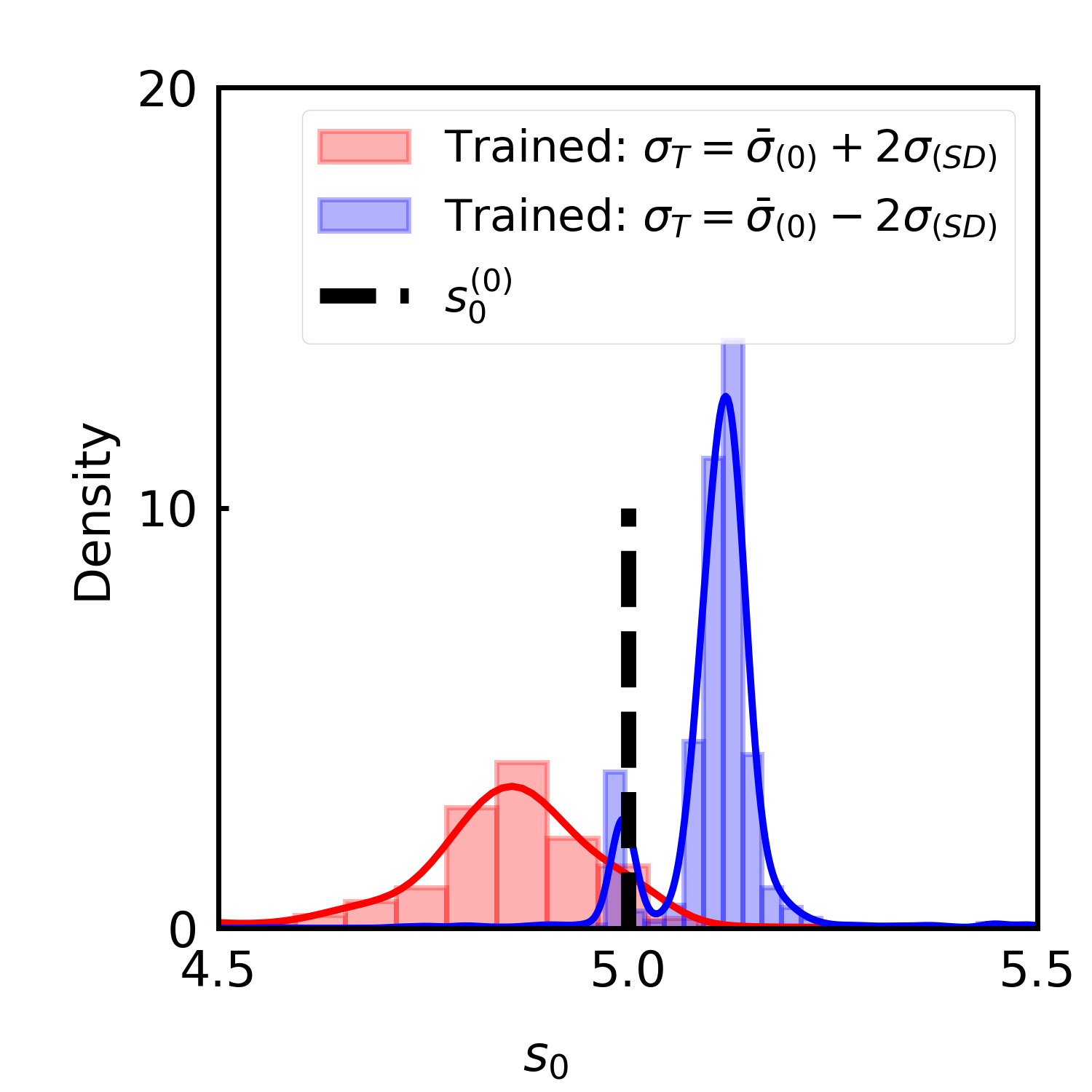}
        \caption{}
        \label{fig:1e}
    \end{subfigure}
    \vfill
    \begin{subfigure}{0.19\linewidth}
        \includegraphics[width=\linewidth]{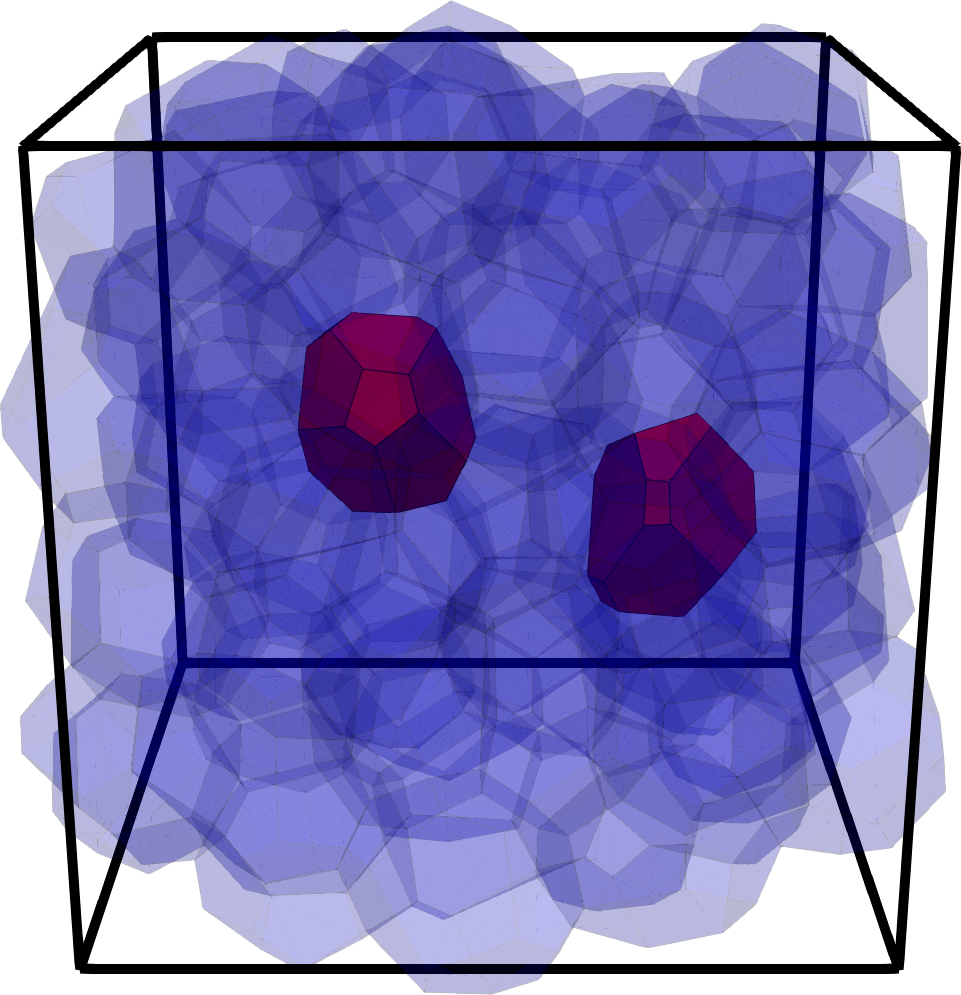}
        \caption{}
        \label{fig:1f}
    \end{subfigure}
    \begin{subfigure}{0.19\linewidth}
        \includegraphics[width=\linewidth]{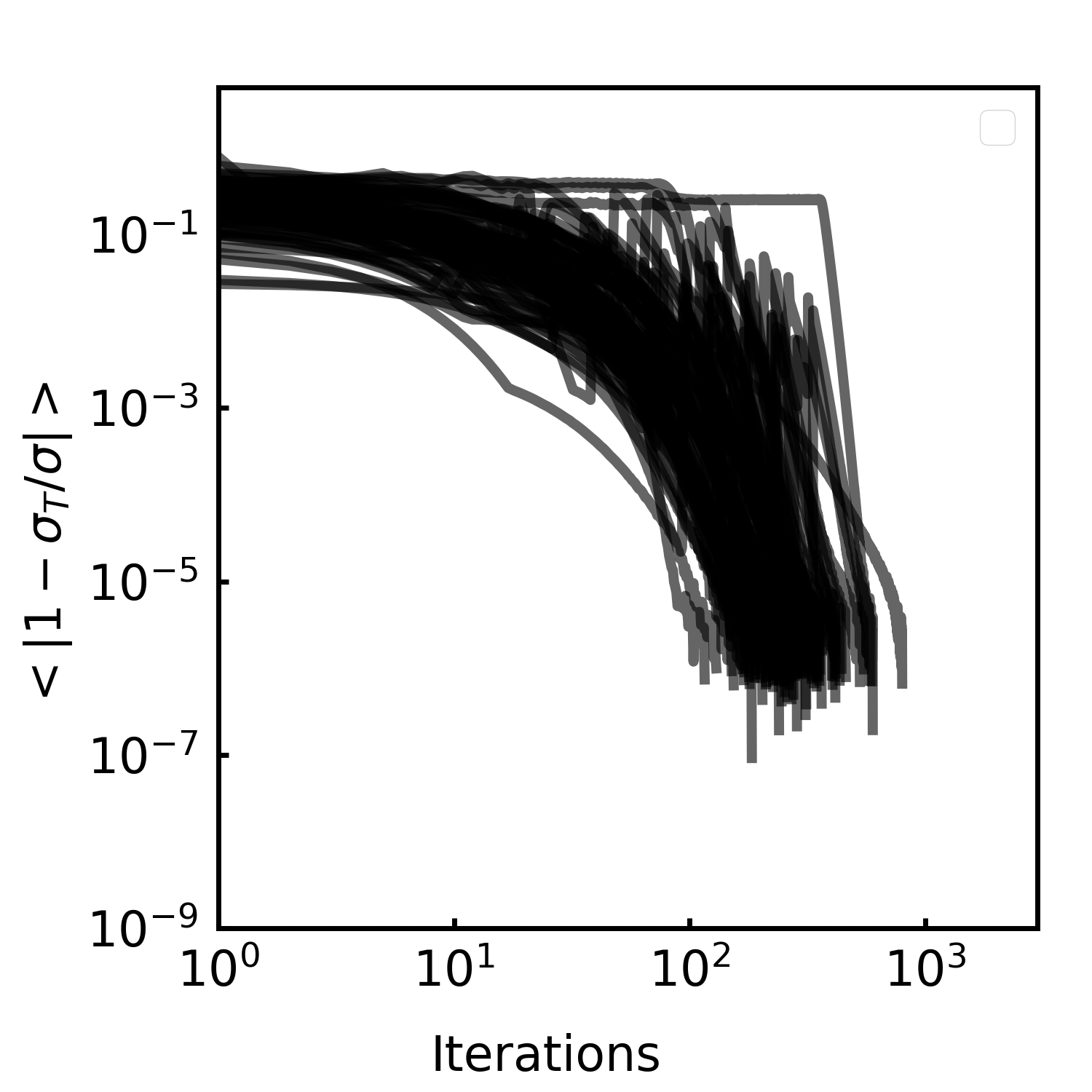}
        \caption{}
        \label{fig:1g}
    \end{subfigure}
    \begin{subfigure}{0.19\linewidth}
        \includegraphics[width=\linewidth]{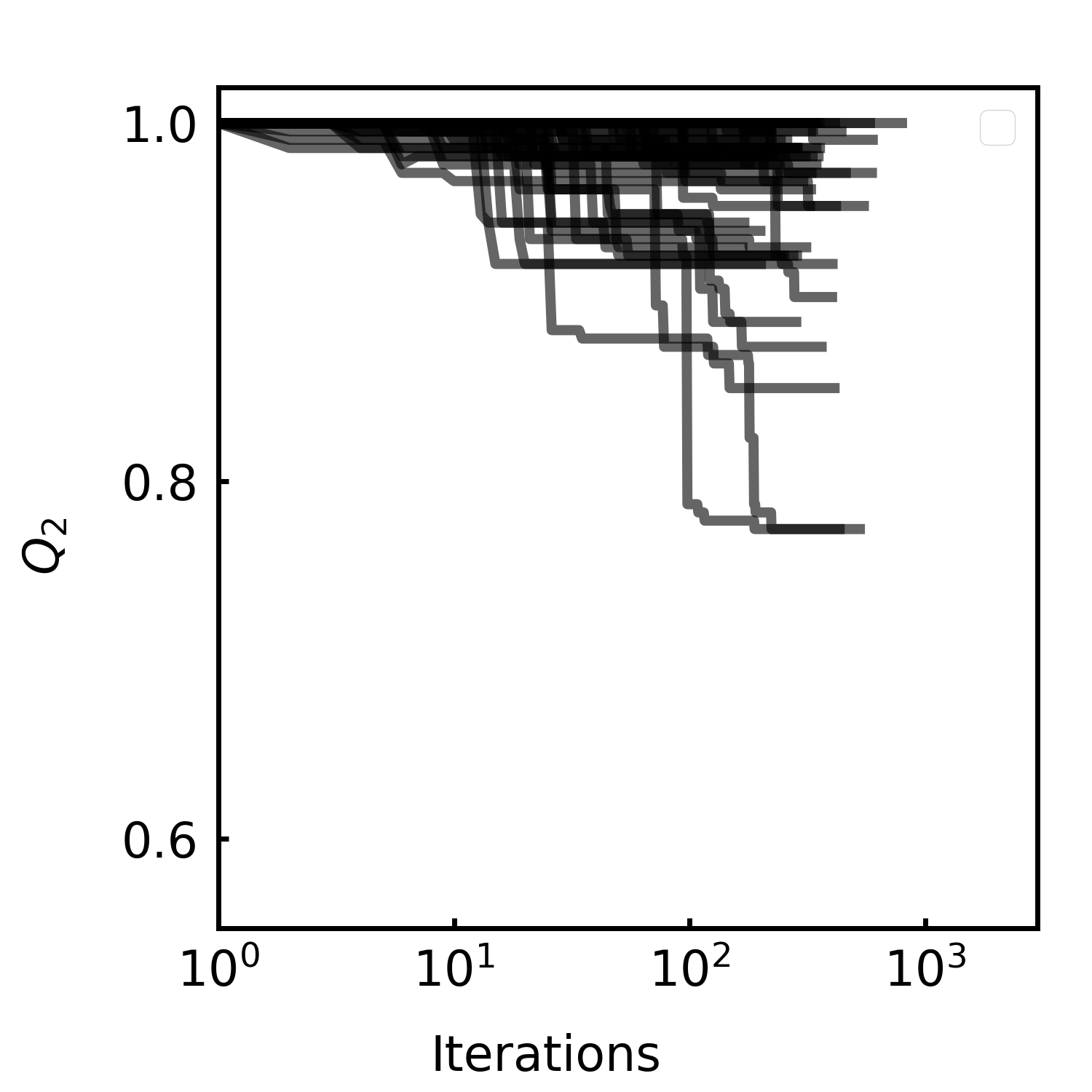}
        \caption{}
        \label{fig:1h}
    \end{subfigure}
    \begin{subfigure}{0.19\linewidth}
        \includegraphics[width=\linewidth]{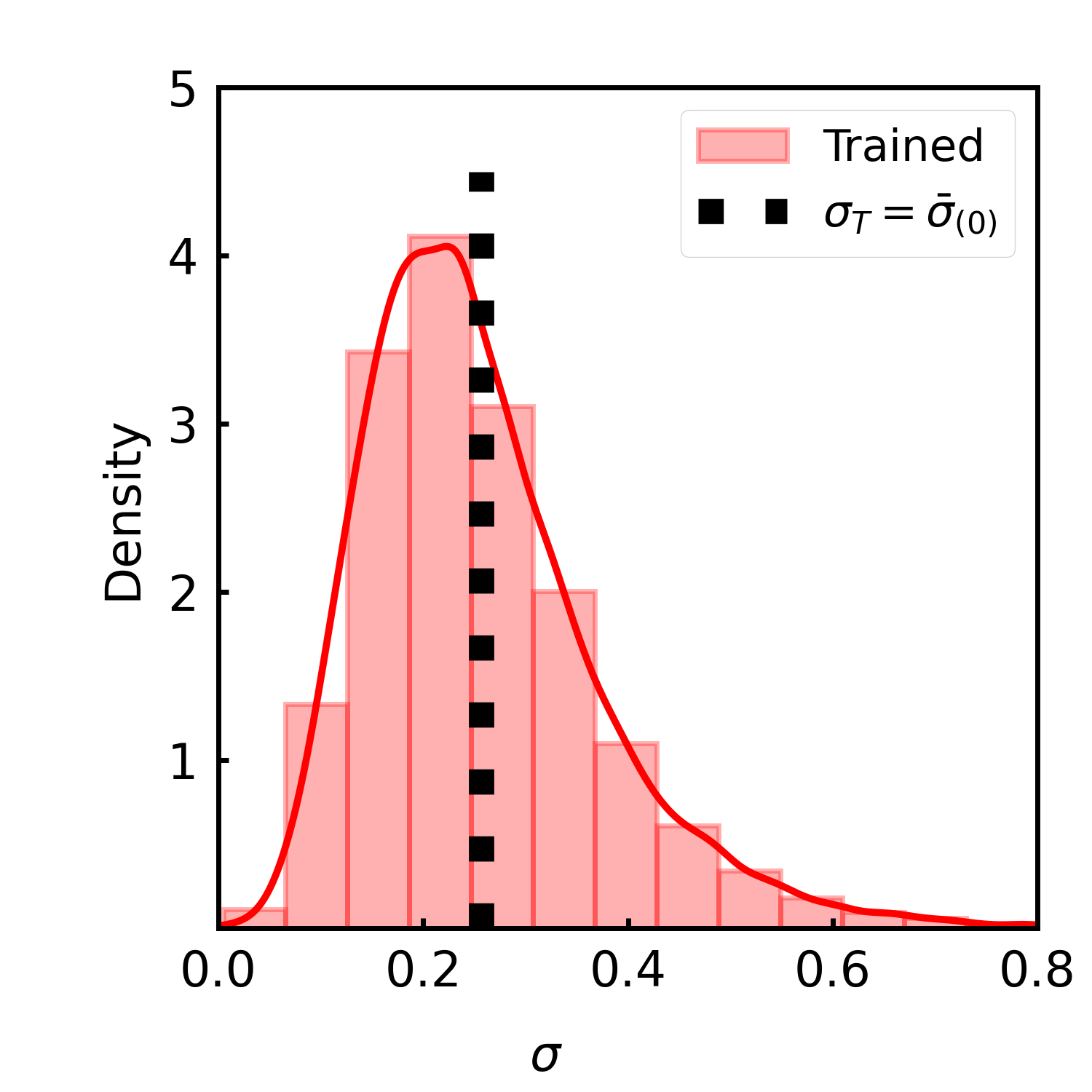}
        \caption{}
        \label{fig:1i}
    \end{subfigure}
    \begin{subfigure}{0.19\linewidth}
        \includegraphics[width=\linewidth]{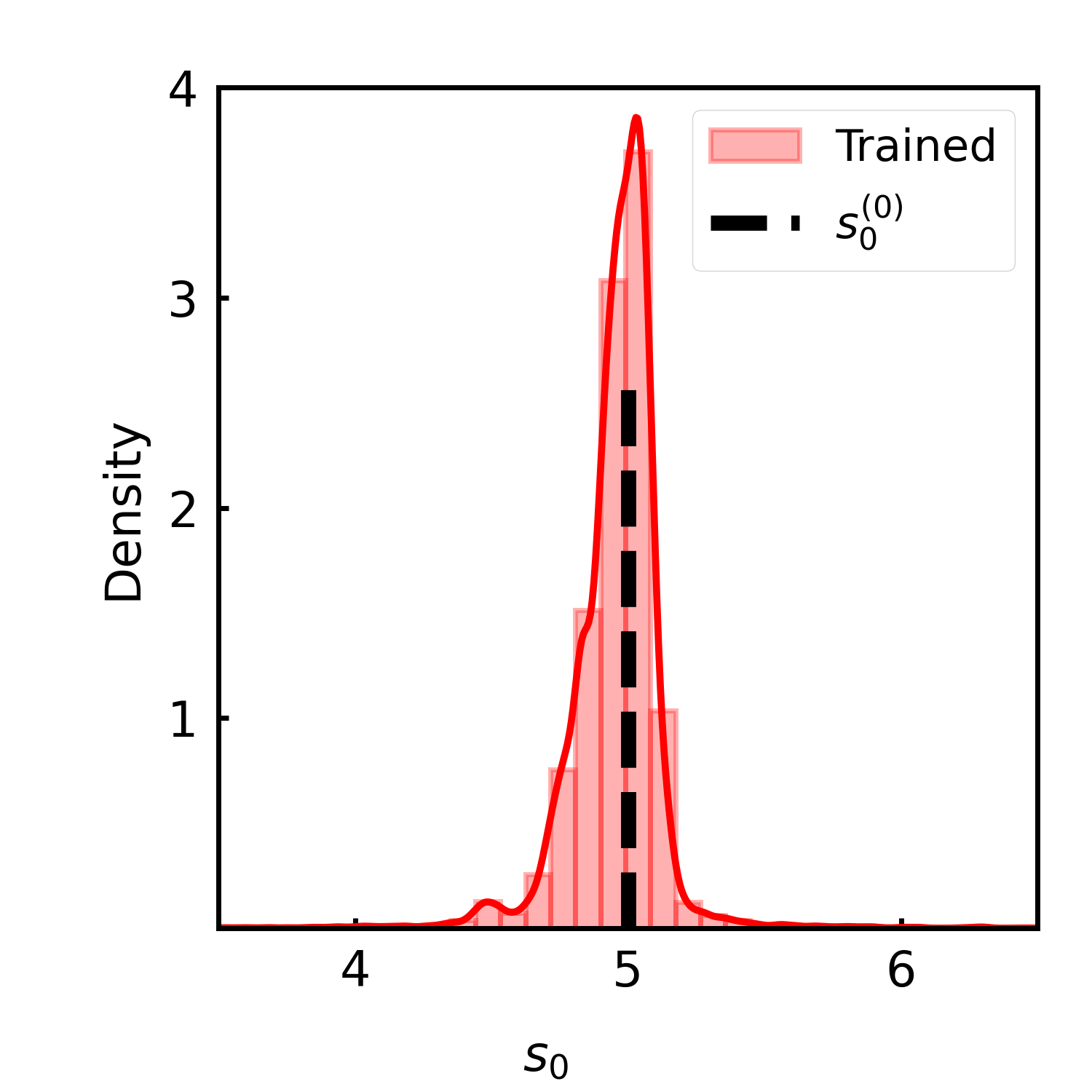}
        \caption{}
        \label{fig:1j}
    \end{subfigure}
    \vfill
    \begin{subfigure}{0.19\linewidth}
        \includegraphics[width=\linewidth]{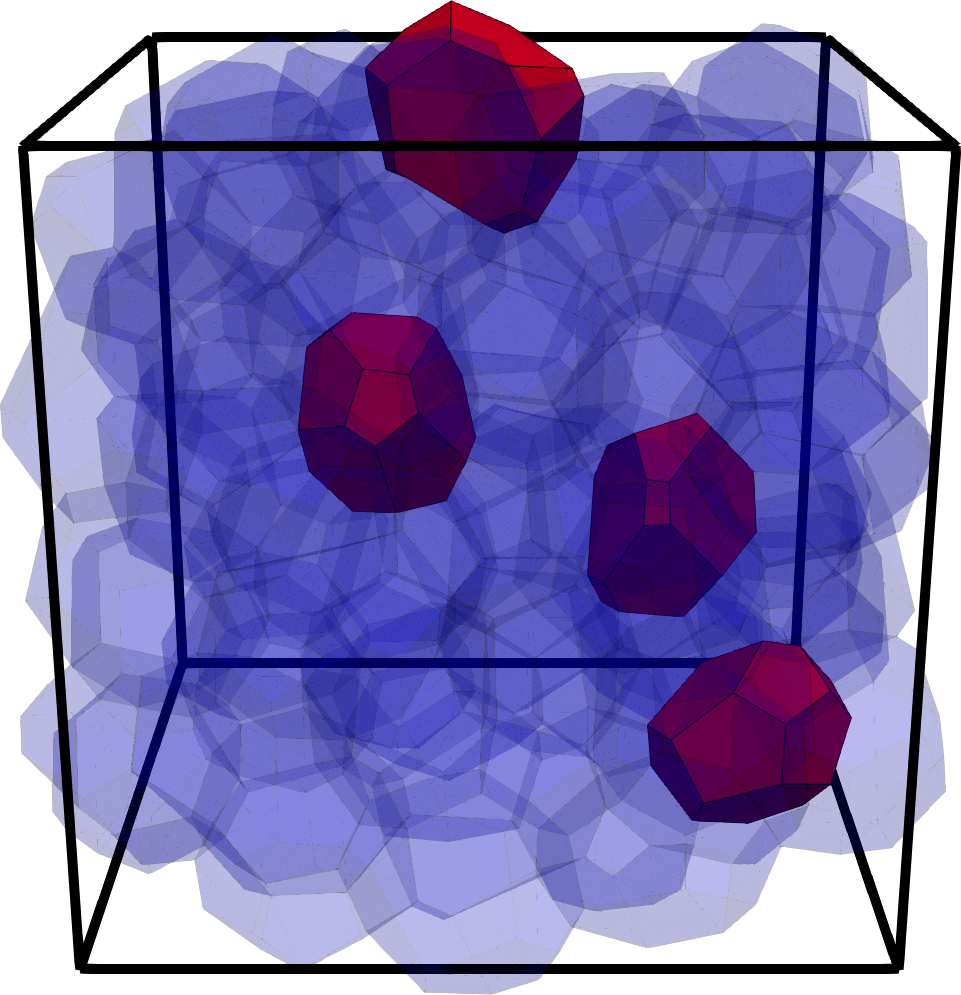}
        \caption{}
        \label{fig:1k}
    \end{subfigure}
    \begin{subfigure}{0.19\linewidth}
        \includegraphics[width=\linewidth]{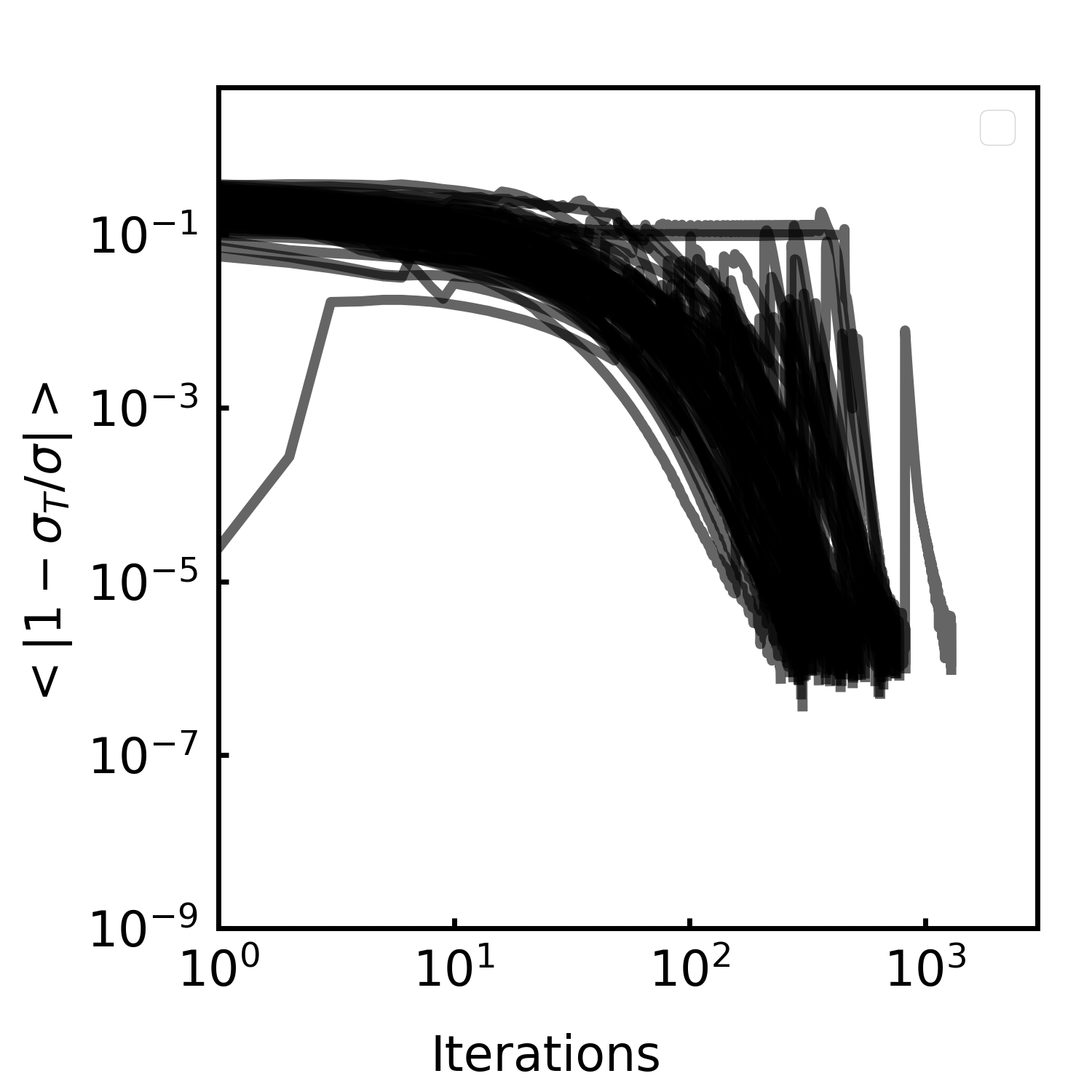}
        \caption{}
        \label{fig:1l}
    \end{subfigure}
    \begin{subfigure}{0.19\linewidth}
        \includegraphics[width=\linewidth]{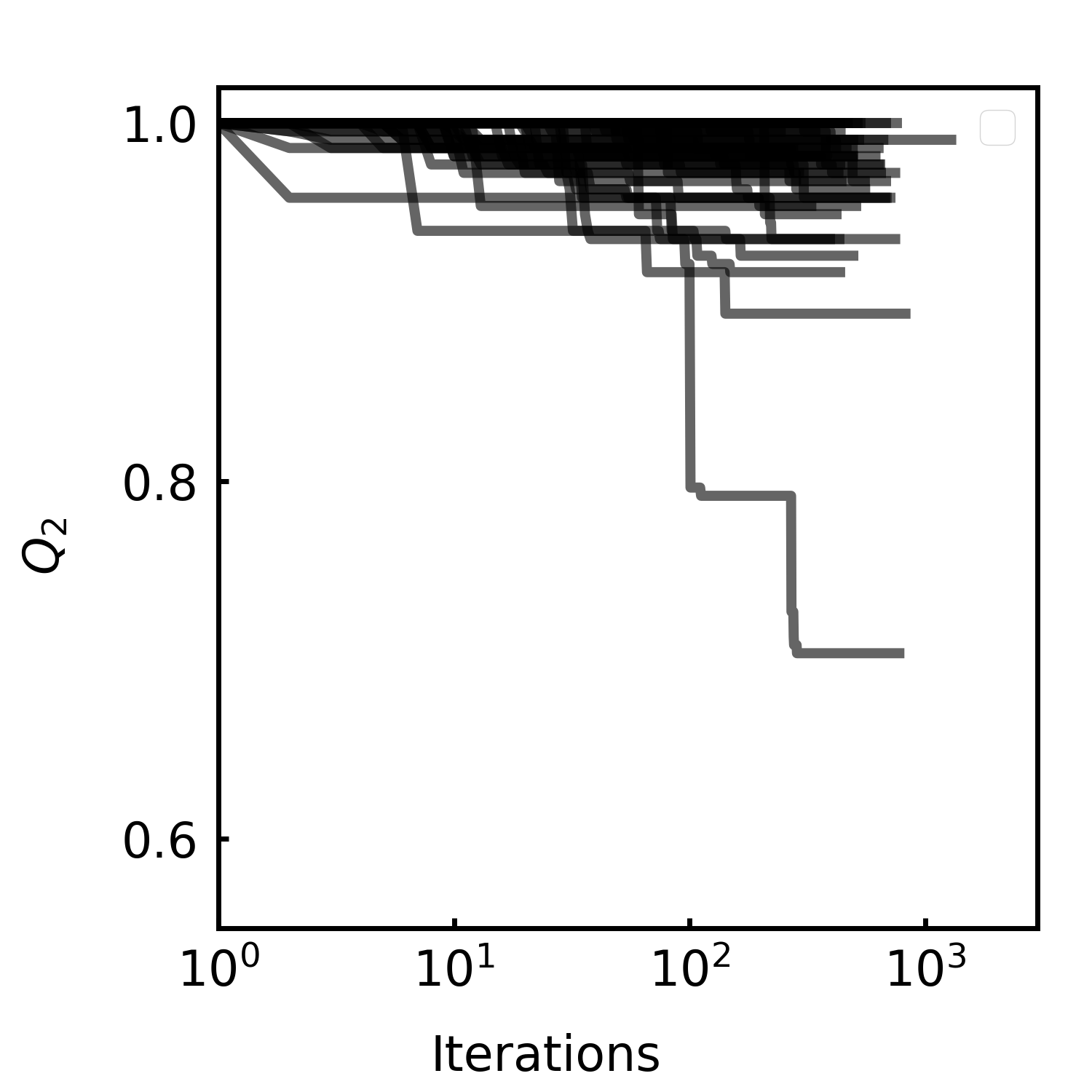}
        \caption{}
        \label{fig:1m}
    \end{subfigure}
    \begin{subfigure}{0.19\linewidth}
        \includegraphics[width=\linewidth]{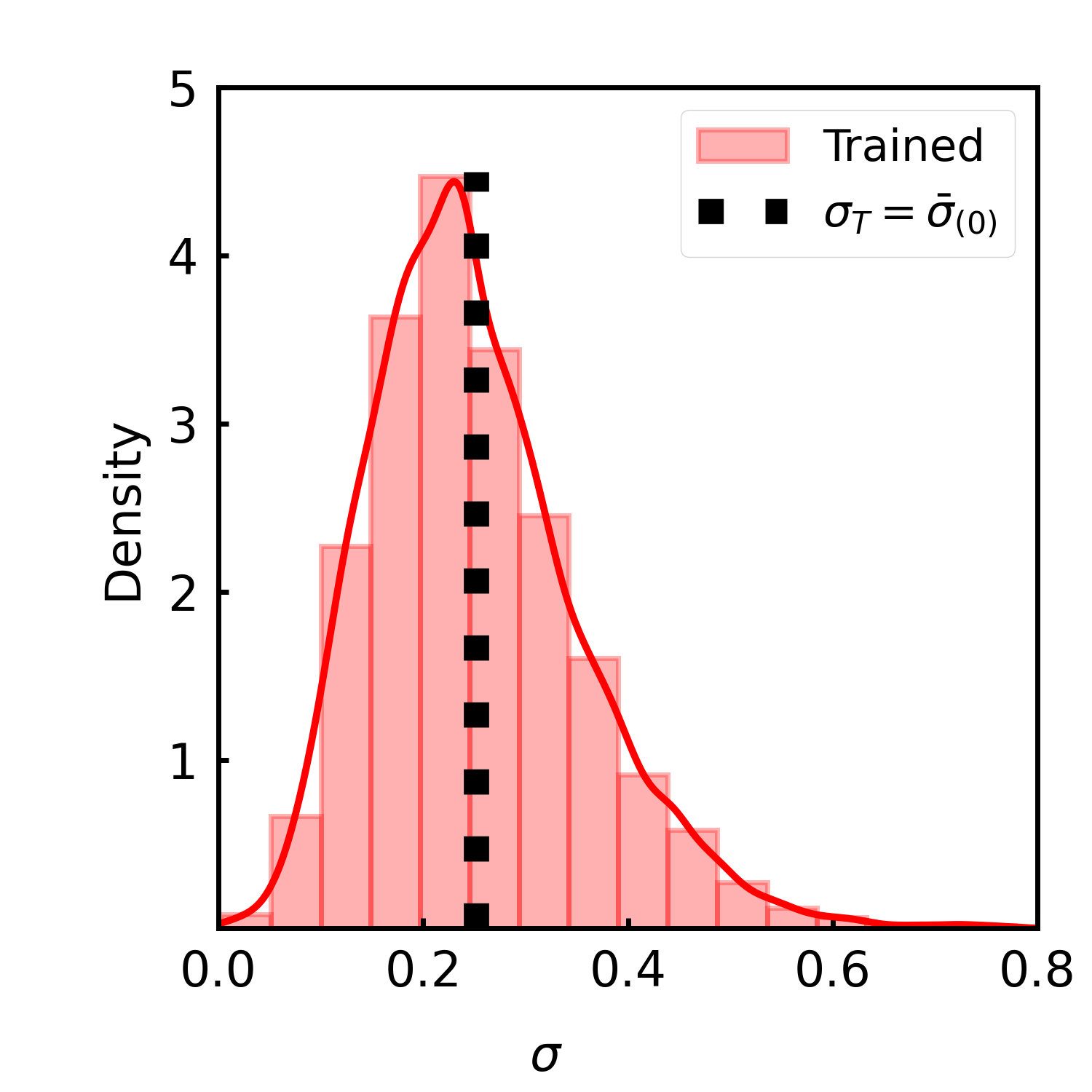}
        \caption{}
        \label{fig:1n}
    \end{subfigure}
    \begin{subfigure}{0.19\linewidth}
        \includegraphics[width=\linewidth]{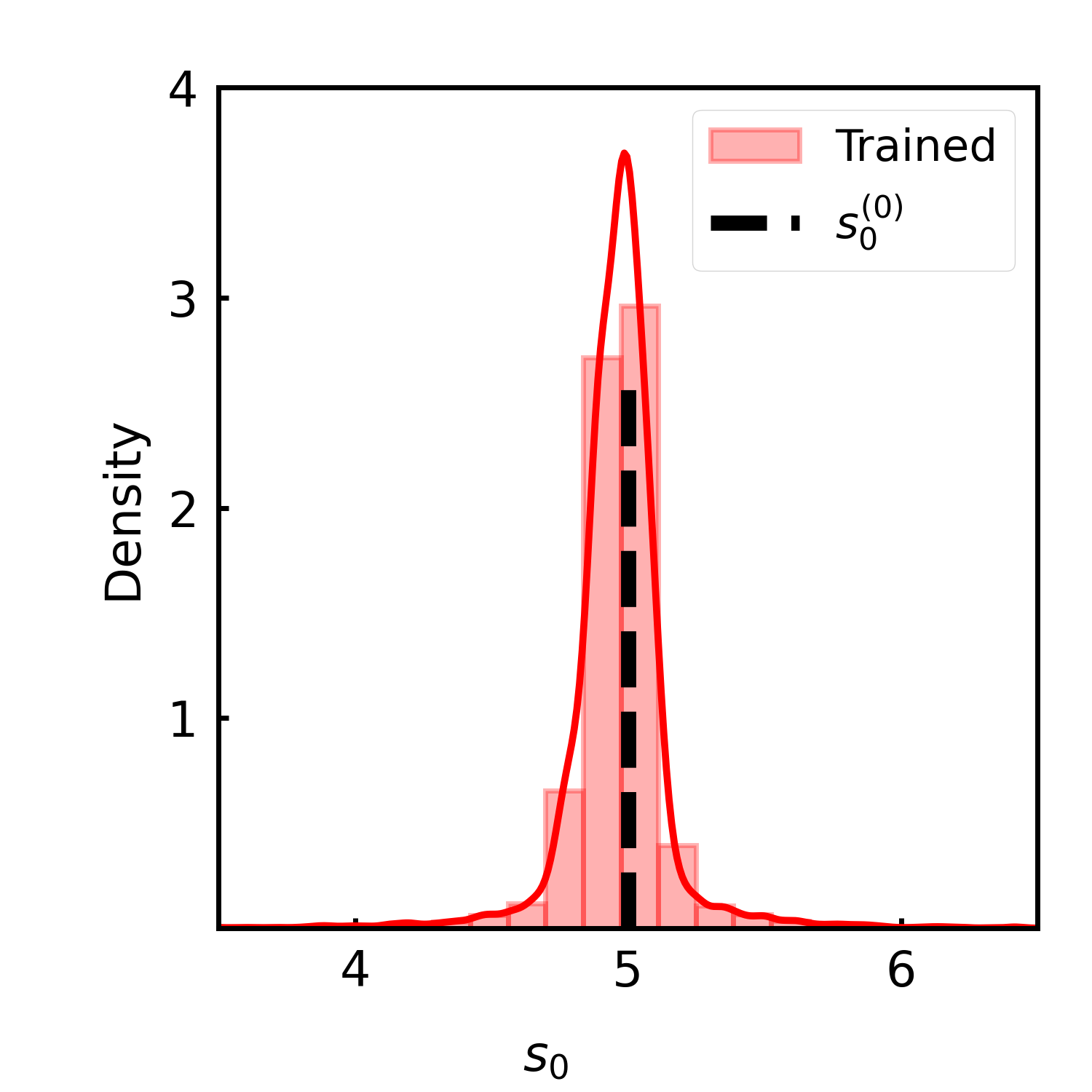}
        \caption{}
        \label{fig:1o}
    \end{subfigure}
    \caption{{\it Learning single-cell stresses reveals asymmetry, global coordination, and slower lowering with more target cells.} We train randomly chosen target cells on 100 initial configurations with 216 total cells and periodic boundary conditions. (a)-(e) Training one target cell to high or low target stresses, defined as 2 standard deviations above (red) and below (blue) the initial mean stress. (b) The error in cell stress with respect to the target stress is tracked with training iterations for every successful training instance for both these single cell training tasks. (c) The overlap of the topology of the cell packing with the initial configuration, tracked for each successful training instance, showing that stress reduction task generally needs more fluidization.  (d) Cell stress histograms, computed from cells in all successfully trained configurations for both task, showing the shift of the histogram towards the corresponding high (red) or low (blue) target stress value, marked with red and blue vertical lines, respectively. (e) Target cell shape index $(s_0)$ histograms with the initial $s_0=5.0$ (dashed black line),  computed from cells in all successfully trained configurations for the high and low target stress, showing a corresponding shift of the system towards rigidification and fluidization respectively. (f)-(j) Training 2 target cells to the initial mean stress with the corresponding error and overlap as a function of training iterations, followed by the final stress distribution of the hidden cells and the corresponding distribution of the hidden parameters. (k)-(o) The same as (f)-(j) but with $N_T=4$. Overall, increasing $N_T$ leads to slower learning, greater variability, and broader hidden-parameter distributions, reflecting the increasing difficulty of satisfying multiple constraints simultaneously.}
  \label{fig:1}
\end{figure*}

The main training algorithm seeks target maximum shear stresses on a specified set of target cells by iteratively updating the target shape indices of the hidden cells. First, the loss
function defined above is evaluated in the free state, where the target cells retain their original target shape indices. If the loss exceeds the prescribed tolerance, the hidden-cell shape indices in the free state, denoted by \(\{s^{H}_{\mathrm{free}}\}\), are recorded. The target cells are then clamped using the clamping subroutine discussed below so that their stresses are driven toward the prescribed target values, and the corresponding hidden-cell shape indices in this clamped state, \(\{s^{H}_{\mathrm{clamped}}\}\), are measured.

The hidden-cell target shape indices are then updated according to the contrastive learning rule
\begin{align}
\Delta s^{H}_{0}=\gamma\left(s^{H}_{\mathrm{free}}-s^{H}_{\mathrm{clamped}}\right),
\end{align}
where \(\gamma\) is the learning rate. After this update, the target cells are unclamped by resetting their target shape indices to their original values, and the full cellular configuration is re-minimized. The loss is then re-evaluated in the new free state. This cycle of free-state evaluation, clamping, contrastive update, unclamping, and minimization is repeated until the loss falls below the desired tolerance. In words, the algorithm compares the hidden-cell states in the free and clamped configurations and uses that contrast to adjust the hidden-cell target shape indices so that, after relaxation, the free state of the tissue better realizes the desired target-cell stress pattern. The overall motivation is similar to that of contrastive learning with spring networks. Namely, we apply changes to the clamped state of each hidden cell that help nudge it closer to the free state counterpart. Thus, when the clamped surface area is smaller than the free dimensionless surface area of a hidden cell, we increase $s_0$.  Conversely, if the clamped dimensionless surface area exceeds the free state area, we should reduce $s_0$.

\textbf{Clamping Subroutine}: In a spring network, the clamping of a linear spring (such that it obtains a particular clamped stress) can be achieved by solving for the corresponding spring length and fixing or ``clamping'' the nodes of the spring by solving a linear equation, should the deformations remain in the linear regime. In the vertex model, solving for vertex positions as a function of maximum shear stress is less straightforward. Instead, the clamping of a target cell to a target value of maximum shear stress is realized as an iterative subprocess within each iteration of the overall training algorithm. For each iteration of the clamping subprocess, the $\{s_0^{T,clamped}\}$ which produces the target clamped stress for the cell, given its current geometric configuration, is solved for numerically. The solved target shape indices $\{s_0^{T,clamped}\}$ are then temporarily applied to the target cells and the configuration is optimized. Since the optimization changes the geometric configuration of the cell, the maximum shear stress of the cell is now different from the target stress, but may be nudged closer. The iteration is then repeated up to a maximum number of clamping iterations, or until the target clamped stress is achieved. We restate the clamping procedure below in algorithmic form, noting our application of an overshoot factor $\lambda$ to set overshot target stresses $\sigma^T_{tmp}$ at each iteration of this subprocess.

\begin{figure}[htbp]
\centering
\begin{minipage}{0.95\columnwidth}
\hrule
\hrule
\vspace{0.5em}
\textsc{\textbf{Algorithm 3} Clamping Subroutine}
\vspace{0.5em}
\hrule
\begin{algorithmic}[1]
\Require  $TOL$, $\{\sigma^{T}\}$, $\{s_0^T\}$, $\lambda>0$, $i_{max}$
\State $\{s_0^{T,cl}\}\gets\{s_0^T\}$
\For{$i = 1$ to $i_{\max}$}
\State Calculate clamped target cell stresses  $\{\sigma_{cl}^{T}(s_0^{T,cl})\}$
\If {$\abs{\sigma_{cl}^{T} - \sigma^{T}} < TOL  $ for all T  }
\State \textbf{break}
\Comment{All target cells clamped}
\Else
\For{$T: \abs{\sigma_{cl}^{T} - \sigma^{T}} > TOL$}
\State $\sigma^{T}_{tmp} \gets \sigma^{T} + \lambda(\sigma^{T}-\sigma_{cl}^{T})$
\If{$\sigma^{T}_{tmp} <0$}
\State $\sigma^{T}_{tmp} \gets \sigma^{T}$ \Comment{No Overclamping}
\EndIf
\State Solve (numerically)  $s_0^{T,cl}: \sigma_{cl}^{T}(s_0^{T,cl}) = \sigma^{T}_{tmp} $
\EndFor
\State $\{s_0^T\}\gets\{s_0^{T,cl}\}$
\State Minimize configuration
\EndIf
\EndFor
\end{algorithmic}

\vspace{0.5em}
\hrule
\end{minipage}
\end{figure}


\subsection{Measurements}

 To study the impact of the training algorithm, we will measure the error in the loss function as a function of iterations of the algorithm. We will also measure what is known as the overlap parameter, $Q_2$, which tells us the fraction of cells that have changed neighbors from the initial configuration. 

Consider a configuration of $n$ cells, where each cell is labeled with consecutive integers. A $n$-dimensional vector $\vec{s_0}$ can be assigned to list the target cell shape indices, ordered by the integer labels of the cells. The difference between two target cell shape index assignments A, B on the same configuration can be quantified using a parameter-space separation $\mathcal{D}$, which we define as:

\begin{align}
    \mathcal{D}(A,B) = \frac{1}{\sqrt{n}}\norm{\vec{s_0}_A-\vec{s_0}_B}
\end{align}

The normalization factor $\frac{1}{\sqrt{n}}$ is chosen so that if one of the assignments is homogeneous and the other is non-homogeneous but with the same arithmetic mean, then D becomes the standard deviation of the components in the non-homogeneous assignment. For example if $\vec{s_0}_B = s_0(1,...,1)$ and the components of $\vec{s_0}_A$ are such that $\frac{1}{n}\sum_{i=1}^n {{s_0}_A}^i = s_0$, then D(A,B) becomes the standard deviation of the components of  $\vec{s_0}_A$.
We note that, since multiple energy-minimized configurations of the n-cells may be possible for a given assignment of  target shape indices $\vec{s_0}$, $\mathcal{D}=0$ does not necessarily imply that the two states being compared are geometrically identical. Furthermore, since the definition depends on an ordered labeling of cells, a comparison between two different $n$-cell configurations is not meaningful.

\begin{figure*}[t]
    \centering
    \begin{subfigure}{0.24\linewidth}
    \includegraphics[width=\linewidth]{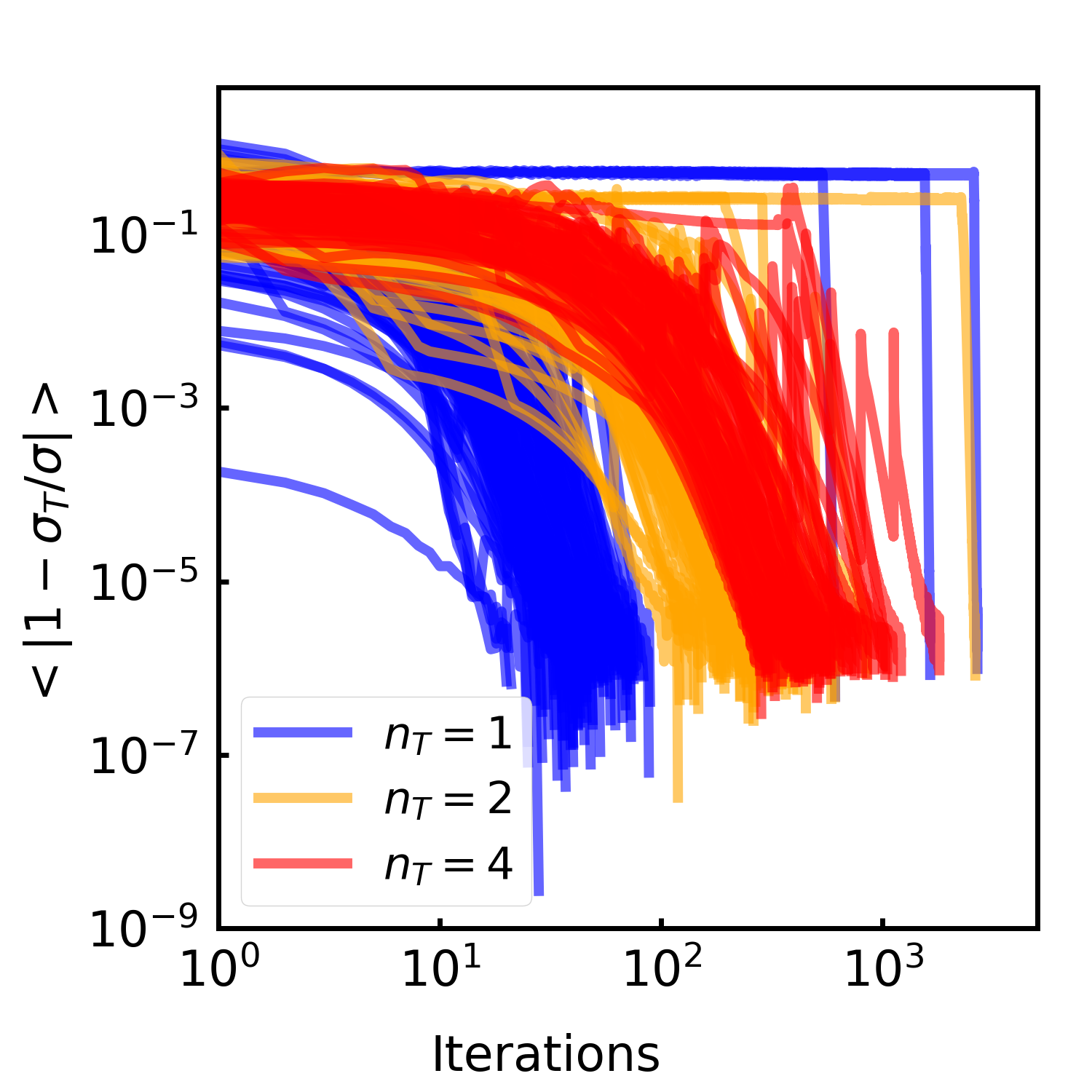}
        \caption{}
        \label{fig:2a}
    \end{subfigure}
    \begin{subfigure}{0.24\linewidth}
        \includegraphics[width=\linewidth]{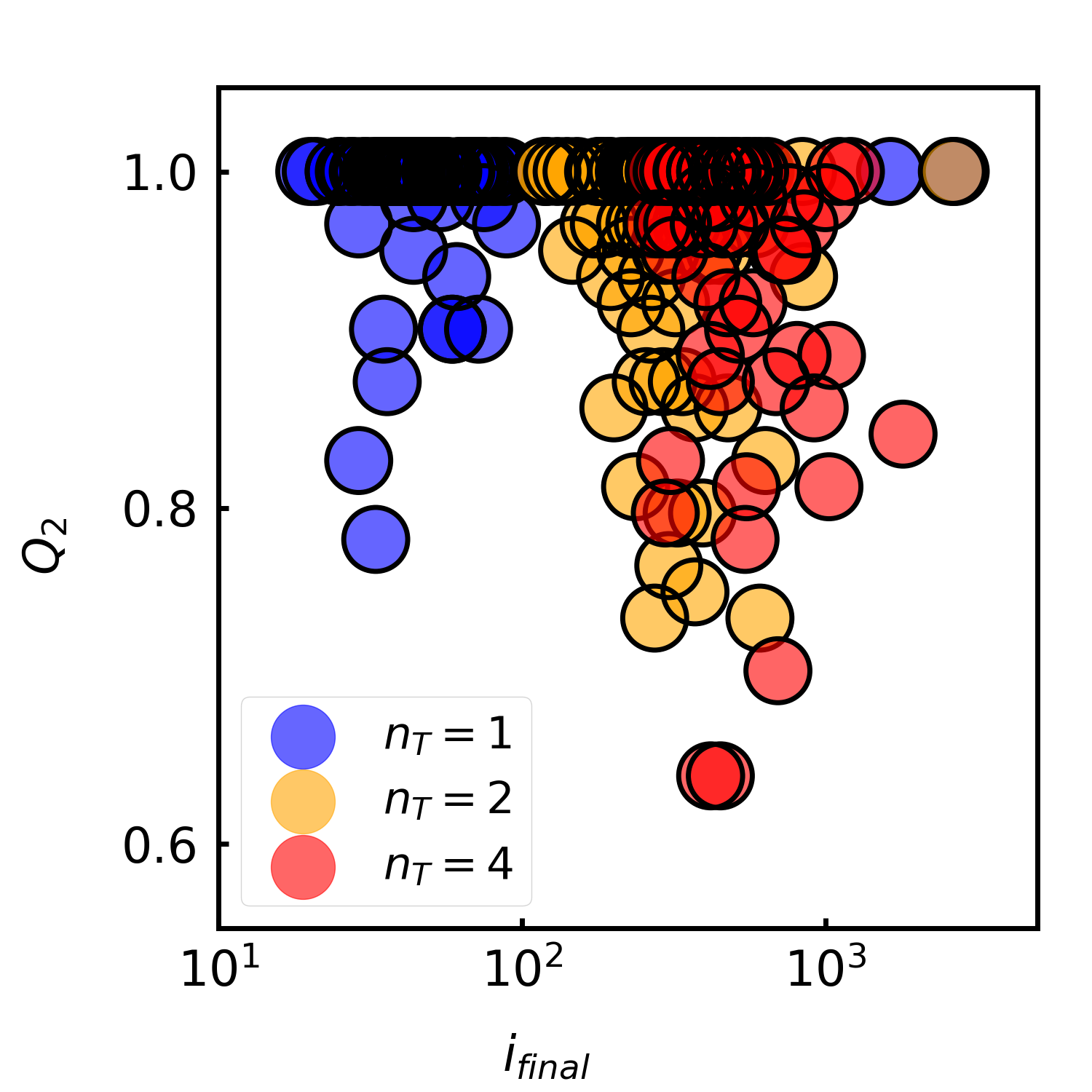}
        \caption{}
        \label{fig:2b}
    \end{subfigure}
    \begin{subfigure}{0.24\linewidth}
        \includegraphics[width=\linewidth]{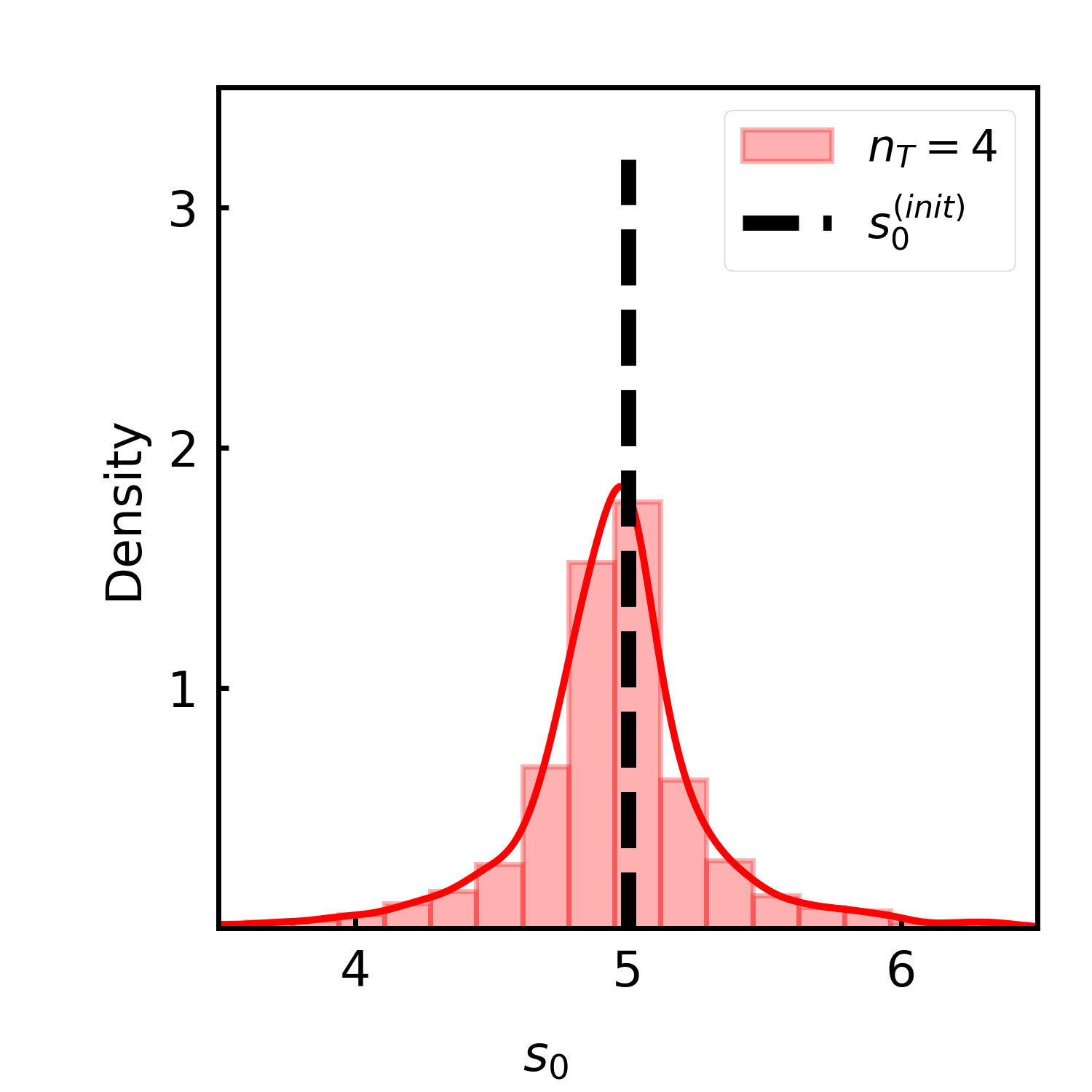}
        \caption{}
        \label{fig:2c}
    \end{subfigure}
    \begin{subfigure}{0.24\linewidth}
        \includegraphics[width=\linewidth]{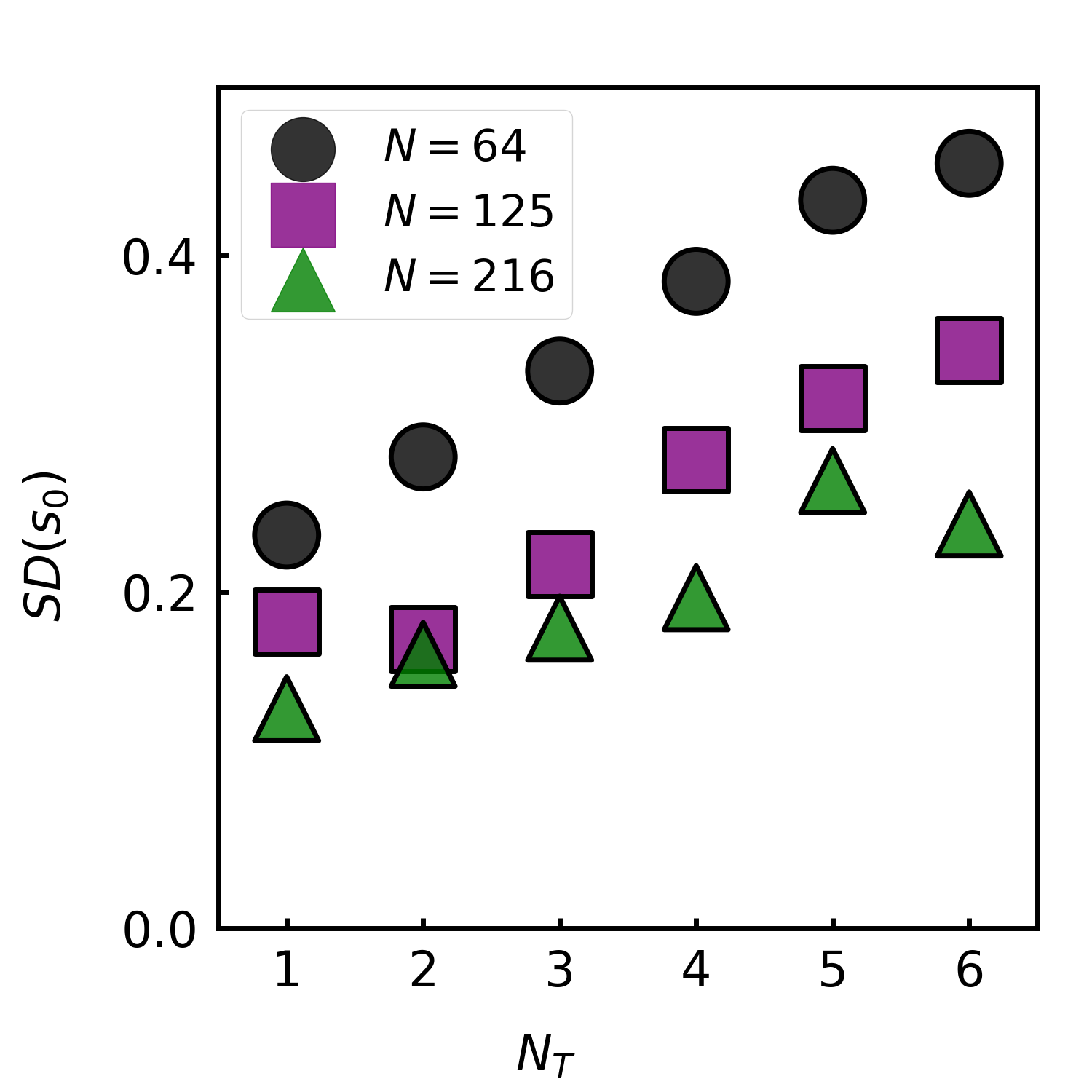}
        \caption{}
        \label{fig:2d}
    \end{subfigure}

  \caption{{\it Learning with different constraint loads and system sizes.}  (a) Error vs iteration for 100 runs, for constraint load $N_T = 1,2,4$ with $N=64$. (b) Scatter plot: overlap of trained configuration with initial configuration vs. final iteration number, for $N_T = 1,2,4$ with $N=64$. (c) Distribution of final $s_0$ values for the hidden cells with $N_T=4$ and, again, $N=64$ (d) Standard deviation for the final $s_0$ distribution as a function of constraint load for different system sizes.}
  \label{fig:2[]}
\end{figure*}

\section{Results}

Given that we can measure the maximum shear stress for each
cell, we will use the quantity as a target quantity for
training $N_T$ cells.  Ensembles of 100 energy-minimized
configurations with $\{s_0^i\}_{i=1}^{l^3} = 5.0$ are used
as initial configurations for all the training tasks for
each system size. The distribution of initial cellular
maximum shear stresses is obtained from these initial
configurations. This  distribution is confirmed to
approximately be independent of system size. We denote
$\bar{\sigma}_{(0)}$ and $\sigma_{SD}$ as the mean and
standard deviation of the initial cellular stresses,
respectively. The ratio of target to total number of cells in the packing, $N_T/N$, defines a constraint load, which acts an effective capacity parameter analogous to under- and overparameterized regimes in conventional learning systems.

\subsection{Teaching a Single-Cell Stress Pattern}

We begin with the simplest task: training a single randomly
chosen cell to achieve a prescribed target maximum shear
stress, $\sigma_T$. To probe how the mechanical state of
the packing influences learnability, we consider two
complementary tasks: (i) increasing the stress of a cell
initially below the mean stress $\bar{\sigma}(0)$ by at
least two standard deviations, and (ii) decreasing the
stress of a cell initially above the mean by a comparable
amount. We find a pronounced asymmetry between these two
tasks. Training a cell to a lower target stress typically
requires significantly fewer iterations than training a
cell to a higher target stress, with the latter also
exhibiting a broader distribution of training times across
realizations (Fig.~1(b)). This asymmetry is consistently
observed across system sizes. To understand this behavior,
we examine the role of cellular rearrangements during
training. As shown in Fig.~1(c), the reduction of target
stress is accompanied by a larger number of neighbor
exchanges, indicating that the packing explores a broader
region of configuration space. In contrast, increasing the
target stress typically involves fewer rearrangements,
with the system remaining closer to its initial topology.

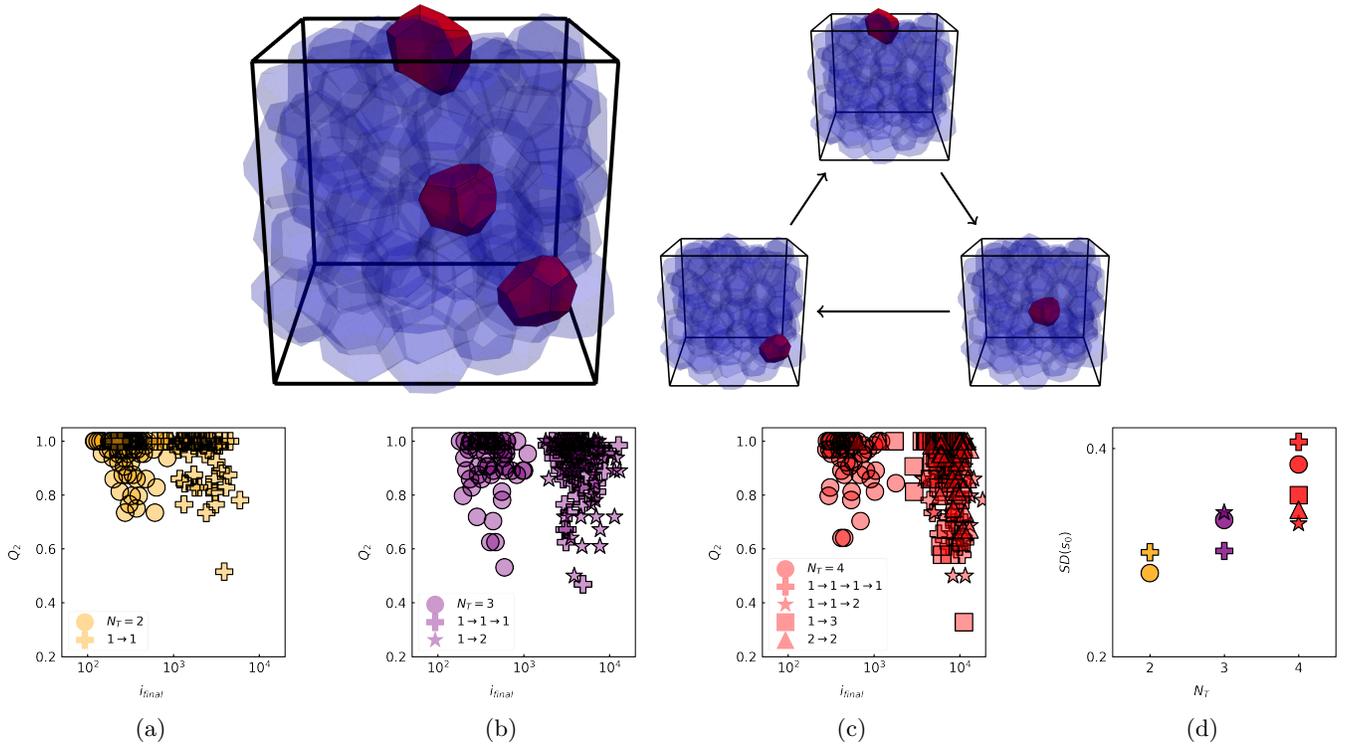
\begin{figure*}[t]
    \centering
    \begin{tikzpicture}
    \node (main) at (0,0) {\includegraphics[width=5cm]{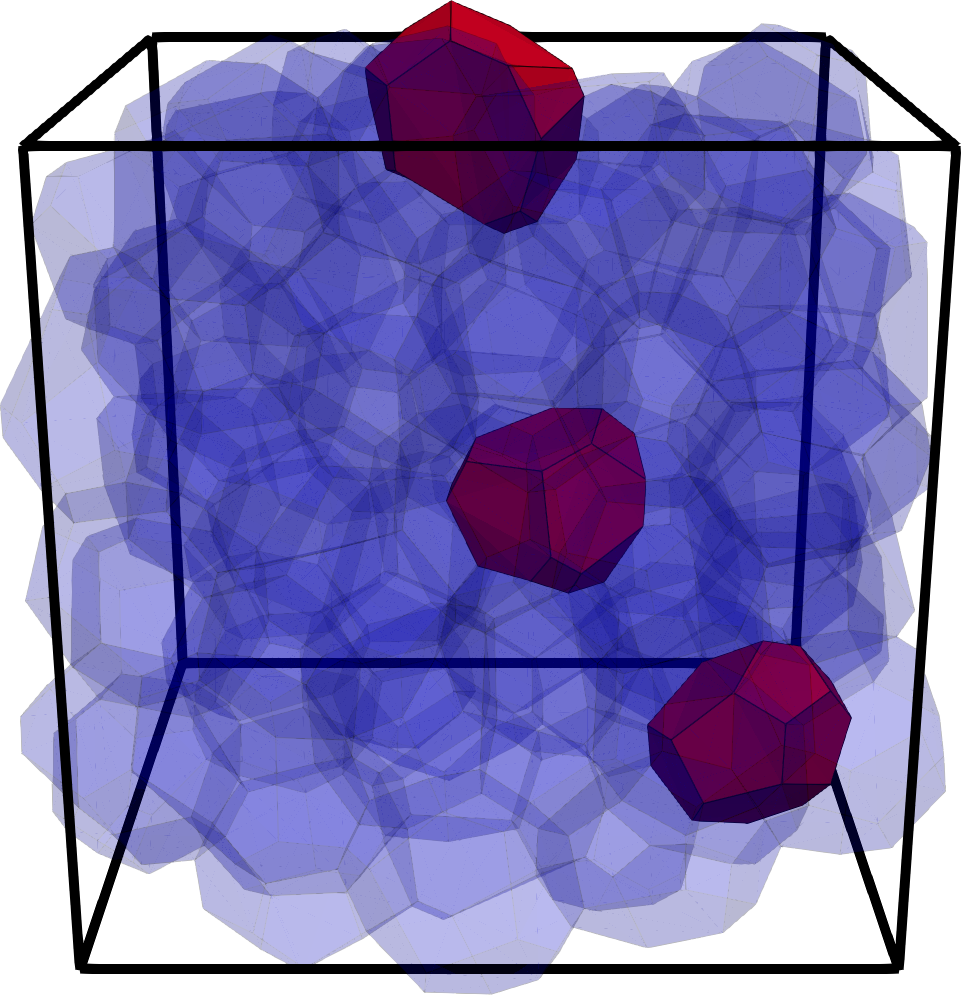}};
    \node (A) at (6,1.5) {\includegraphics[width=2cm]{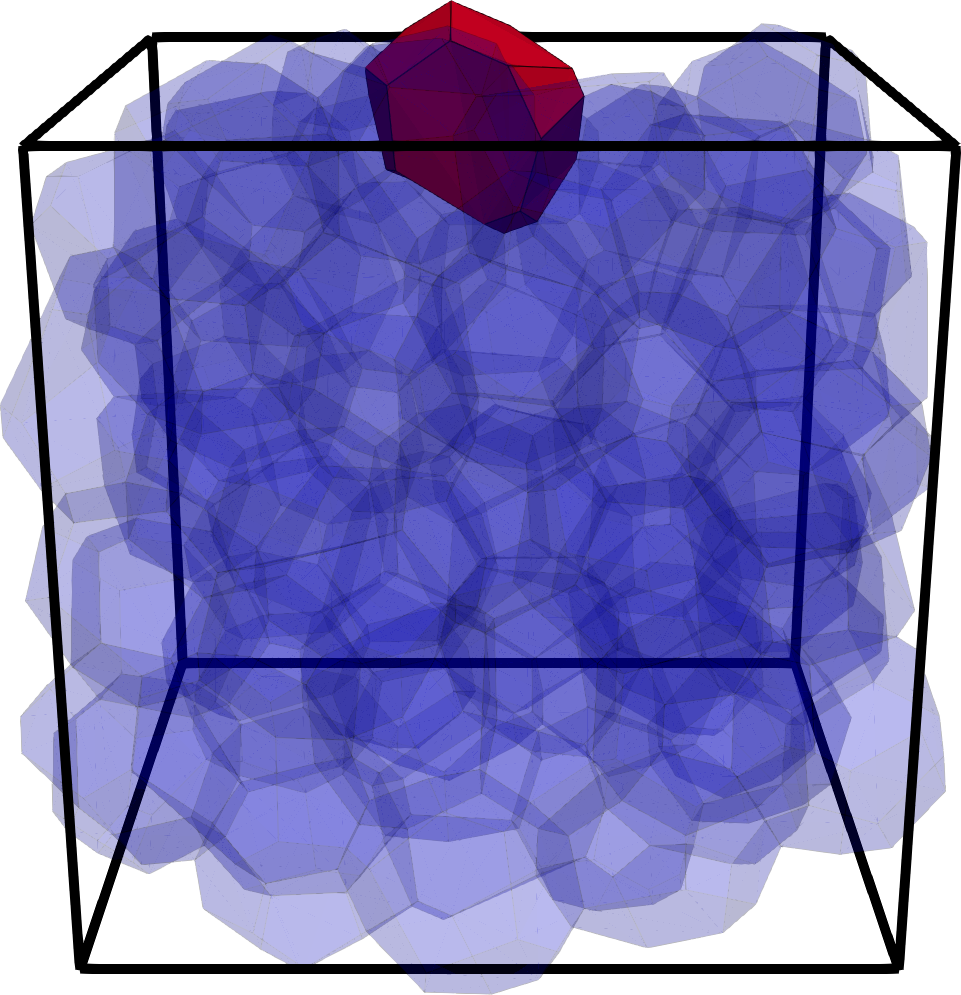}};
    \node (B) at (8,-1.5) {\includegraphics[width=2cm]{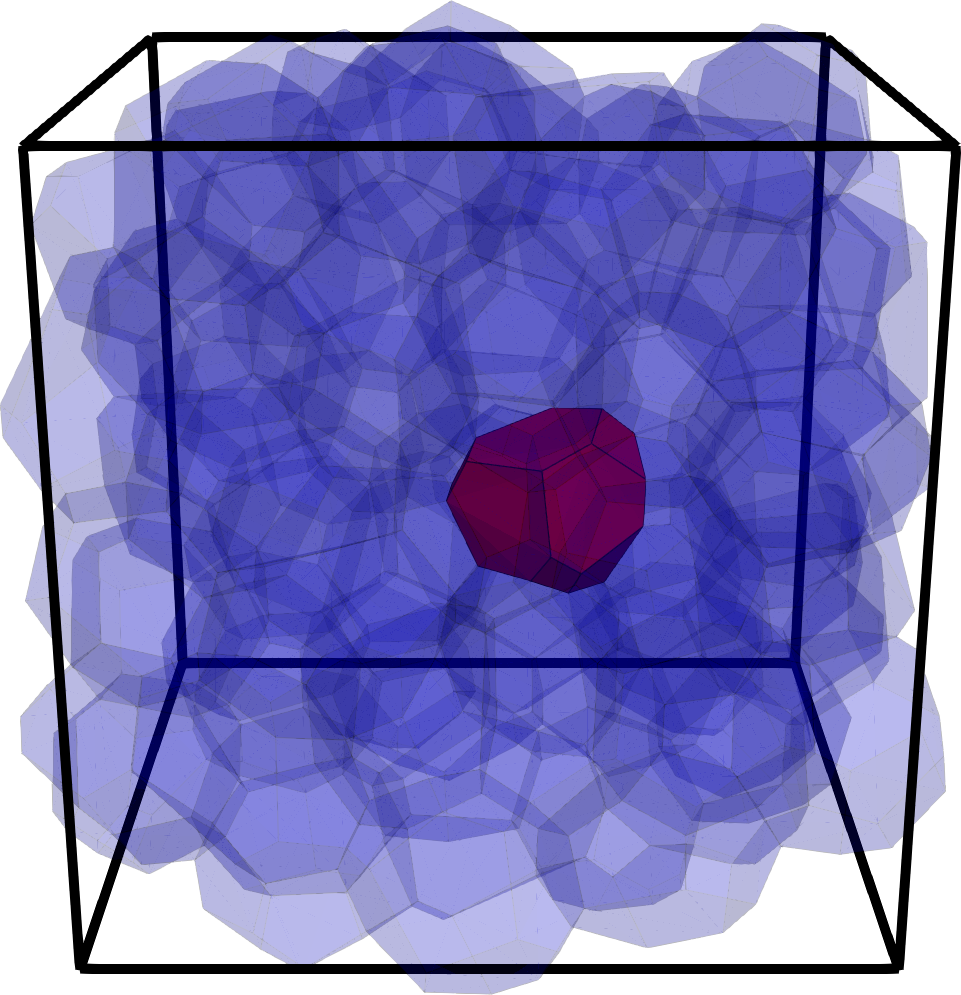}};
    \node (C) at (4,-1.5) {\includegraphics[width=2cm]{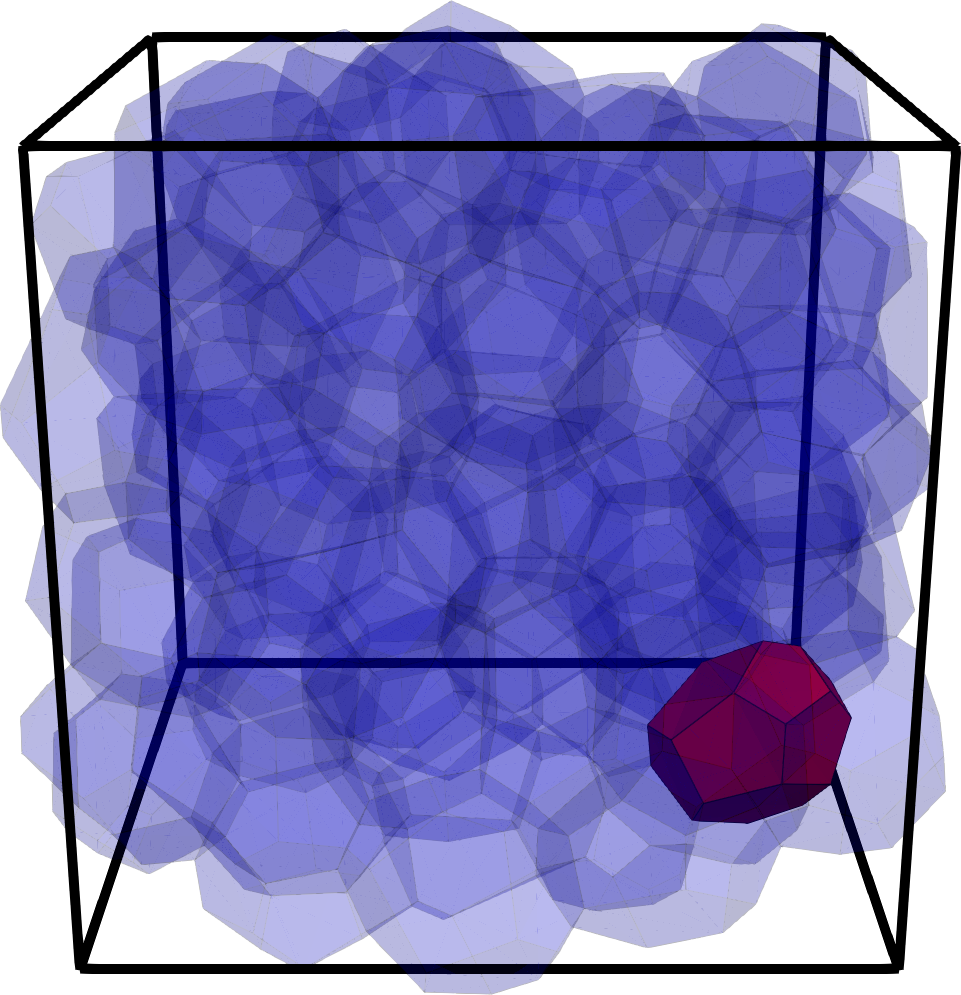}};
    
    \draw[->, thick] (A) -- (B);
    \draw[->, thick] (B) -- (C);
    \draw[->, thick] (C) -- (A);
    \label{}
    \end{tikzpicture}
\vfill

    \begin{subfigure}{0.22\linewidth}
        \includegraphics[width=\linewidth]{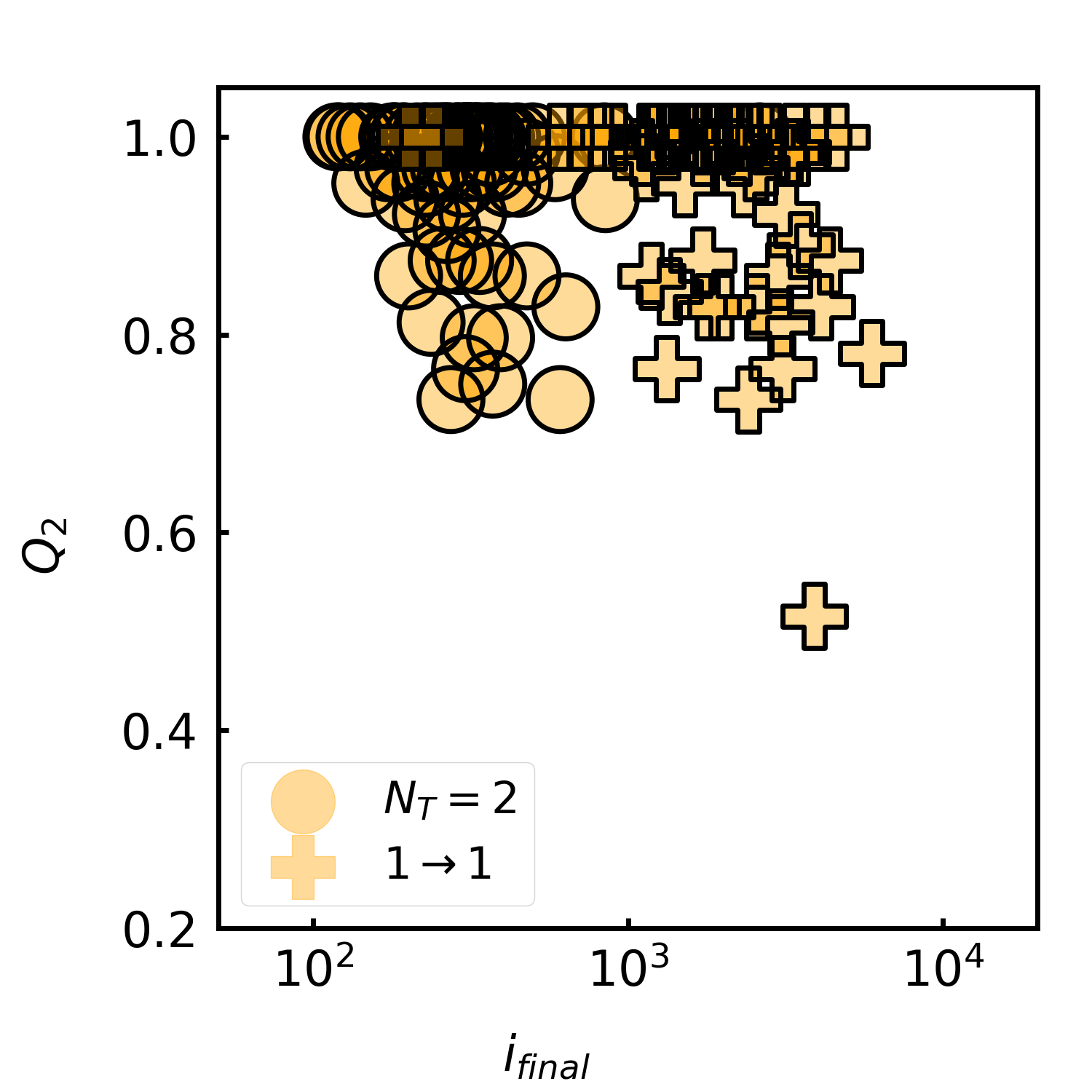}
        \caption{}
        \label{fig:3a}
    \end{subfigure}
    \hfill
    \begin{subfigure}{0.22\linewidth}
        \includegraphics[width=\linewidth]{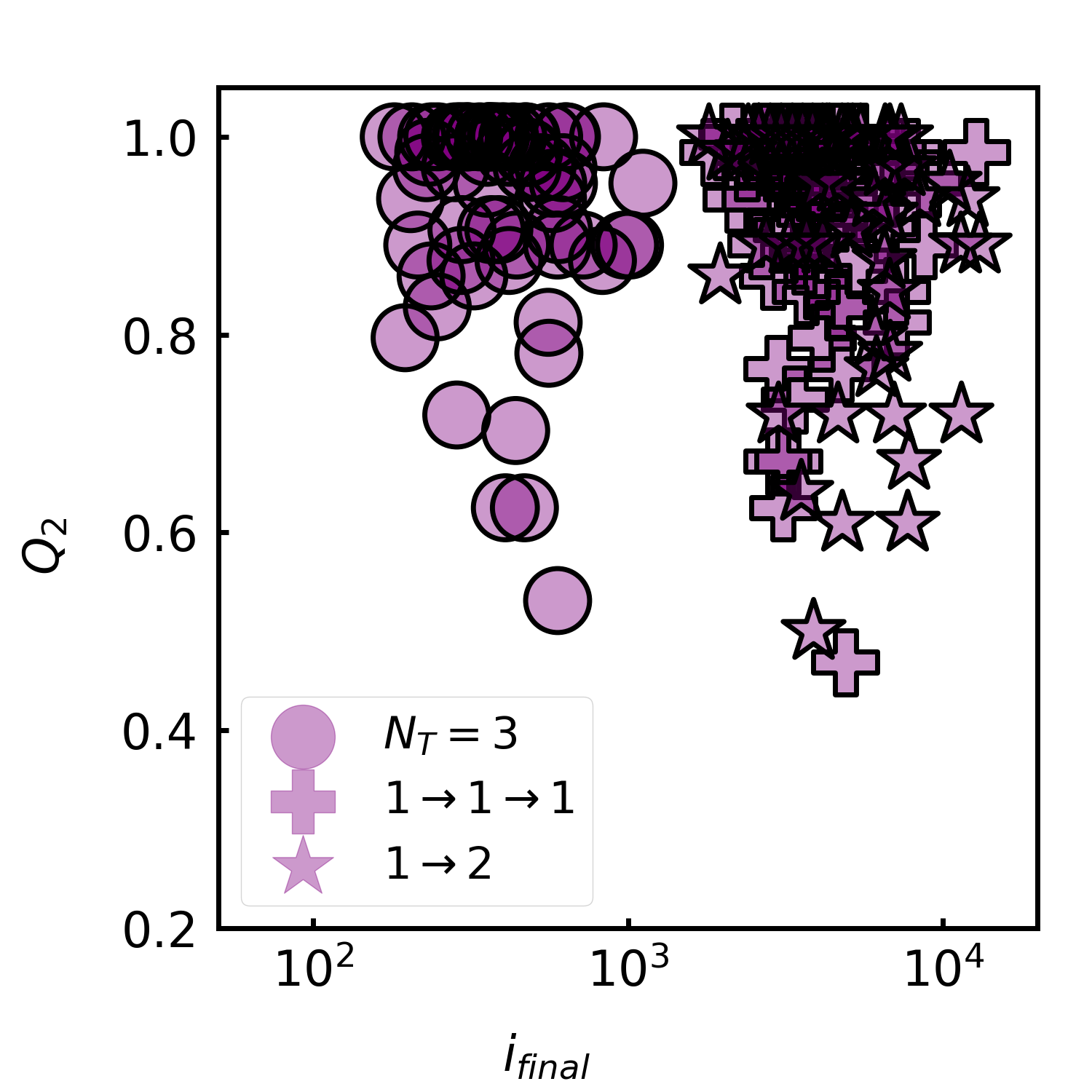}
        \caption{}
        \label{fig:3b}
    \end{subfigure}
    \hfill
    \begin{subfigure}{0.22\linewidth}
        \includegraphics[width=\linewidth]{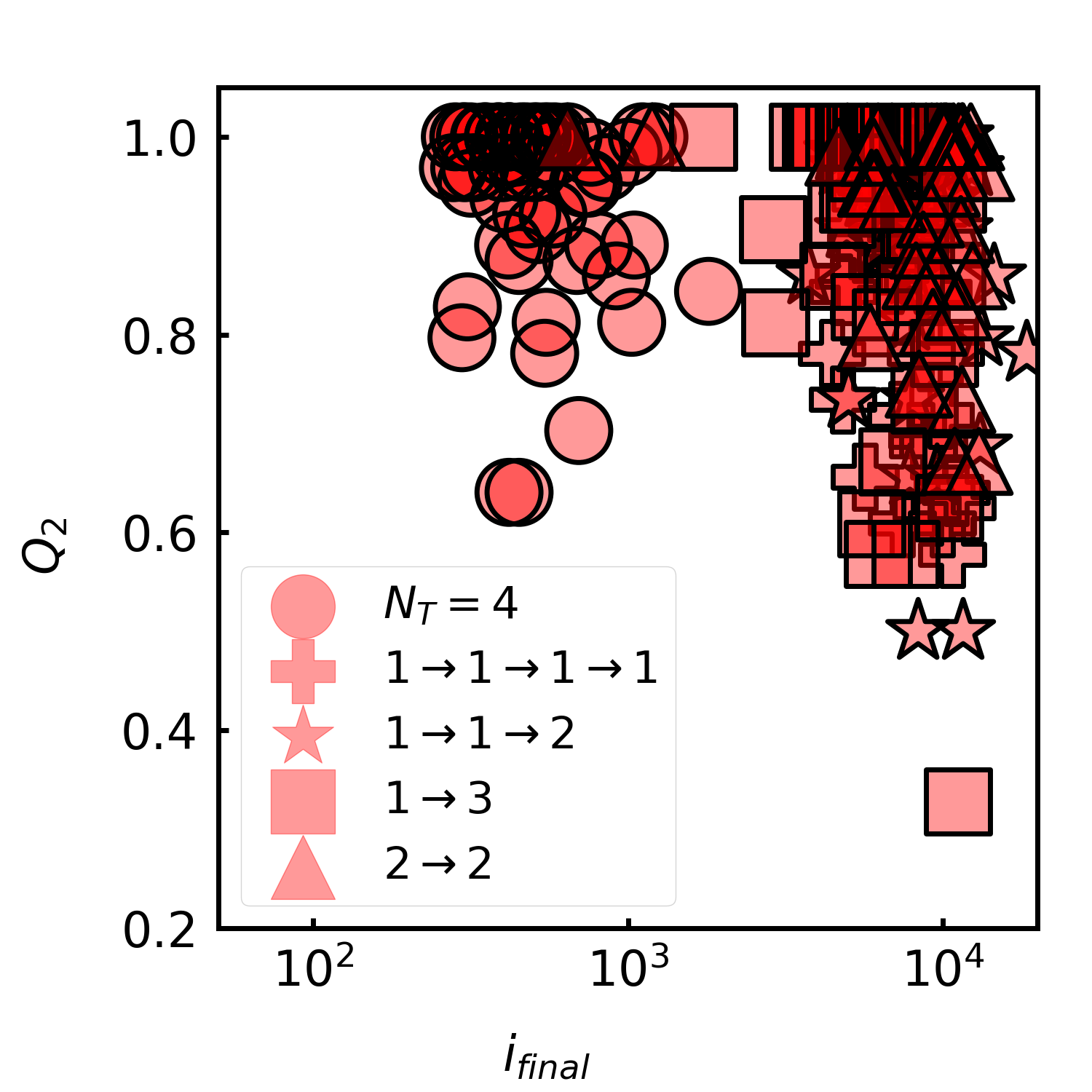}
        \caption{}
        \label{fig:3c}
    \end{subfigure}
    \hfill
    \begin{subfigure}{0.22\linewidth}
        \includegraphics[width=\linewidth]{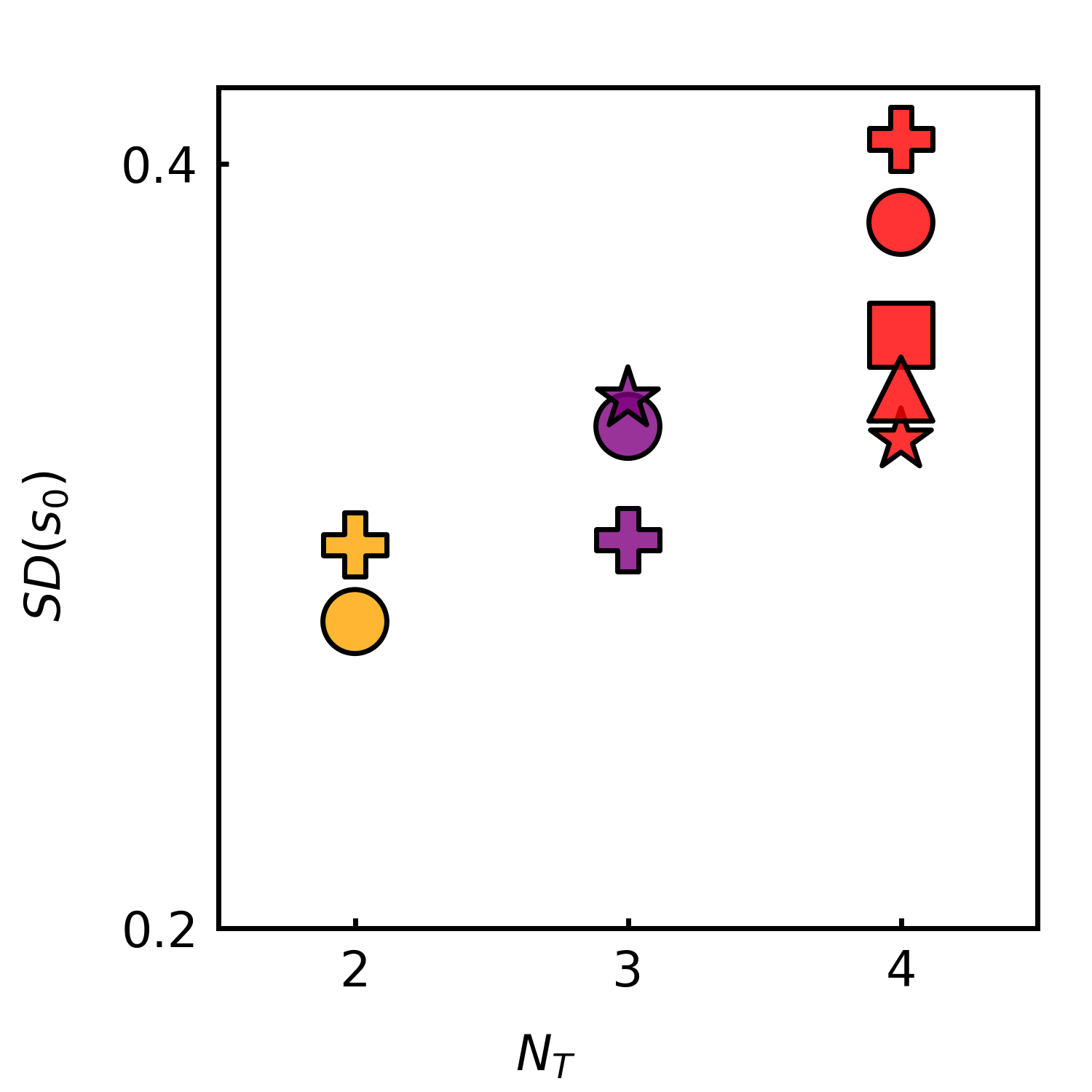}
        \caption{}
        \label{fig:3d}
    \end{subfigure}
    
  \caption{{\it Training target cells in sequence.} The schematic shows how a 3-cell pattern can be addressed in a $1\rightarrow1\rightarrow1$ sequence. Training (a) $N_T=2$ (b) $N_T=3$ (c) $N_T=4$ target cells with $N=64$ for various sequential patterns denoted by different symbols. (d) A comparison of the final standard deviation hidden cell parameter distribution as a consequence of the choice of sequence.} 
  \label{fig:3}
\end{figure*}

\begin{figure*}[t]
  \centering
    \begin{subfigure}{0.3\linewidth}
    \includegraphics[width=\linewidth]{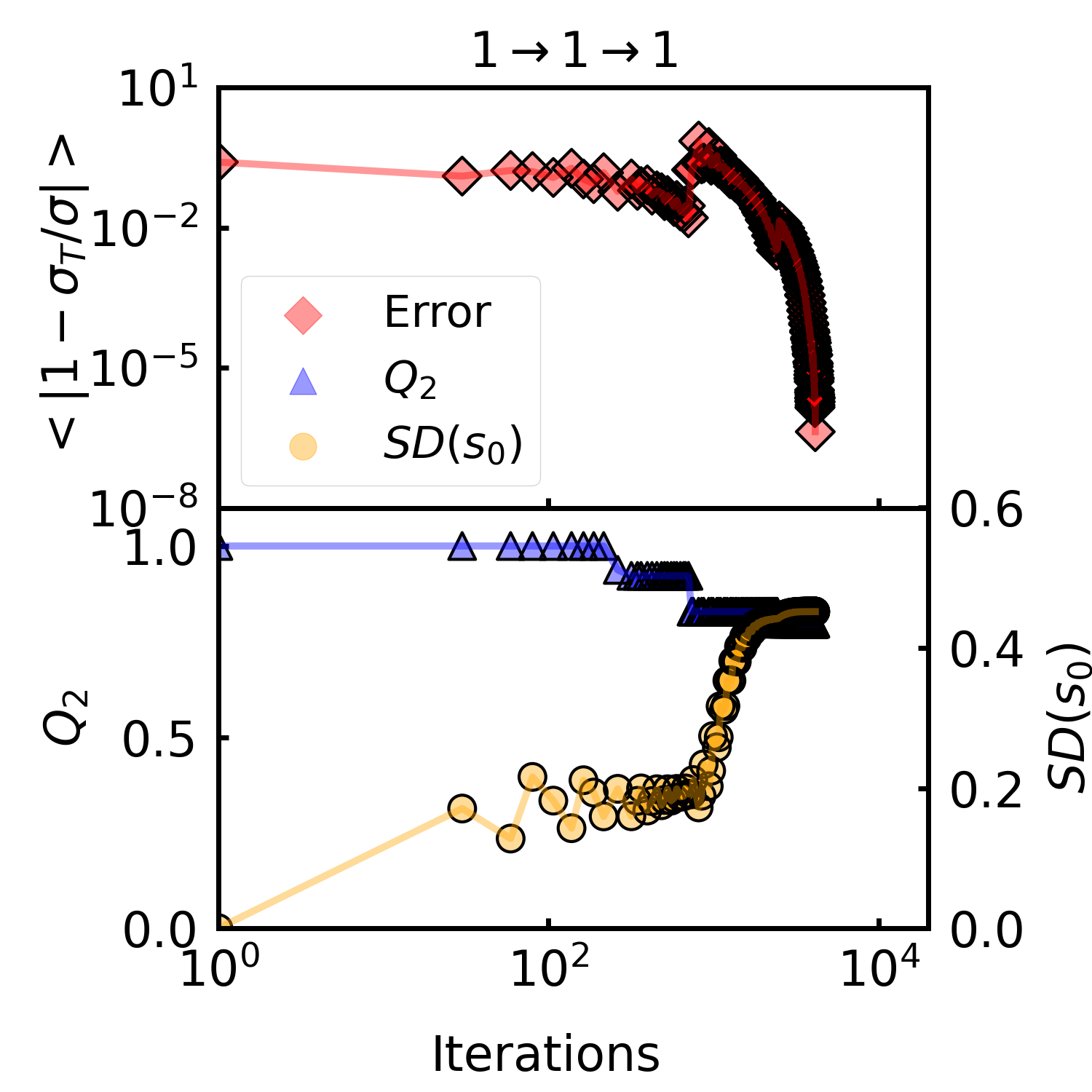}
        \caption{}
        \label{fig:4a}
    \end{subfigure}
    \hfill
    \begin{subfigure}{0.3\linewidth}
        \includegraphics[width=\linewidth]{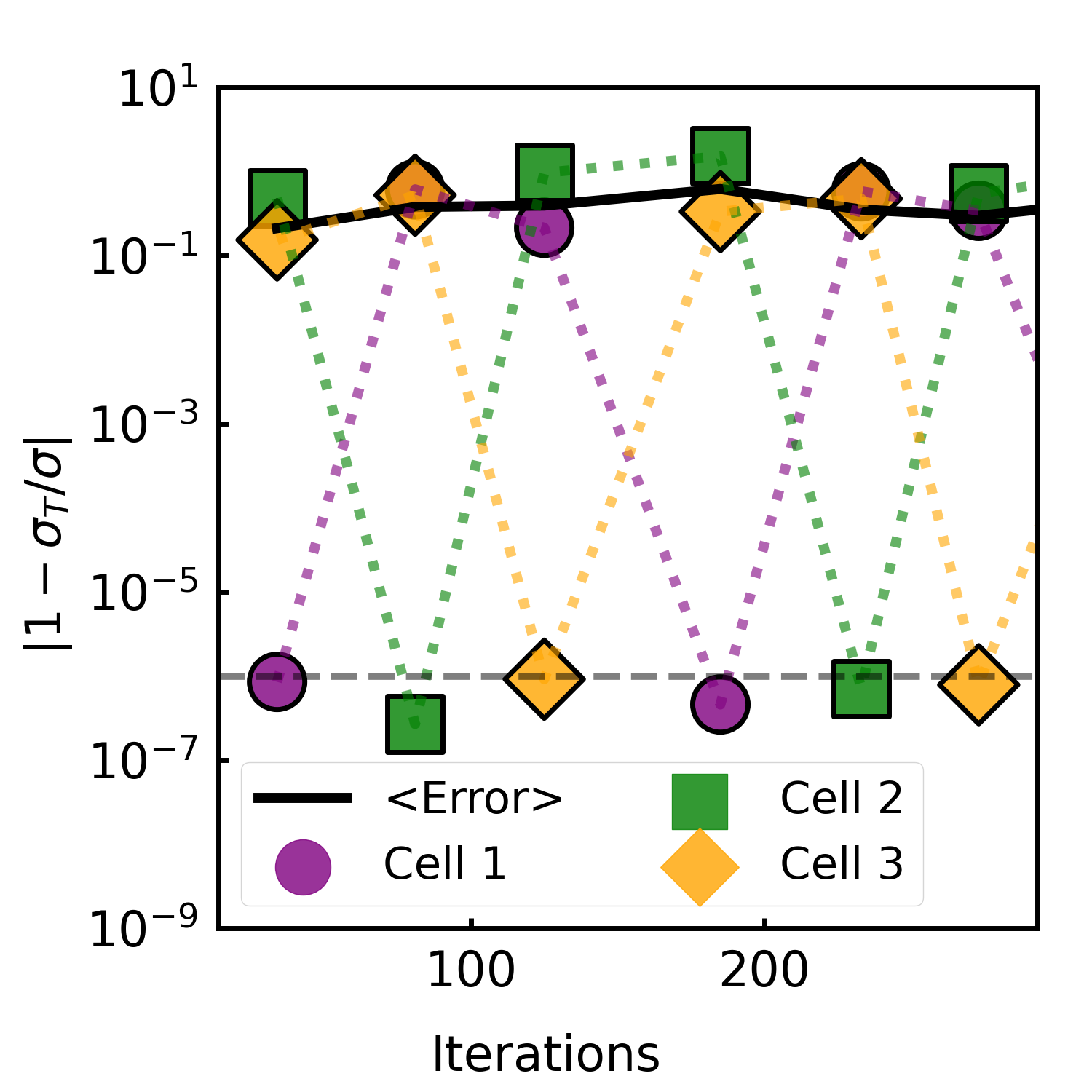}
        \caption{}
        \label{fig:4b}
    \end{subfigure}
    \hfill
\begin{subfigure}{0.3\linewidth}
        \includegraphics[width=\linewidth]{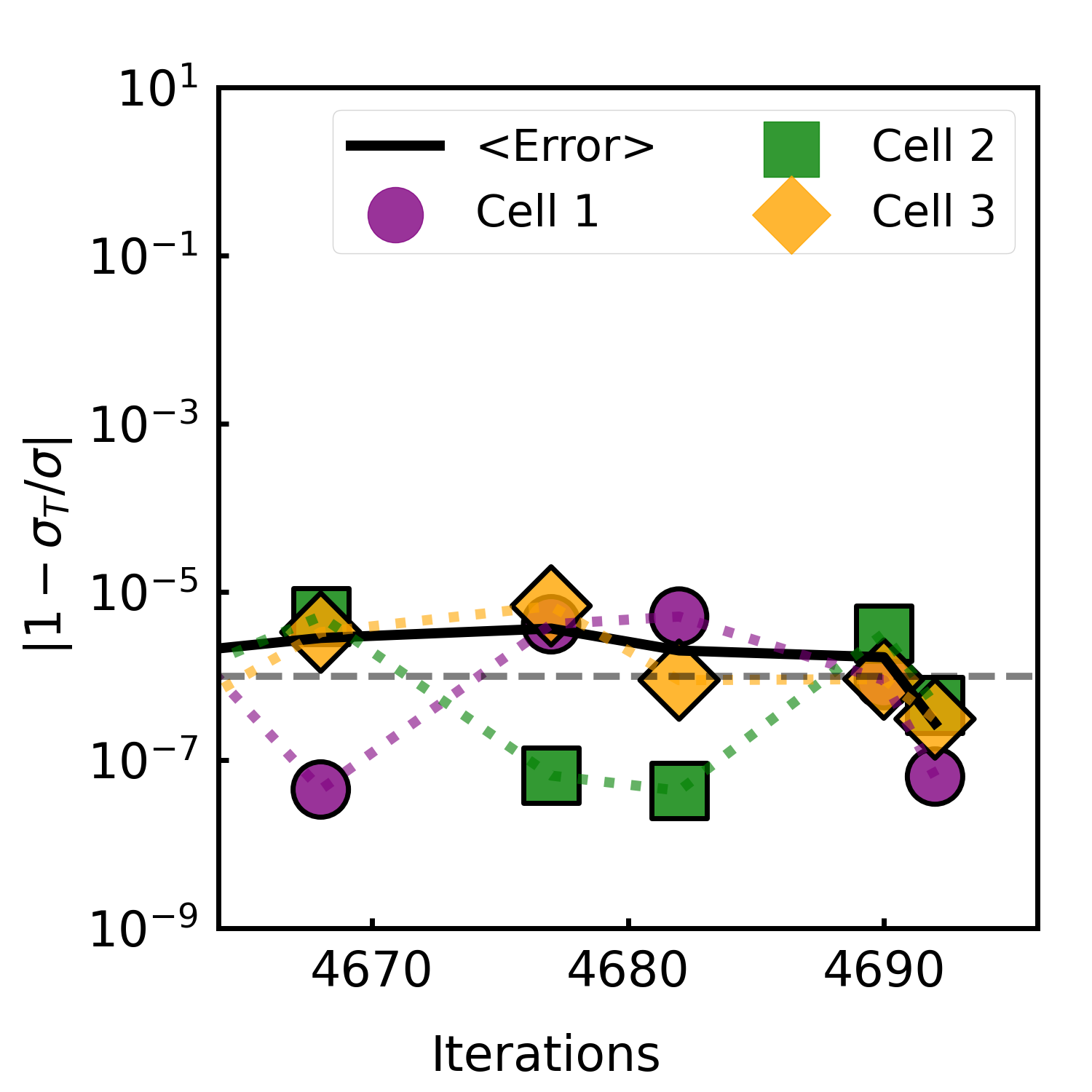}
        \caption{}
        \label{fig:4c}
    \end{subfigure}

  \caption{ {\it Compatibility and forgetting in sequential learning.} A single instance of training a $1\rightarrow1\rightarrow1$ sequence with $N=64$.  (a) Error, overlap, and standard deviation of the hidden parameter distribution vs. training iterations.  (b)  Tracking errors on each cell in the early stages of training. (c) Tracking errors on each cell in the later stages of training.}
  \label{fig:4}
\end{figure*}

This difference can be understood in terms of the
mechanical state of the tissue. Lowering the stress drives
the packing toward a more fluid-like regime, where
rearrangements are more readily accessible. These
rearrangements allow the system to escape locally
constrained configurations and explore alternative
pathways in the energy landscape. In this sense, the
increased number of rearrangements facilitates learning by
enabling access to configurations that would otherwise be
inaccessible. By contrast, increasing cell stress drives
the packing deeper into a rigid regime, where
rearrangements are suppressed. In this regime, the system
must rely primarily on continuous deformations of cell
shapes, which can be insufficient to overcome geometric
constraints, leading to longer and more variable training
times. This asymmetry can thus be interpreted as a shift in the exploration–exploitation balance: fluid-like states promote exploration through frequent rearrangements, while rigid states restrict dynamics and force the system to exploit a limited set of accessible configurations.

Further support for this interpretation is provided by the 
results in Figs.~S1(a)-(b), where the volume stiffness
$K_V$ is reduced by an order of magnitude. Lowering $K_V$ weakens volume constraints and effectively reduces the
geometric frustration associated with maintaining cell
volumes. In this regime, the asymmetry between increasing
and decreasing stress is diminished, and training
generally proceeds more efficiently. This observation is consistent with the picture that geometric constraints, rather than the learning rule itself, limit the accessibility of solutions in the rigid regime.

We also examine the distributions of cellular stresses and target shape indices after training (Figs.~1(d),(e) and Figs. S1(c),(d)). Increasing the stress of a single cell leads to a broadening of the hidden cell stresses while a narrowing of the distribution when training for a lower stress. Moreover, training for higher target stress is achieved through a coherent shift of the hidden-cell parameters toward lower target shape indices $s_0$, corresponding to an overall rigidification of the tissue. Conversely, decreasing the stress leads to a shift toward higher $s_0$, consistent with a more fluid-like state. Notably, these shifts are global rather than localized, indicating that training even a single-cell stress requires coordinated adjustments across the entire packing. Finally, the minimal overlap between the resulting $s_0$ distributions for high-stress and low-stress training tasks (Fig.~1(e) and Fig.~S1(d)) suggests that these tasks correspond to distinct regions of parameter space. This observation points to an inherent incompatibility between certain stress patterns, motivating the use of consistent target stresses when training multiple cells simultaneously.

\subsection{Teaching Multicellular Stress Patterns in Parallel}

We now consider the task of training multiple target cells simultaneously to a common prescribed stress. Motivated by the incompatibility observed between high- and low-stress single-cell tasks, we focus on training all target cells to the mean initial stress, $\sigma_T = \bar{\sigma}(0)$. In this setting, the target assignment does not introduce a global bias toward either rigidification or fluidization, allowing us to isolate the effects of increasing task complexity. 

We begin by examining representative training trajectories and resulting distributions, as shown in Figs.~1(e)--(o). As the number of target cells increases from $N_T = 2$ and $N_T = 4$, several trends emerge. First, the error decreases more slowly with iteration number (Figs.~1(g),(l)), indicating that training becomes progressively more difficult as additional constraints are imposed. Second, the cumulative overlap parameter quantifying the fractional change in cell rearrangments between the trained and initial configurations decreases more substantially on average for the $N_T=2$ as compared to $N_T=4$, (Figs.~1(h),(m)), reflecting a decreased number of cellular rearrangements and a lesser exploration of configuration space in the larger $N_T$ task. This result suggests that, for these target assignments, the system accommodates multiple constraints primarily through coordinated adjustments of hidden-cell parameters rather than through large-scale topological rearrangements. Third, while the distribution of cellular stresses remains centered near $\bar{\sigma}(0)$ (Figs.~1(i),(n)), the distribution of hidden-cell target shape indices broadens relative to its initial delta-function form (Figs.~1(e),(j),(o)). This broadening indicates that learning requires increasingly heterogeneous adjustments of hidden parameters as more target cells are trained simultaneously.
\begin{figure*}[t]
    \centering
    \begin{subfigure}{0.24\linewidth}        \includegraphics[width=\linewidth]{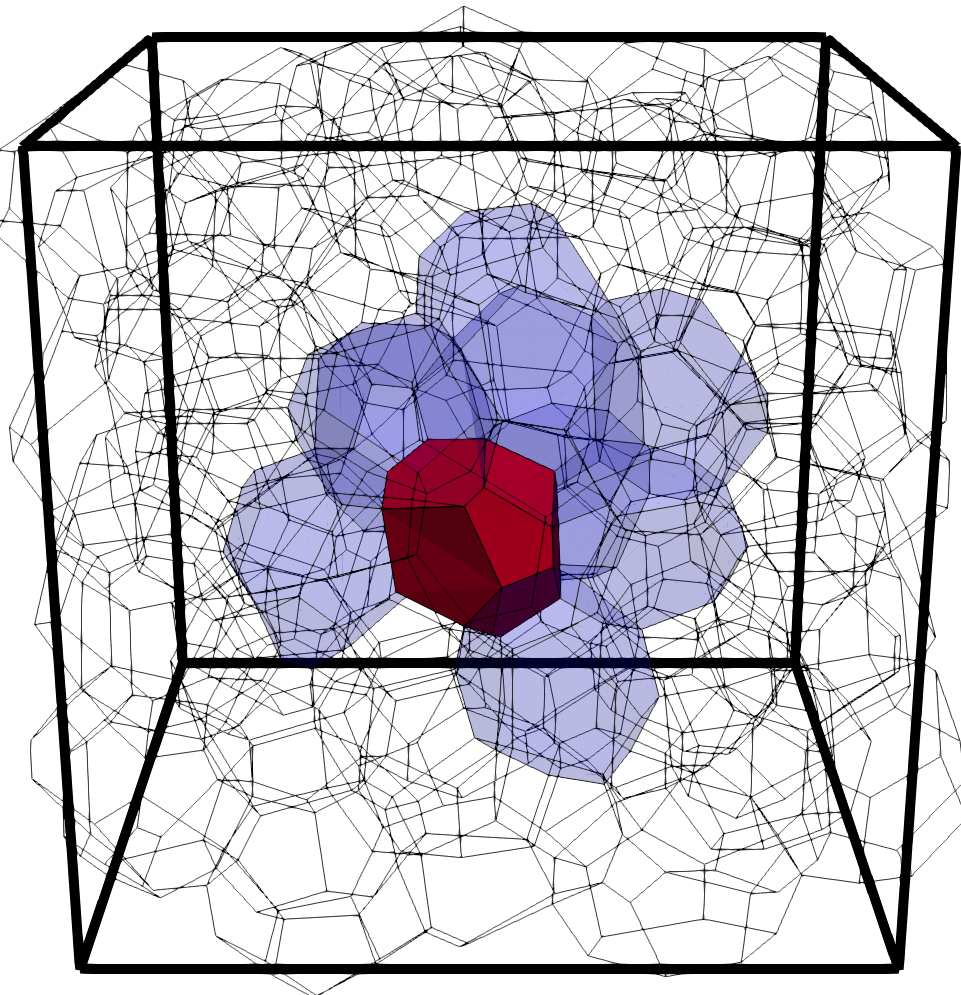}
        \caption{}
        \label{fig:5a}
    \end{subfigure}
    \begin{subfigure}{0.24\linewidth}
        \includegraphics[width=\linewidth]{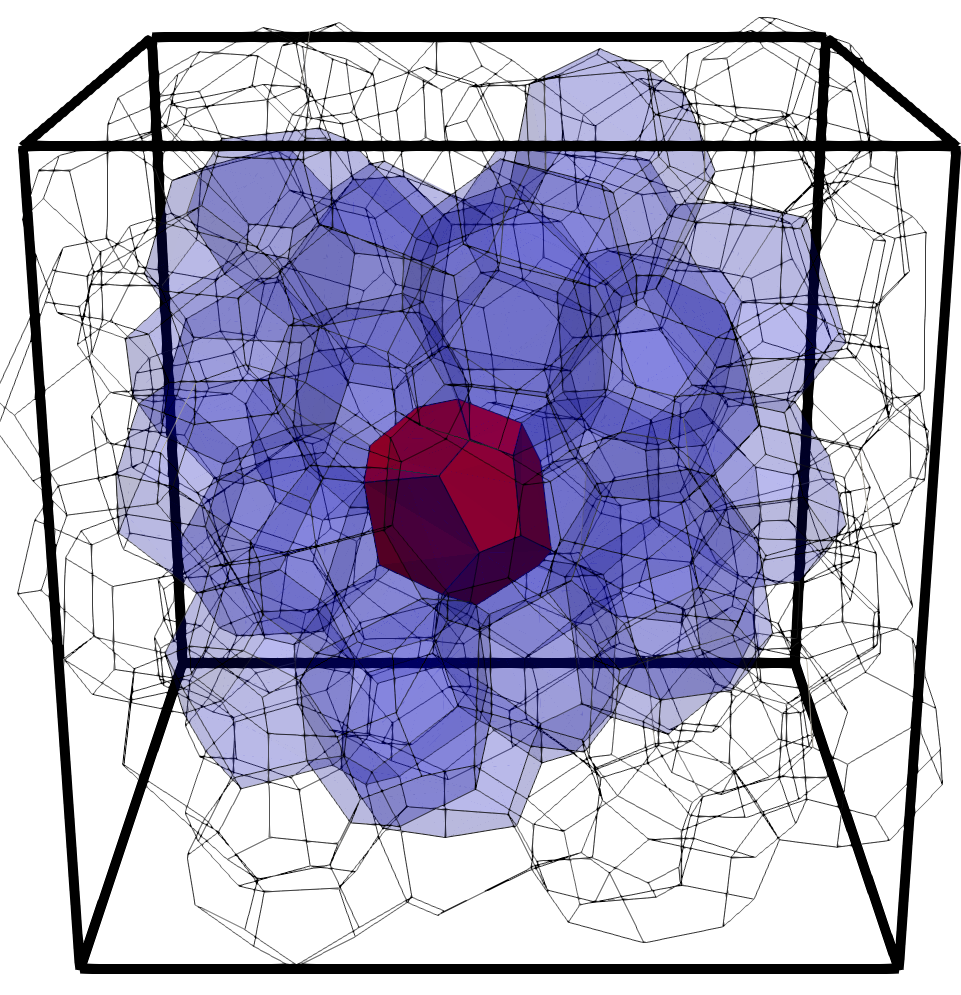}
        \caption{}
        \label{fig:5b}
    \end{subfigure}
    \vfill
    \begin{subfigure}{0.24\linewidth}
        \includegraphics[width=\linewidth]{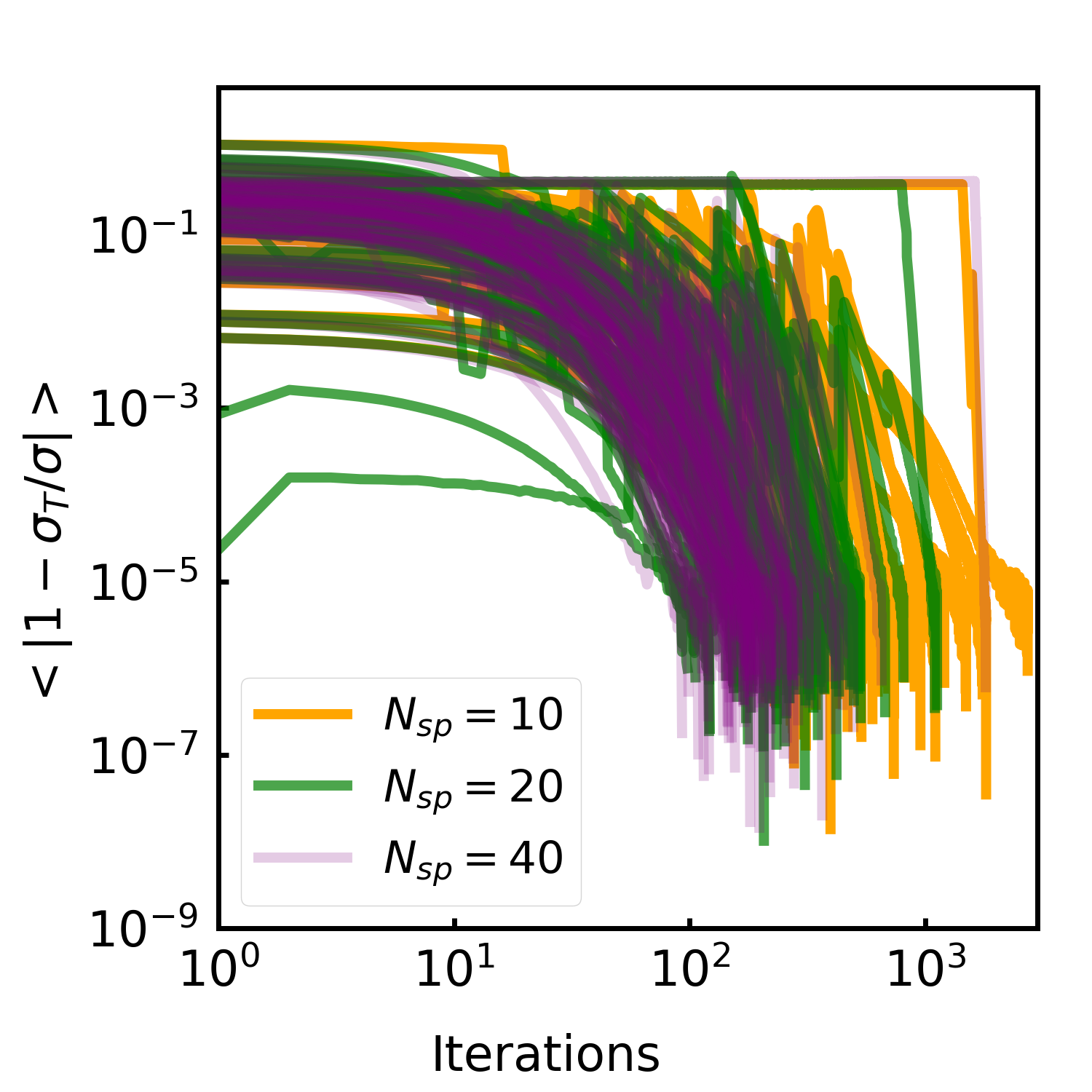}
        \caption{}
        \label{fig:5c}
    \end{subfigure}
    \begin{subfigure}{0.24\linewidth}
        \includegraphics[width=\linewidth]{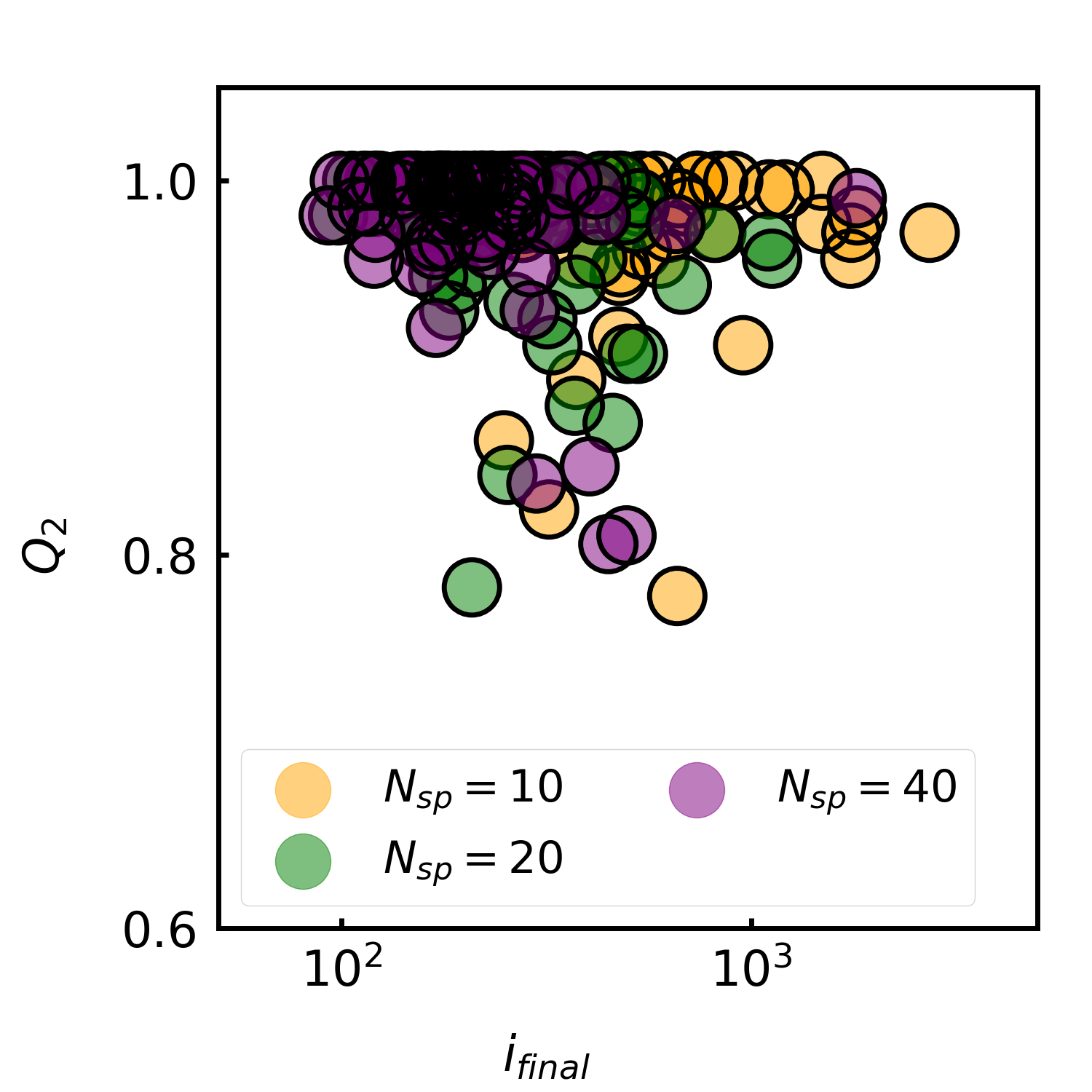}
        \caption{}
    \label{fig:5d}
    \end{subfigure}
    \begin{subfigure}{0.24\linewidth}
        \includegraphics[width=\linewidth]{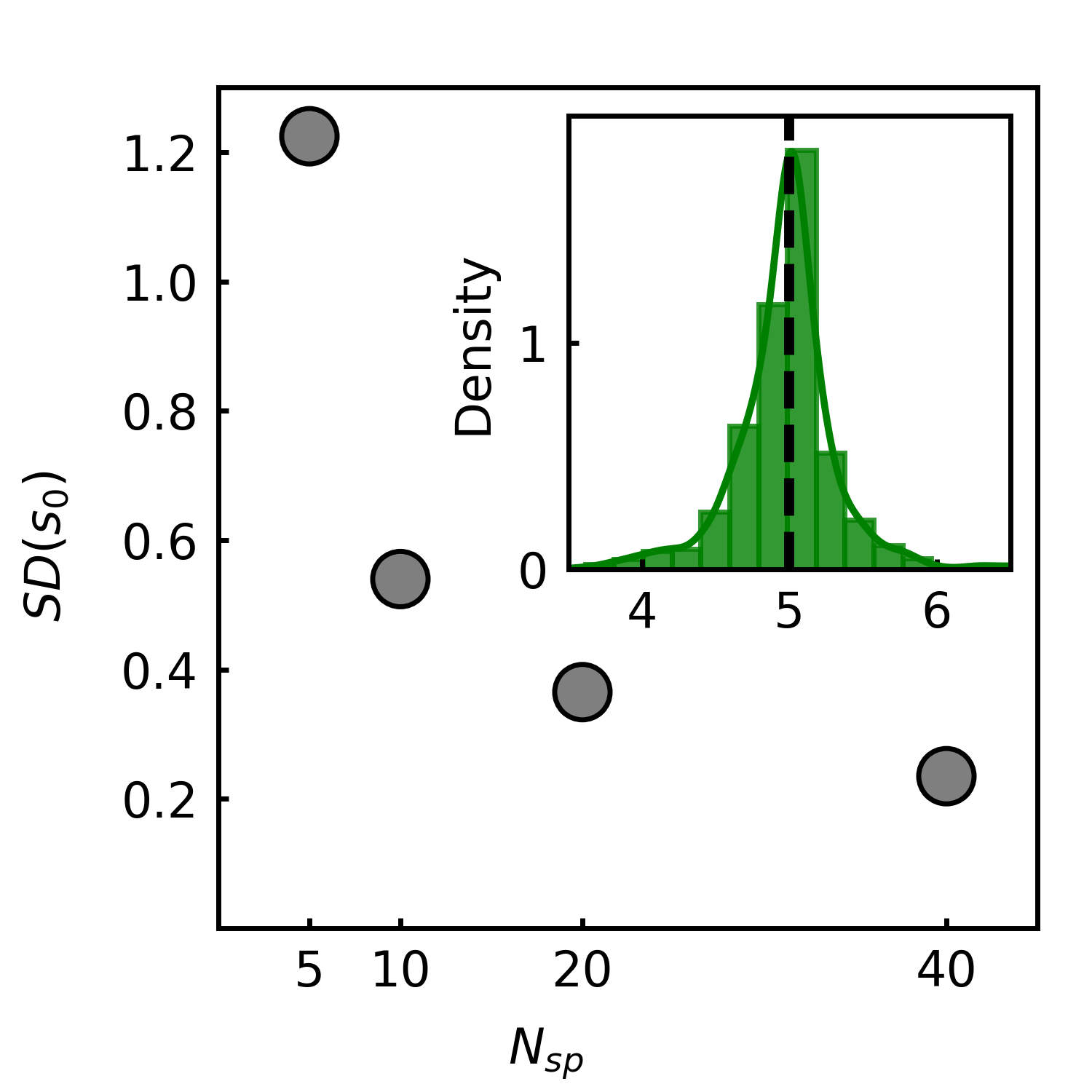}
        \caption{}
        \label{fig:5e}
    \end{subfigure}
    \begin{subfigure}{0.24\linewidth}
        \includegraphics[width=\linewidth]{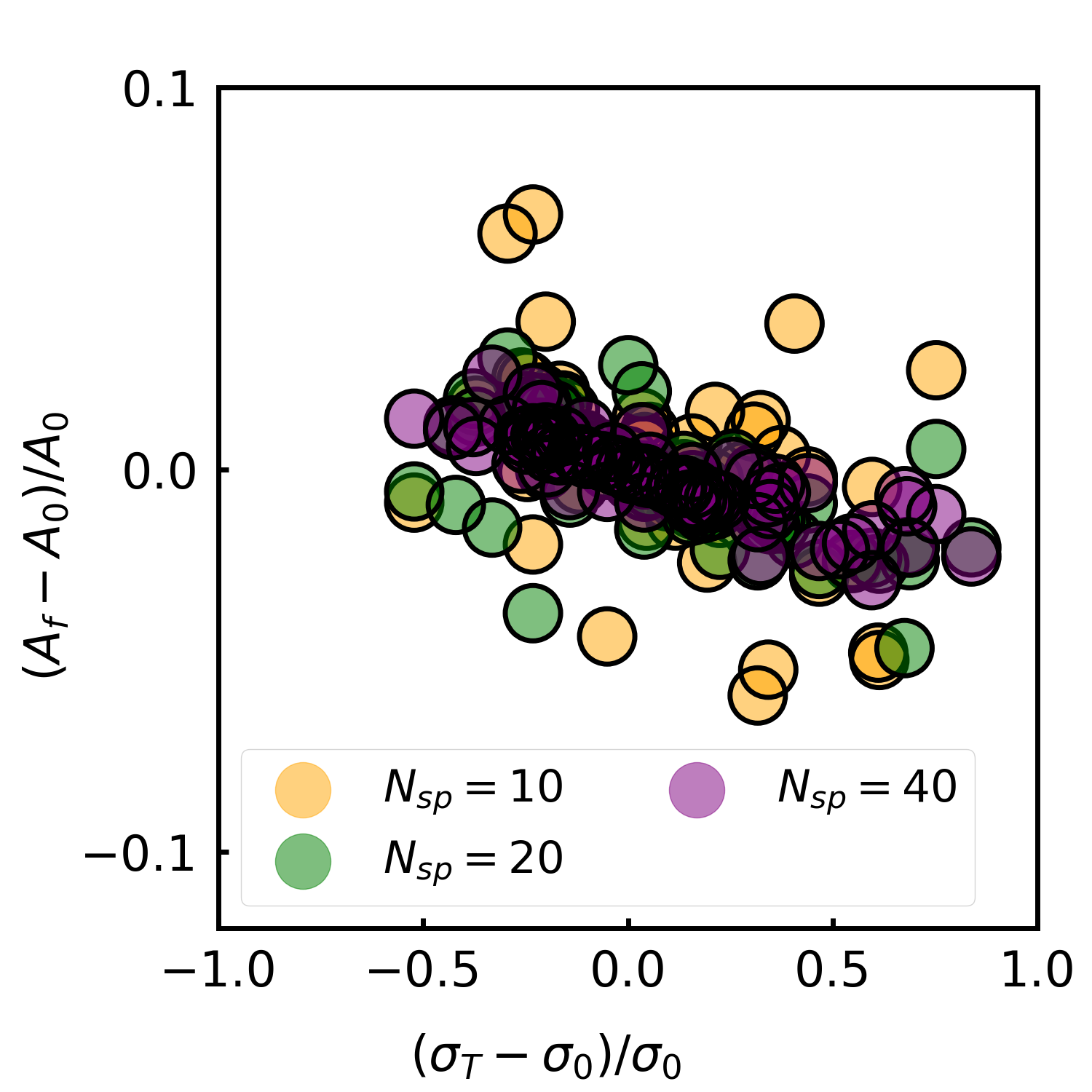}
        \caption{}
        \label{fig:5f}
    \end{subfigure}
  \caption{{\it Training cells stresses in spheroids that can impact spheroid shape.} A single target cell is picked at the center of a spheroid of $N_{sp}$ cells, and trained for a target stress $\sigma_T = \bar{\sigma}_0$, or the initial mean stress.  (a) A spheroid with $N_{sp}$ = 10 and (b) a spheroid with $N_{sp}$ = 40 cells. The single target cell at the center is shown in red. (c) Error versus interation number for different spheroid sizes demonstrates that increasing the spheroid size for the single-cell training task reduces the number of iterations needed. (d) Scatter plot of final overlap parameter and final iteration number. Generally, increasing spheroid size also reduces the number of cell neighbor changes needed. (e) Standard deviation of the final hidden parameter distribution for different spheroid sizes. Corresponding to the increased number of hidden parameters when spheroid size increases, the amount change in hidden parameters enacted through the learning rule is also decreased. (f) The fractional change in spheroid surface area between the initial and trained configuration is negatively correlated to the change in stress. The correlation becomes stronger with increasing spheroid size.}
  \label{fig:5}
\end{figure*}

These trends already suggest that increasing the number of target cells introduces competing constraints that must be resolved through a combination of parameter updates and topological rearrangements. In particular, the simultaneous satisfaction of multiple target stresses requires coordinated, system-wide adjustments of the hidden-cell shape indices. This coordination may involve cellular rearrangements. Certainly, there is an outlying example for $N_T=4$ in which approximately 25 percent of the cell contact network has changed from the initial topology and those rearrangements allowed the system to reach a new local minimum, i.e., a discontinuous jump, in the cost function landscape such that the task was achieved. Thus, rearrangements can impact training performance. 

To further quantify these observations, we analyze ensembles of training runs across system sizes, as shown in Fig.~2. We train $N_T = 1, 2, 4$ randomly chosen target cells in systems with fixed total cell number, and track the evolution of error, overlap, and hidden-parameter distributions. As $N_T$ increases, we observe a systematic shift toward larger numbers of iterations required for convergence (Fig.~2(a)), along with a broader distribution of convergence times across realizations. At the same time, the cumulative overlap between the trained and initial configurations decreases and exhibits increased variability (Fig.~2(b)), indicating that more complex training tasks, on occasion, require the system to traverse more distant regions of configuration space and undergo more frequent topological rearrangements. Consistently, the standard deviation $\mathrm{SD}(s_0)$ increases with $N_T$ (Fig.~2(c)), reflecting the need for larger and more heterogeneous adjustments of hidden-cell parameters to accommodate multiple constraints.

These observations can be understood in terms of increasing constraint density,if you will. Each target cell imposes a condition on the system that must be satisfied simultaneously in the trained state. As the number of such constraints grows, the space of hidden-cell parameters that satisfies all constraints becomes progressively smaller, requiring more substantial parameter updates and, in some cases, topological rearrangements to reach a compatible configuration.

We also investigate how system size influences learnability by increasing $n_{\mathrm{total}}$ while keeping $N_T$ fixed (Fig.~2(d)). In this case, we find that the broadening of the hidden-parameter distribution decreases with system size. This indicates that larger systems, with more hidden cells, can accommodate the same set of target constraints with smaller parameter adjustments. In other words, increasing the number of hidden degrees of freedom reduces the effective constraint density, allowing the system to more efficiently distribute the learning task across the packing. 

To complete this sub-story, we should address yields in learning
the stress pattern. For one or two target cells, the training yield
remains high ($80-90$\%) for the given tolerance across all system sizes. 
However, as $N_T$ increases from 3 to 6, the yield decreases
systematically with, for example, $37$\% for $N_T=6$. This decrease reflects the increasing difficulty of simultaneously satisfying multiple target conditions as $N_T$ grows.  The dependence of the yield on system size is less systematic.  While larger systems provide more hidden degrees of freedom, this does not translate into a monotonic increase in yield across the system sizes studied, suggesting that geometric constraints and the accessibility of compatible configurations play a comparable role to the number of hidden degrees of freedom. Moreover, there are different types of failure modes. Failure occurs in realizations that remain elastic, with no cell rearrangements. Failure also occurs in realizations where some fraction of the cells rearrange, noting that rearrangements can also lead to a decrease in the error such that training is successful for the chosen tolerance.   

Collectively, these results indicate that multicellular learning in this system is controlled by a balance between the number of constraints imposed by target cells and the number of available hidden degrees of freedom. In regimes where the number of hidden cells is large compared to the
number of targets, the system behaves analogously to an overparameterized model, in which multiple solutions exist and learning proceeds efficiently. As the number of target cells/ constraint load increases, the system approaches an underparameterized regime in which solutions become sparse, convergence slows, and learning increasingly depends on rearrangements that enable exploration beyond local minima.
So while cellular rearrangements are not the primary
mechanism enabling successful training in typical
realizations, they serve as essential exploration events in outlying cases, enabling discontinuous transitions between solution basins when continuous parameter updates alone are insufficient to achieve the target stresses.

\subsection{Teaching Multicellular Patterns Sequentially}

A pattern consisting of multiple target cells trained in parallel can always be broken down into an ordered sequence of subpatterns. As an example, in Figure 3 we illustrate breaking down a pattern of $N_T=3$ cells into a $1\rightarrow1\rightarrow1$ sequence - by which we denote an ordered list of three distinct single-cell subpatterns. We investigate sequences generated by breaking up the main pattern of $N_T$ target cells into P ordered, non-overlapping subpatterns $n_{T_1}\rightarrow n_{T_2}\rightarrow\cdots\rightarrow n_{T_P}$ such that  $n_{T_1}+n_{T_2}+\cdots+n_{T_P} = N_T$

\begin{figure*}[t]
\centering
        \includegraphics[width=\linewidth]{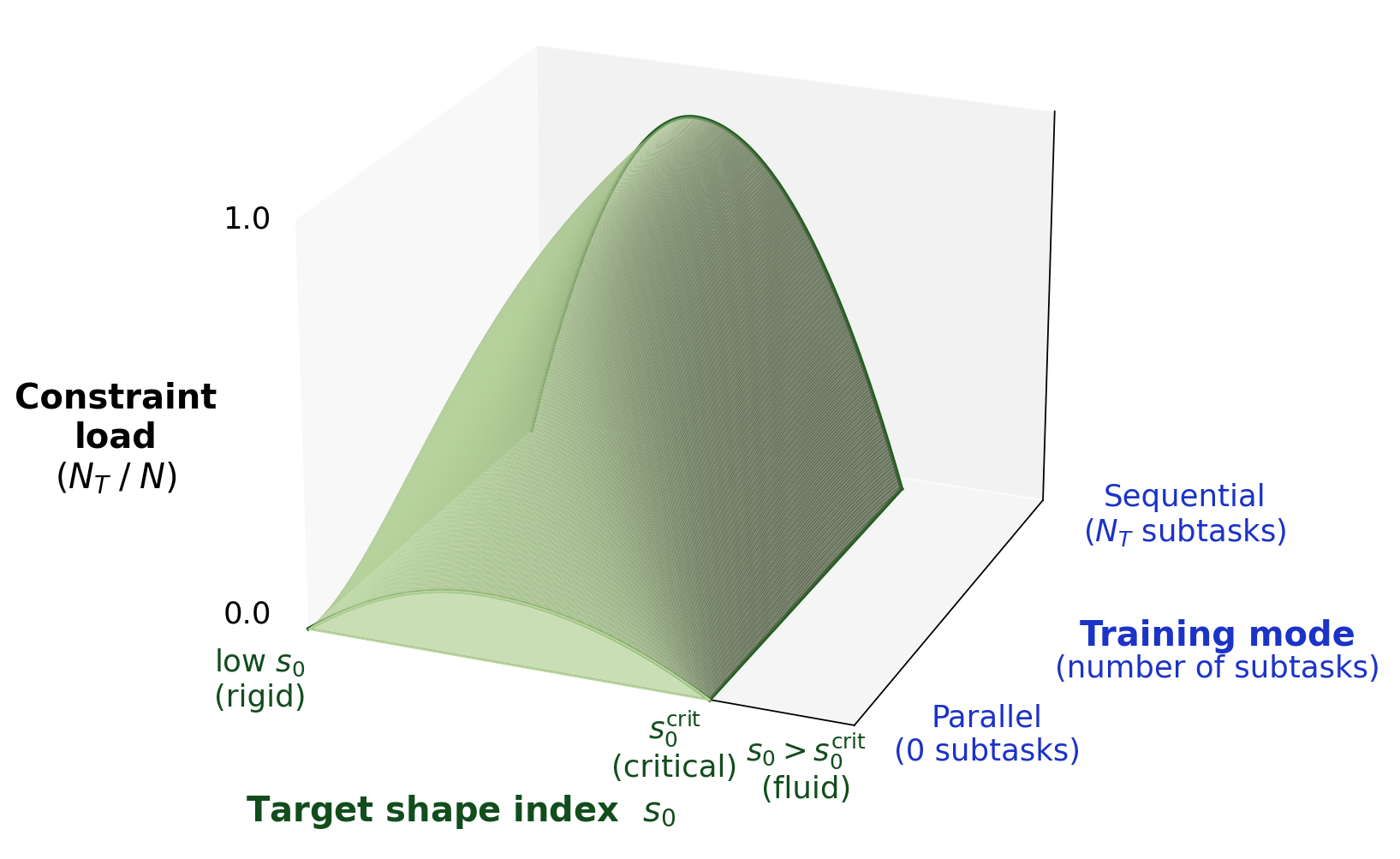}
        \label{learning_phase_diagram}
    \caption{{\it A conceptual learning phase diagram for training cell 
stresses.} The trainable region (green) is shown as a function of constraint load (ratio of target to hidden cells), cell packing rigidity (controlled by the
target shape index), and training mode. Rigidity governs an effective
exploration–exploitation balance: fluid-like regimes enhance exploration through
cellular rearrangements but do not necessarily lead to successful learning, as
increased rearrangements can also elevate error and destabilize target stress
patterns, while rigid regimes constrain accessible configurations and favor exploitation of existing states. Constraint load sets an effective capacity, with low values corresponding to overparameterized regimes that admit many solutions, and high values to constrained regimes in which the space of compatible configurations is reduced. The training protocol further shapes learnability: decomposing a task into subtasks can expand the set of accessible solutions by enabling intermediate configurations, but may increase training time due to incompatibility between subtasks. }
\end{figure*}

We can describe the overall training protocol for such a sequence of subpatterns as being structured in terms of epochs. Every training epoch commences by addressing each subpattern in sequence. The target cells in the subpattern being addressed are trained for target stress, while target cells in all other subpatterns are added to the set of hidden cells. Upon training each subpattern (up to training tolerance), we calculate the average error from \emph{all} target cells in the main pattern. The training terminates when, upon the training of some subpattern to training tolerance, the resulting average error from all target cells in the main pattern is reduced below training tolerance. We present the algorithm below:
\begin{figure}[htbp]
\centering
\begin{minipage}{0.95\columnwidth}
\hrule
\hrule
\vspace{0.5em}
\textsc{\textbf{Algorithm 4} Sequential Learning Algorithm}
\vspace{0.5em}
\hrule
\begin{algorithmic}[1]
\Require {$TOL$, $\{\sigma^{T}\}$, $\{s_0^{T_1}\}$,$\cdots$,$\{s_0^{T_N}\}$,$\{s_0^H\}$,$\{s_0^P\}$, $\gamma$}
\State $C \gets C(\{\sigma^T\},\{\sigma^T_{free}\})$ \Comment{Initial main pattern error}
\While {$C>TOL$} \Comment{training epochs}
\For{$i = 1$ to $N$}
 \Comment{loop on sub-patterns}
\State $\{s_0^H\} \gets \{s_0^H\}\cup\{s_0^{T_1}\}\cup\cdots\cup\{s_0^{T_N}\}\setminus\{s_0^{T_i}\}$ 
\State Let $\{\sigma^{T_i}\}$ be the target stresses corresponding to current sub-pattern
\State Train sub-pattern using $\{\sigma^{T_i}\}$, $\{s_0^{T_i}\}$, $\{s_0^H\}$, $\{s_0^P\}$, $\gamma$ in Algorithm 1
\State $C \gets C(\{\sigma^T\},\{\sigma^T_{free}\})$ \Comment{Main pattern error}
\If{$C<TOL$}
\State \textbf{break}
\EndIf
\EndFor
\EndWhile
\end{algorithmic}
\vspace{0.5em}
\hrule
\end{minipage}
\end{figure}


We first compare sequential and parallel training across different decompositions of the target pattern (Figs.~3(a)-(c)). For $N_T = 2, 3, 4$ in systems with $n_{\mathrm{total}} = 64$ cells, sequential training is consistently less efficient than parallel training, requiring more than an order of magnitude more iterations to converge. This behavior is robust across different choices of subpattern decomposition, including $1 \rightarrow 1 \rightarrow 1$ and $1 \rightarrow 2$ sequences, indicating that the inefficiency is not tied to a particular ordering, but is instead a general feature of sequential learning in this regime.

To understand the origin of this inefficiency, we examine the evolution of the error during training (Fig.~4). Within a given epoch, training one subpattern typically increases the error of previously trained subpatterns. That is, fitting one subset of target cells drives the system away from configurations that satisfy others. As successive subpatterns are trained, the system undergoes repeated cycles of improvement and degradation, reflecting an inherent interference between subpatterns when trained sequentially, where interference is the increase in error of previously trained target cells induced by updates that reduce the error of a different subpattern. Increases in interference reflect the geometric incompatibility of the corresponding solution sets in parameter and configuration space. However, as training progresses across epochs, the magnitude of this interference decreases, and the system eventually converges to a configuration that satisfies all subpatterns simultaneously. This behavior can be understood in terms of the geometry of the solution space. Each subpattern defines a set of configurations in hidden-parameter space that satisfies its corresponding constraints. Sequential training amounts to moving between these sets, rather than directly approaching their intersection. As a result, the system repeatedly departs from configurations that partially satisfy the full task, leading to cycles of forgetting and relearning. 

We also examine the distribution of hidden-cell target shape indices resulting from different sequential decompositions (Fig.~3(d)). While the standard deviation $\mathrm{SD}(s_0)$ increases systematically with the total number of target cells $N_T$, indicating that larger tasks require broader adjustments of hidden parameters, we do not observe a consistent dependence on the specific decomposition of the task into subpatterns. In particular, decompositions into smaller subpatterns (e.g., $1 \rightarrow 1 \rightarrow 1 \rightarrow 1$) do not lead to systematically smaller variance compared to coarser decompositions (e.g., $2 \rightarrow 2$). This suggests that the overall constraint imposed by the full target pattern, rather than the manner in which it is partitioned during training, primarily determines the extent of parameter variation. Fig. 3(d) In this sense, task decomposition affects learning dynamics but not the structure of the solution.

Further insight into the underlying training mechanisms is provided by examining individual training trajectories (Fig.~S2). Two distinct classes of behavior are observed: 

I. Spring network-like: In the first row of Fig. S2 we present instances where the entire training commences and concludes successfully without any change in cell packing topology. The changes in the hidden parameters across each epoch of training can be monotonic or more complicated, as can the change in error over the epochs.

II. Training with rearrangements: In the second and third rows of Fig. S2 we present instances where updates to hidden parameters followed by energy minimization in the training algorithm lead to minimizations. The effect of reconnections on the error is complicated. In the second row we generally try to present instances where reconnections immediately lead to reductions in error; while in the third row we show instances where error increases immediately after a reconnection. But in a given training instance both can happen.

Thus, in some realizations, training proceeds without any change in cell packing topology, and convergence is achieved through continuous adjustments of hidden-cell parameters alone. In other cases, training involves discrete cellular rearrangements, which can either decrease or increase the error depending on the configuration. These observations indicate that rearrangements are not uniformly required for learning, but instead play a contingent role that depends on the specific realization. In particular, these results suggest that cellular rearrangements are not the dominant mechanism enabling successful training in typical cases, as was the case for parallel training. Rather, most realizations converge through coordinated adjustments of hidden-cell parameters with minimal topological change. However, in outlying cases where geometric constraints prevent simultaneous satisfaction of multiple subpatterns, rearrangements become important by enabling the system to escape locally constrained configurations and access new regions of configuration space.

All in all, the inefficiency of sequential training in the present setting can be attributed to the relatively low constraint density of the task. As discussed in Sec.~IV.B, when the number of hidden cells is large compared to the number of targets, the system possesses a broad manifold of solutions that simultaneously satisfy all constraints. In this regime, parallel training can directly converge to a compatible configuration, whereas sequential training introduces unnecessary path dependence by repeatedly fitting individual subpatterns. This behavior can also be interpreted in terms of a bias--variance tradeoff~\cite{Geman1992,Hastie2009,Advani2020}. Training a given subpattern in isolation reduces bias for that subpattern, but increases variance with respect to others by disrupting previously learned configurations. As training progresses across epochs, the system searches for configurations that balance these competing effects. In the present regime, where compatible solutions are readily accessible, this search process is less efficient than direct parallel optimization. Therefore, sequential training functions less as an efficient optimization strategy and more as a search process over configuration and topology space. While it is less effective than parallel training in regimes of low constraint density, it provides a mechanism for exploring solutions in more constrained or frustrated settings, where direct convergence may be hindered. At present, our systems may be too small to observe the benefits of sequential training in the typical case, but we hypothesize this trend.  
 
\subsection{Spheroids}
 
To move beyond fully periodic systems, we consider training within a localized subset of cells embedded in a larger packing. Specifically, we define an embedded spheroidal training region by selecting a connected set of cells around a chosen center. This region, referred to here as a spheroid, contains one central target cell and \(N_{\mathrm{sp}}-1\) surrounding hidden cells, so that the spheroid contains \(N_{\mathrm{sp}}\) trainable cells in total. All remaining cells in the packing are treated as passive. Passive cells retain fixed target shape indices \(s_0 = 5.0\), but are allowed to change position and participate in rearrangements during energy minimization. In this way, passive cells act as a surrounding medium that imposes geometric and mechanical constraints on the trainable subsystem.

From an initial minimized configuration, we construct embedded spheroids of varying size by selecting cells according to their distance from a chosen center. See Figs.~5(a) and (b). We investigate spheroids with \(N_{\mathrm{sp}} = 5, 10, 20,\) and \(40\) cells embedded within a larger system of \(216\) cells. 

We find that it is possible to successfully train the stress of the central target cell even in the presence of passive cells (Fig. 5(c)). However, the difficulty of the task depends strongly on the relative size of the spheroid. As the spheroid size increases, the number of hidden cells available to accommodate the target constraint also increases, while the relative influence of the passive surroundings decreases. Correspondingly, we observe that the number of training iterations required for convergence decreases with increasing spheroid size (Fig.~5(c)), and the number of cellular rearrangements required is also reduced (Fig.~5(d)) as does the standard deviation of the hidden target shape index distribution. 

These trends can be understood in terms of constraint load. Passive cells effectively impose external constraints on the spheroid, limiting the accessible configurations of the hidden cells. For small spheroids, the number of hidden degrees of freedom is limited relative to these constraints, making it more difficult to realize the desired target stress. As the spheroid size increases, the number of hidden degrees of freedom grows, allowing the system to distribute the learning task more effectively across the packing and reducing the need for large parameter changes or rearrangements. Consistent with this picture, the magnitude of changes in hidden-cell parameters decreases with increasing spheroid size (Fig.~5(e)), indicating that larger spheroids require less extreme adjustments to achieve the same target stress. This behavior further reinforces the interpretation of hidden cells as effective degrees of freedom: increasing spheroid size increases capacity, allowing the system to distribute constraints more efficiently and reducing the need for large parameter changes or rearrangements.

At the smallest scales, we identify a minimal spheroid size required to successfully learn the task (Fig.~S3). In particular, a spheroid consisting of one target cell and four hidden cells represents the smallest configuration in which the desired stress can be achieved. Below this size, the number of hidden degrees of freedom is insufficient to accommodate the imposed constraint under the geometric restrictions of the surrounding packing as indicated by zero yield for the given number of maximum iterations and error tolerance.  Note that as the size of the spheroid increases from 5 to 40 (for $N=125$), the yield increases from $41$\% to $91$\%. Moreover, as $N$ increases to $N=216$, the yields decrease overall given the increase in passive cells whose target shape index is not being adjusted. 

Prior work has demonstrated correlation between cell target shape indices and/or interfacial tension between a spheroid and its surrounding ECM and spheroid shape with, for example, increasing interfacial tension leading to smaller spheroid area in the fluid phase~\cite{Parker2020,Parker2024}. To demonstrate this effect, see Fig. S4 where there is a gradient in target cell shape indices and so the spheroid contracts near the lower cell shape indices and expands near the higher cell shape indices. 

As for training one cell in the spheroid to a target stress, we observe a clear relationship between the mechanical state of the spheroid and its geometry. The change in the target-cell stress is negatively correlated with the change in the total surface area of the spheroid (Fig.~5(f)). Increasing the stress of the target cell is achieved through a coordinated decrease in the target shape indices $s_0$ of hidden cells, leading to an overall rigidification and compaction of the spheroid. Conversely, decreasing the target-cell stress corresponds to an increase in $s_0$, resulting in a more fluid-like state and an expansion of the spheroid. This correlation demonstrates that training local stresses induces global geometric changes, reflecting the coupled nature of mechanics and structure in cellular packings. These findings highlight the role of the surrounding environment in modulating learnability. Passive cells act as external constraints that can hinder or facilitate learning depending on the availability of hidden degrees of freedom, while the resulting trained states exhibit a tight coupling between local stress and global morphology.

\section{Discussion}

The results presented here demonstrate that three-dimensional cellular packings constitute a new class of trainable matter {\it with localized outputs (and inputs), in which learning emerges not just from abstract weight updates on a fixed architecture, but with a reconfigurable architecture}. By training local cellular stresses, we show that multicellular systems can be programmed to realize prescribed internal patterns through physically grounded learning rules, thereby extending contrastive learning frameworks beyond fixed networks to reconfigurable, living-like materials.

Several key principles emerge. First, learning in cellular systems is intrinsically collective. The training of a even single target cell induces system-wide adjustments of hidden-cell parameters, indicating that information is encoded in distributed mechanical states rather than localized modifications. Second, learnability is strongly shaped by the mechanical state of the system. Lowering stress, which can locally fluidize the tissue, enhances exploration of configuration space and can accelerate convergence, though global fluidization leads to essentially zero cellular stresses and so the system cannot be trained, at least for these tasks. On the other hand, increasing stress drives the system into rigid regimes where geometric constraints limit accessible solutions. Third, each target cell imposes a constraint that must be satisfied simultaneously in the trained state, while hidden cells provide the degrees of freedom available to meet these constraints. When the number of hidden cells is large relative to the number of targets, the system operates in an overparameterized regime with a large manifold of compatible solutions. As the constraint load increases, this manifold shrinks and the system transitions toward an underparameterized regime, where solutions become sparse. In this regime, learning requires larger and more heterogeneous parameter updates and may depend on cellular rearrangements to access otherwise inaccessible regions. The existence of a minimal spheroid size capable of learning further underscores that learnability is governed by a balance between constraints and hidden degrees of freedom, analogous to capacity limits in conventional learning systems. Finally, as stress and shape can be related, we find that training for cell stress can impact spheroid shape which gives us a window into morphogenesis as a tried-and-true learning algorithm. 

 Rearrangements act as a conditional but powerful resource for learning: while not required in typical cases, they provide a mechanism for discontinuous transitions between solution basins, enabling exploration when continuous parameter updates are insufficient. In the typical regime, learning proceeds primarily through smooth adjustments of hidden-cell parameters with minimal topological change, rendering the system effectively spring-network-like despite its potential for reconfigurability. However, rearrangements become essential in two limiting cases. When the system is driven toward more fluid-like states, rearrangements enhance learning by enabling exploration of otherwise inaccessible regions of cost function space. In contrast, in highly constrained or incompatible tasks, rearrangements act as escape mechanisms that allow the system to traverse between disconnected solution basins. At the same time, rearrangements can hinder learning by disrupting partially learned configurations, particularly in sequential protocols where updates to one subset of targets degrade others. Thus, rearrangements introduce a fundamentally new ingredient into learning: a source of discrete, nonlocal transitions that can both expand expressivity and generate incompatibility. This dual role highlights that learning in tissues is governed by the interplay of continuous parameter adaptation and discontinuous structural reconfiguration.

Our findings motivate a learning phase diagram for cellular systems, organized by three control parameters: constraint load (e.g., $N_T$/$N$), training protocol (parallel versus sequential) in terms of the number of subtasks, and the initial dimensional shape index of the cells.  See Fig. 6. For an intermediate $s_0(0)$, in the low-complexity regime, where the number of constraints is small relative to the hidden degrees of freedom, the system exhibits a highly degenerate solution space. In this regime, parallel training is most effective: all constraints can be addressed simultaneously, and the system converges efficiently without requiring extensive exploration. Sequential training, by contrast, leads to cycles of forgetting and relearning and, thus, takes longer. As task complexity increases, the structure of the solution space changes qualitatively. The intersection of constraint manifolds shrinks and may fragment, rendering direct convergence increasingly difficult. In this regime, parallel training becomes less effective, and sequential or partially sequential strategies gain relevance by decomposing the task and enabling exploration through intermediate configurations. Rearrangements become more prominent, reflecting the need to navigate a rugged and potentially disconnected landscape. Although the present system sizes likely probe only the onset of this transition, the observed increase in variability and rearrangement-driven events strongly suggests a crossover from a parallel-dominated regime to one in which sequential strategies—and more generally, history-dependent protocols—become advantageous. For higher $s_0(0)$, the packing fluidizes and there is no learning beyond $s_0(0)=5.4$.  For lower $s_0(0)$ (but bounded below by the shape index of a sphere), the systems becomes more geometrically frustrated and so it is more difficult to train unless there are some rearrangements and/or subtasks are implemented.

Recent work has established that epithelial tissues can behave as self-tuning materials in which internal mechanical parameters—such as junctional tensions or preferred cell shapes—adapt under local, mechanosensitive rules \cite{arzash2025rigidity,arzash2025learning, arzash2025epithelial,pasqui2026vertax}. Across these studies, tissues are shown to learn through repeated deformation or driving, leading to the emergence of bulk material properties, including rigidity transitions, morphogenetic flows, and programmable elastic responses. In other words, learning corresponds to the evolution of internal variables that define the constitutive behavior of the tissue. In contrast, the present work focuses on a complementary regime in which learning is defined at the level of spatially resolved, cell-scale observables, namely stress patterns. Rather than training bulk response functions, we train heterogeneous internal states that require coordinated, system-wide adjustments and, in some cases, topological rearrangements. This distinction highlights two distinct but related modes of learning in tissues: one in which tissues learn what material they are, and another in which they learn what internal patterns they can realize. By enabling the training of detailed, spatially structured outputs, our framework moves us closer to viewing tissues not only as adaptive materials, but as candidate physical AI platforms in which computation and function are encoded directly in the learned internal state.

These findings connect naturally to recent demonstrations of learning in granular systems, where disordered particle packings can encode memories through cyclic driving \cite{guo2026learningassociationsreconfigurableparticle}. In such systems, learning is mediated almost entirely by contact rearrangements, with memory stored in the history-dependent organization of the network. Cellular, and even subcellular~\cite{banerjee2026learning}, systems occupy a richer regime. Like granular materials, they can store information in rearrangements, but they also possess internal degrees of freedom—such as cell shape and contractility—that enable continuous parameter updates. As a result, tissues implement a hybrid form of learning that combines the discrete, history-dependent memory of granular systems with the continuous adaptability of physical neural networks. This suggests a unifying hierarchy: fixed-architecture systems learn through parameter updates alone; granular systems learn through rearrangements; and living tissues integrate both mechanisms. Learning, in this view, emerges from the capacity of matter to reconfigure—continuously or discontinuously—in response to experience.

Finally, these results suggest a reframing of biological function and dysfunction. If tissues can learn, then disease states may be interpreted as forms of \emph{mislearning}, in which maladaptive stress patterns and mechanochemical feedback loops become stabilized. Fibrosis, cancer progression~\cite{shomar2022cancer}, and aberrant morphogenesis may thus reflect learned configurations that are locally stable but globally detrimental. From this perspective, therapy becomes not only a matter of perturbation, but of \emph{retraining}—guiding the system back toward functional regions of its configuration space. At the same time, the ability to train cellular systems at the cellular level opens a path toward entirely new classes of materials. For instance, one can envision \emph{cellular weavers}: living systems that learn to construct and remoddel fibrous networks, organize extracellular matrices, or generate functional architectures through training rather than explicit design. Such systems would operate as embodied, adaptive fabricators, in which structure and function co-emerge through mechanochemical learning. Or perhaps a cell packing can act itself as an AI-like platform, as mentioned above, in which the patterns of cell stress contain information that depends on a pattern of stress inputs. Together, these results point toward a broader paradigm in which learning is a fundamental property of collective matter including sand, cells, and other reconfigurable systems.

\section*{Acknowledgements}
JMS acknowledges financial support from National Science Foundation via PoLS-2412961. Tao Zhang acknowledges financial support from the NSFC/China via award 22303051. We would like to acknowledge useful discussion with Alison Patteson, Michael Blatchley, Pranav Soman, Paul Janmey, and Minh Tranh. 
\section*{Data Availability}
All codes developed and implemented in this manuscript will be made available upon request. 
\bibliography{celltraining}
\clearpage
\onecolumngrid   
\setcounter{secnumdepth}{3}
\setcounter{equation}{0}
\setcounter{figure}{0}
\setcounter{table}{0}

\makeatletter
\renewcommand{\thefigure}{S\@arabic\c@figure}
\renewcommand{\thetable}{S\@Roman\c@table}
\renewcommand{\theequation}{S\@arabic\c@equation}
\makeatother
\section*{Supplementary Figures}

\begin{figure*}[h]
\centering
    \begin{subfigure}{0.19\linewidth}
        \includegraphics[width=\linewidth]{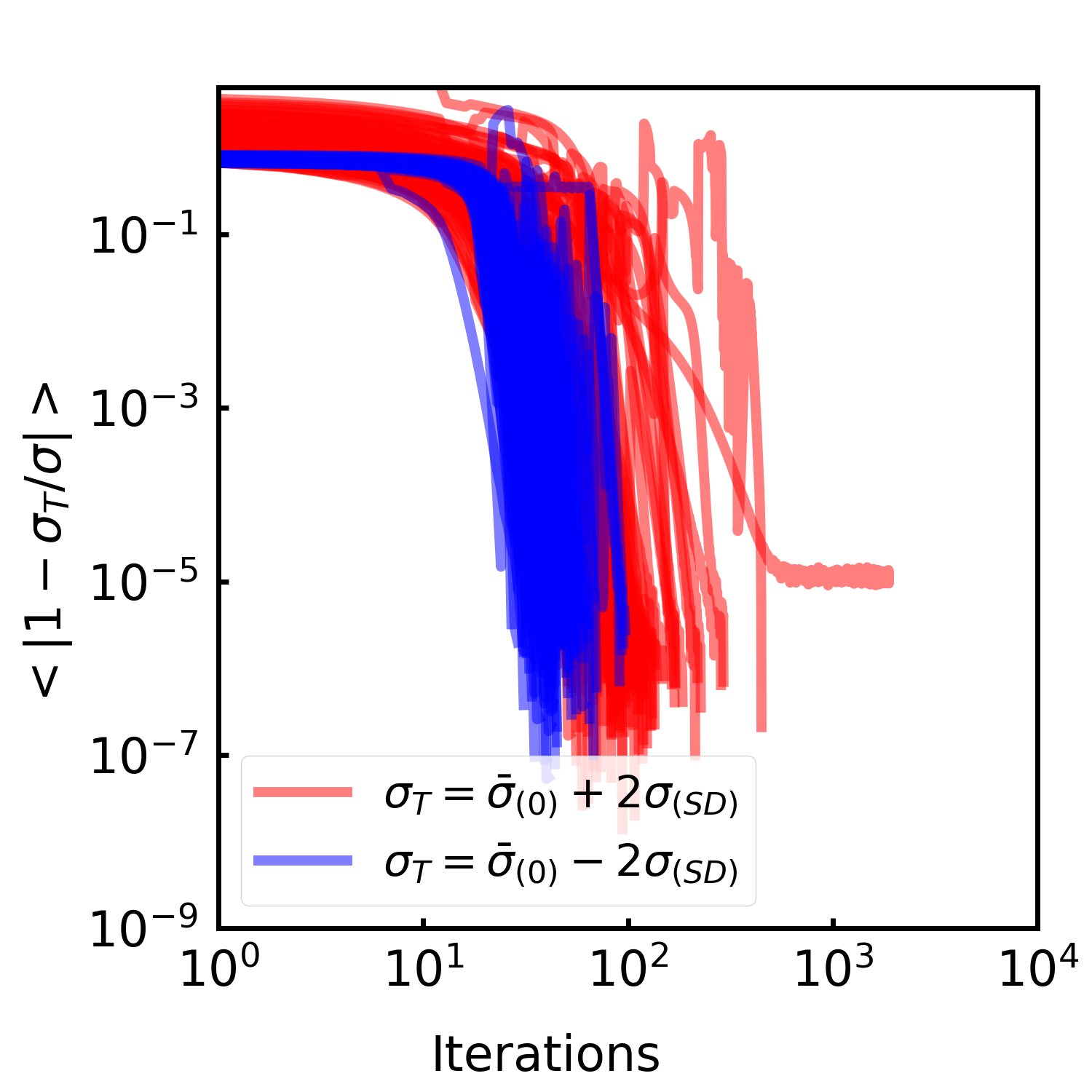}
        \caption{}
        \label{fig:single_cell_errors}
    \end{subfigure}
    \begin{subfigure}{0.19\linewidth}
        \includegraphics[width=\linewidth]{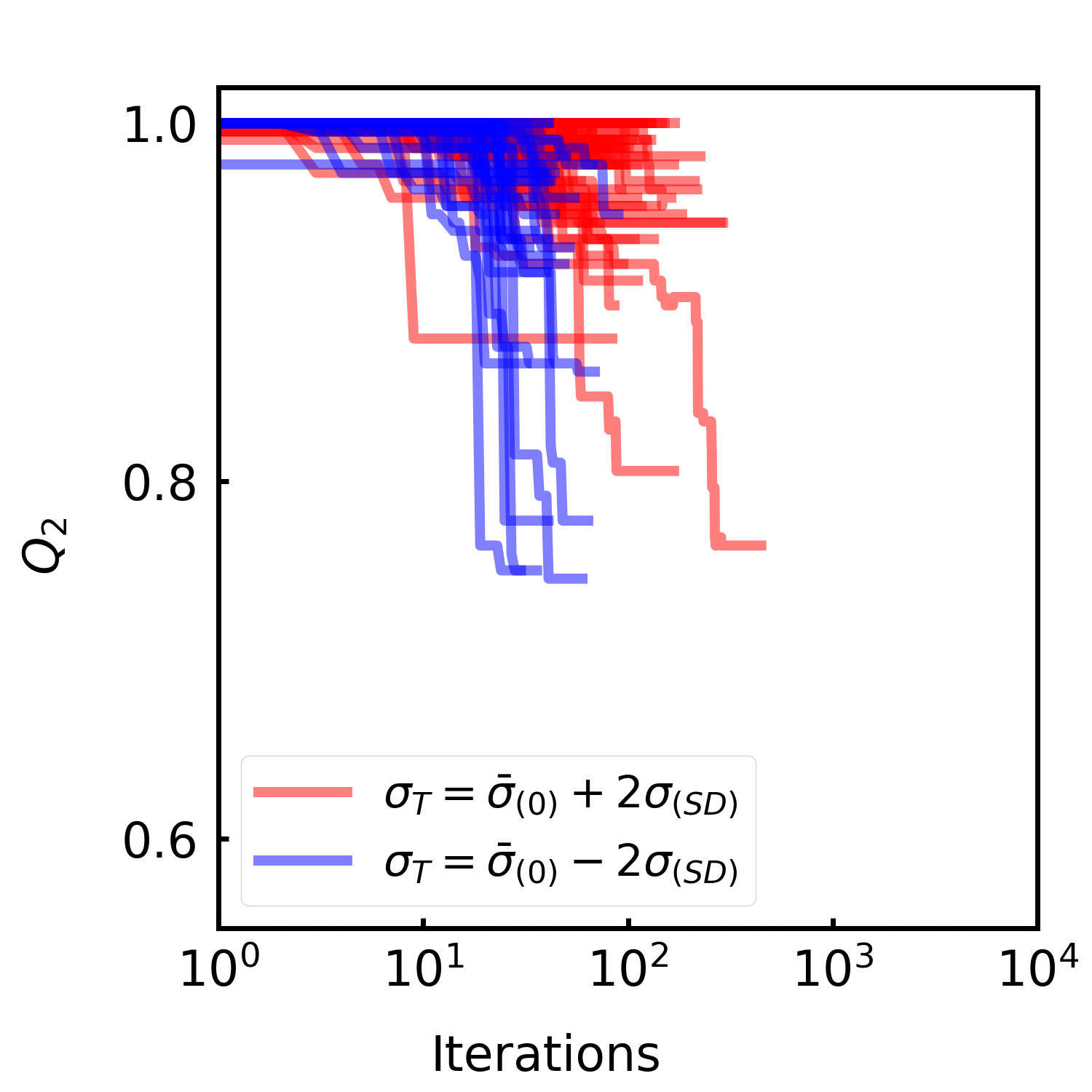}
        \caption{}
        \label{fig:single_cell_overlaps}
    \end{subfigure}
    \begin{subfigure}{0.19\linewidth}
        \includegraphics[width=\linewidth]{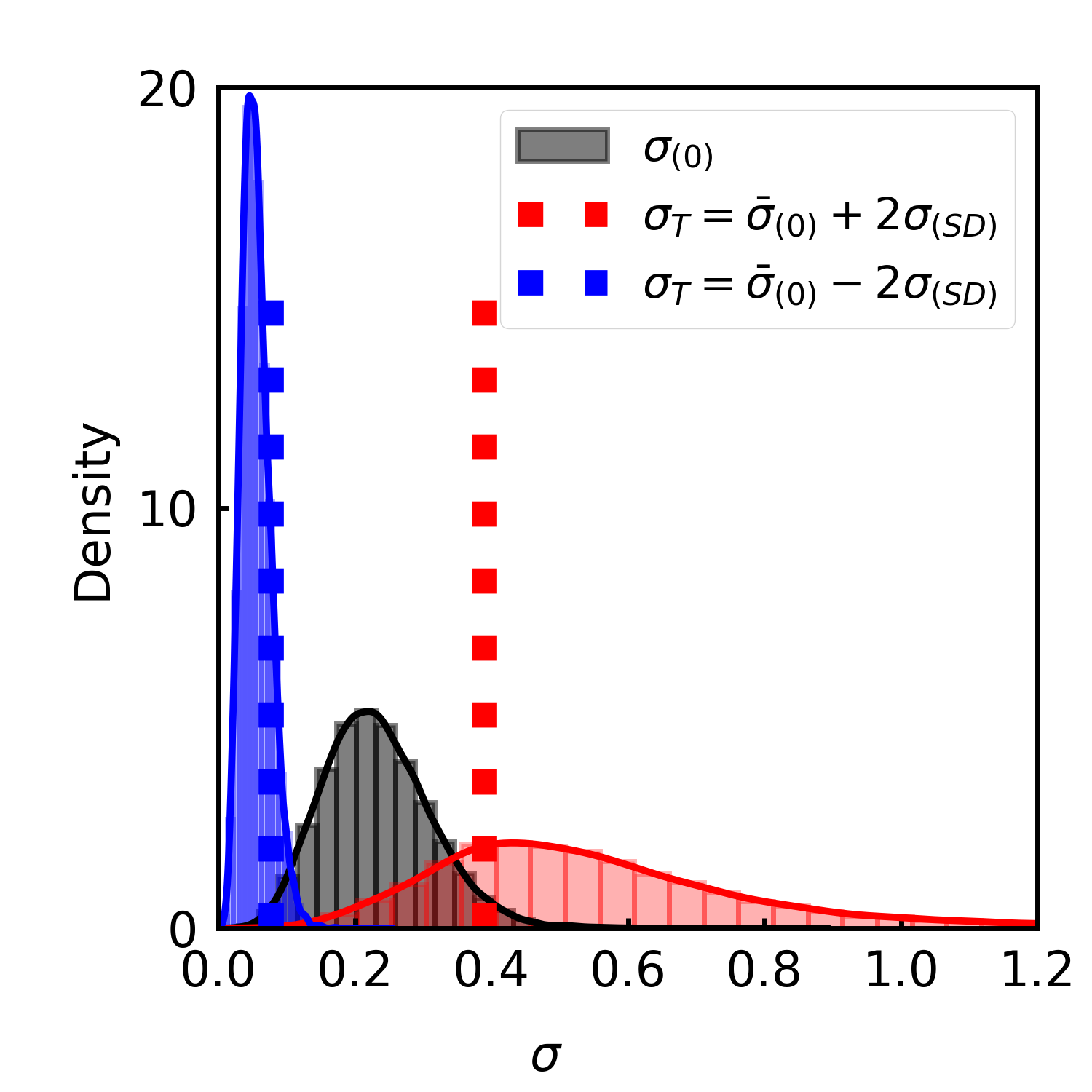}
        \caption{}
        \label{fig:single_cell_sigma_histogram}
    \end{subfigure}
    \begin{subfigure}{0.19\linewidth}
        \includegraphics[width=\linewidth]{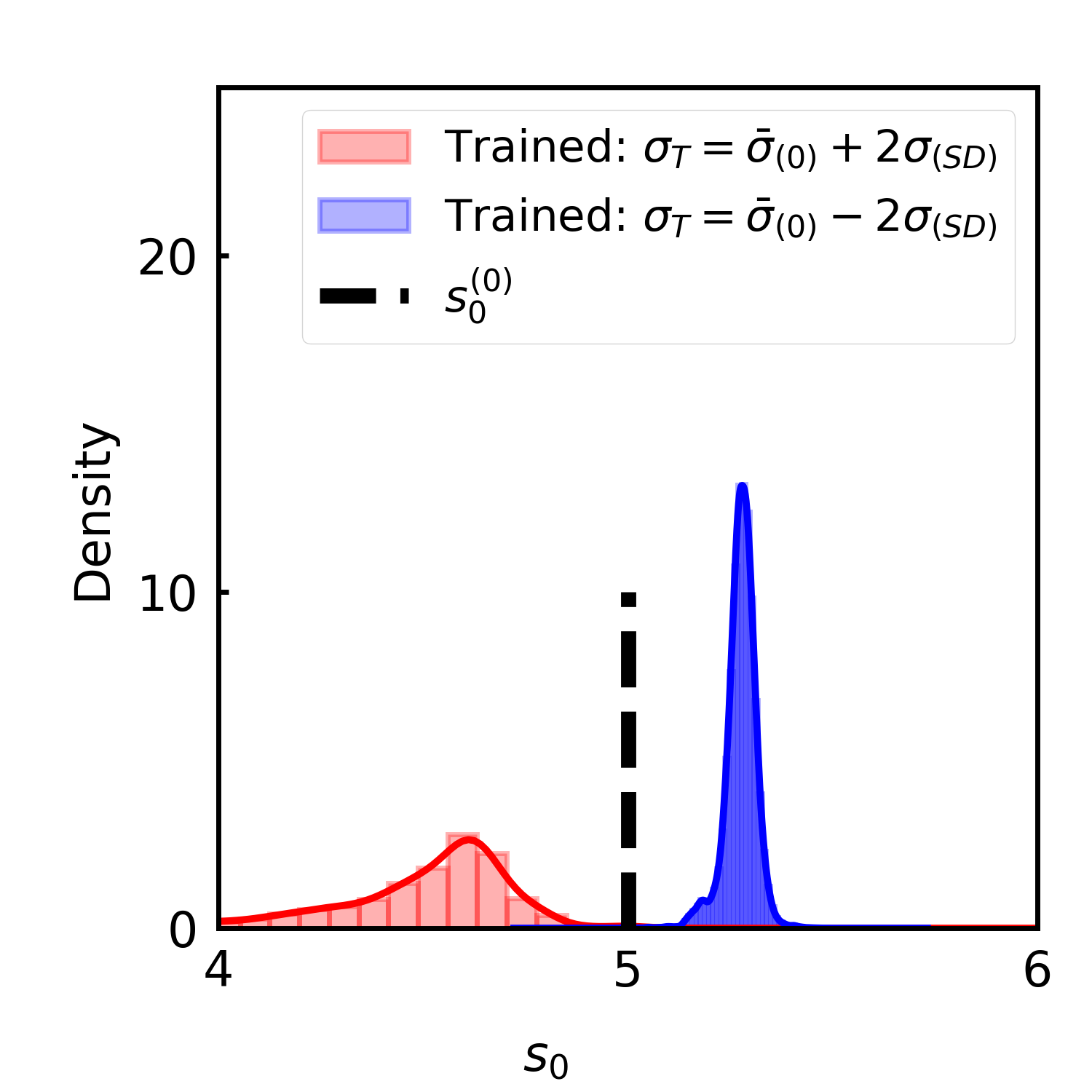}
        \caption{}
        \label{fig:single_cell_s0_histogram}
    \end{subfigure}
    \vfill
    \begin{subfigure}{0.19\linewidth}
        \includegraphics[width=\linewidth]{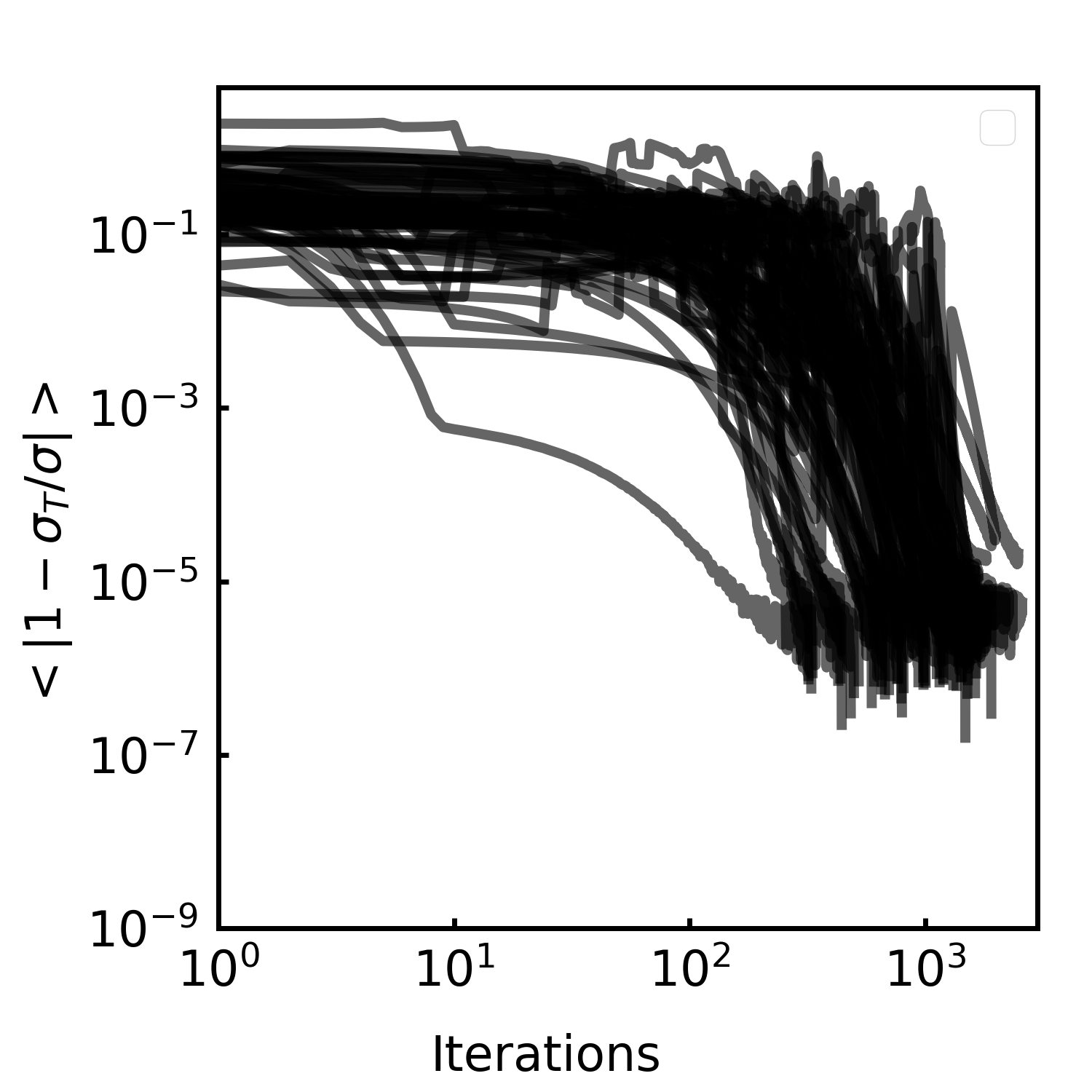}
        \caption{}
        \label{fig:1a}
    \end{subfigure}
    \begin{subfigure}{0.19\linewidth}
        \includegraphics[width=\linewidth]{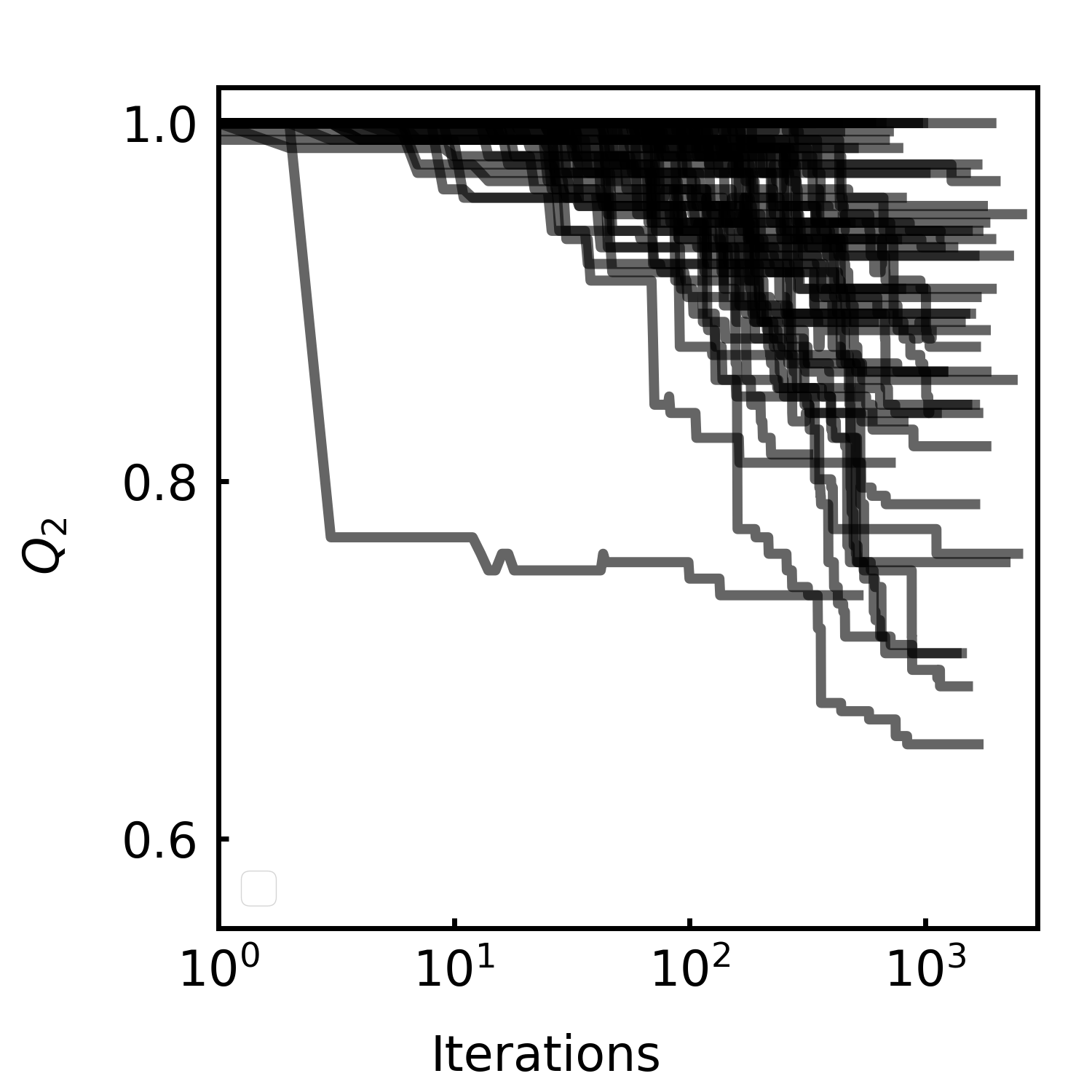}
        \caption{}
        \label{fig:1a}
    \end{subfigure}
    \begin{subfigure}{0.19\linewidth}
        \includegraphics[width=\linewidth]{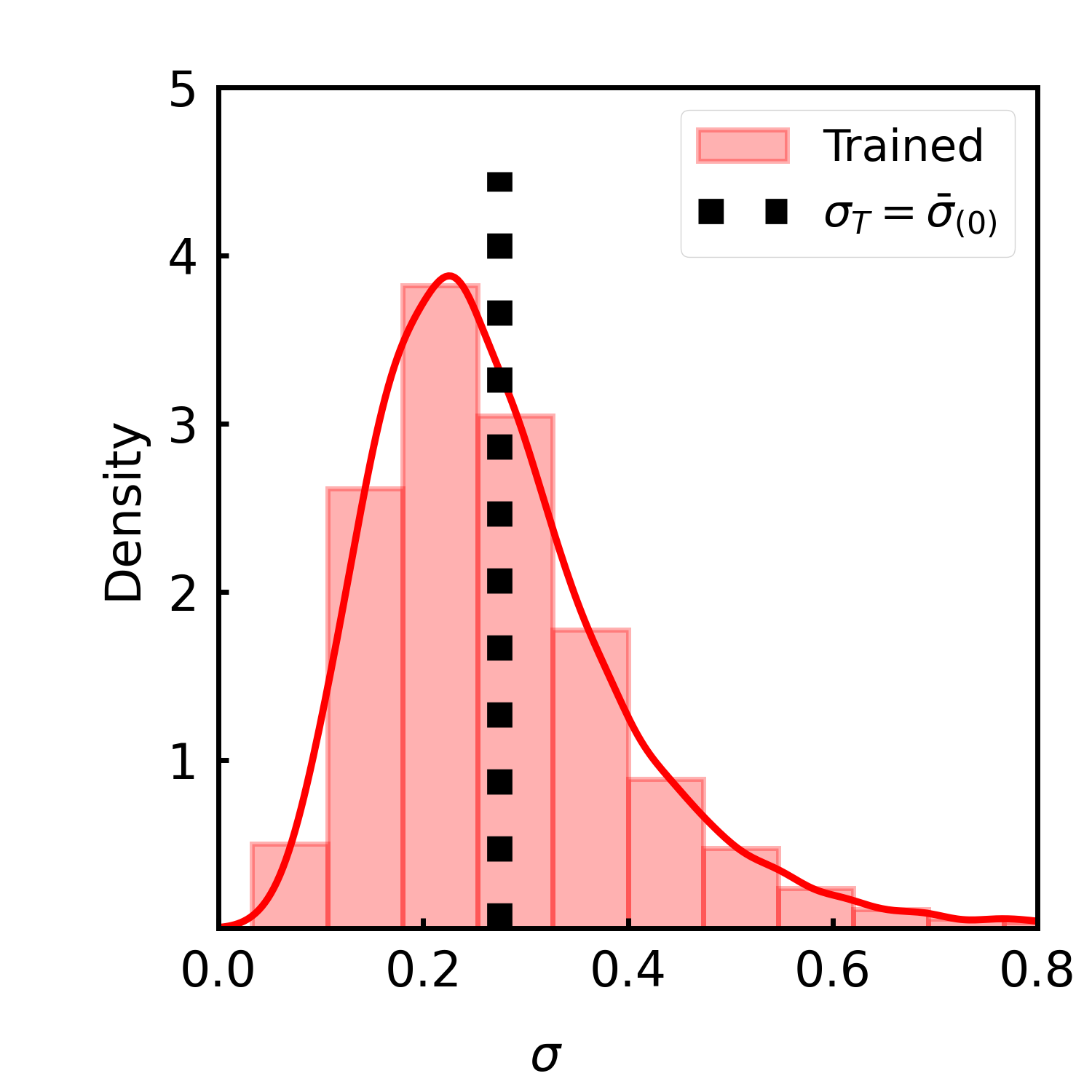}
        \caption{}
        \label{fig:1a}
    \end{subfigure}
    \begin{subfigure}{0.19\linewidth}
        \includegraphics[width=\linewidth]{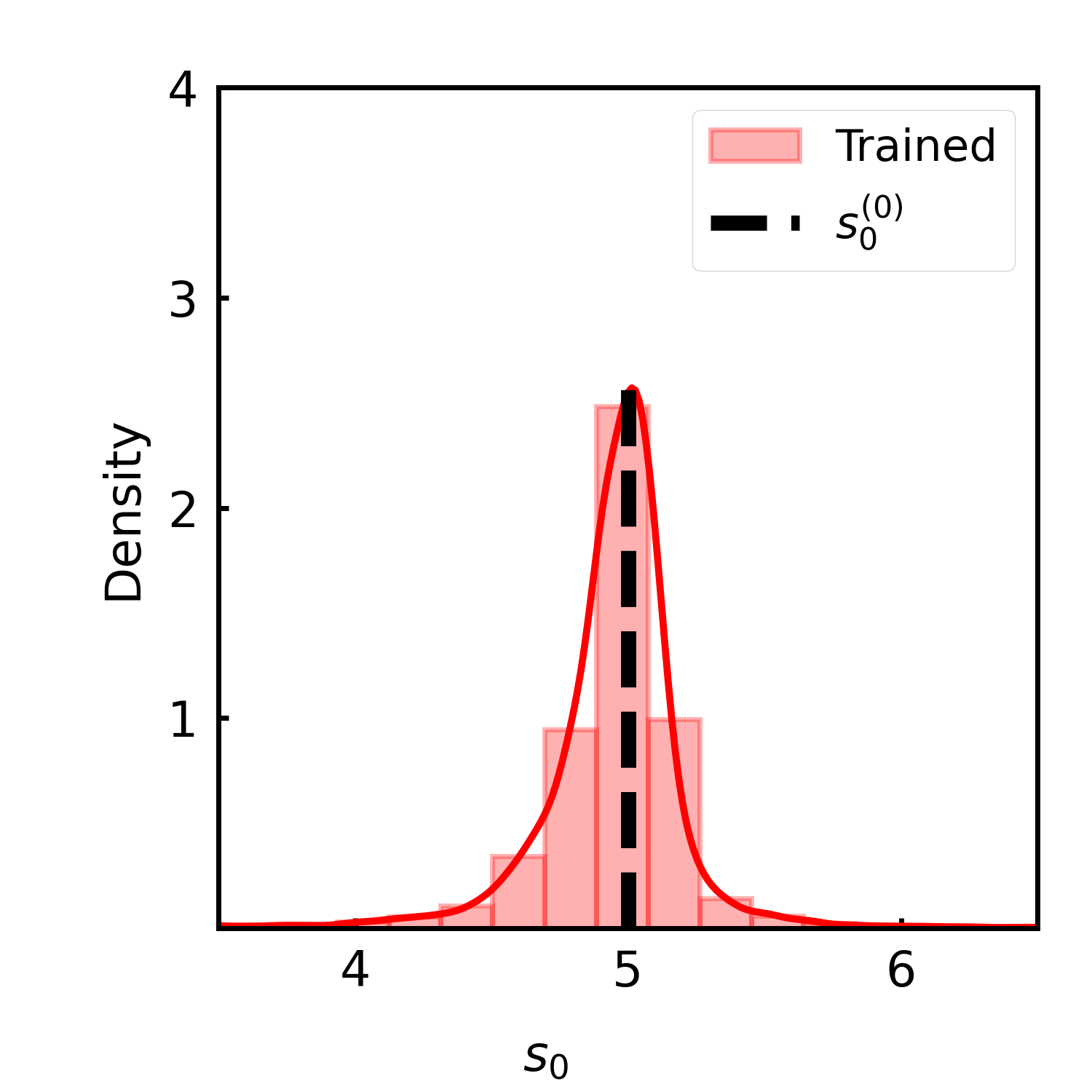}
        \caption{}
        \label{fig:1b}
    \end{subfigure}
    \vfill
    \begin{subfigure}{0.19\linewidth}       \includegraphics[width=\linewidth]{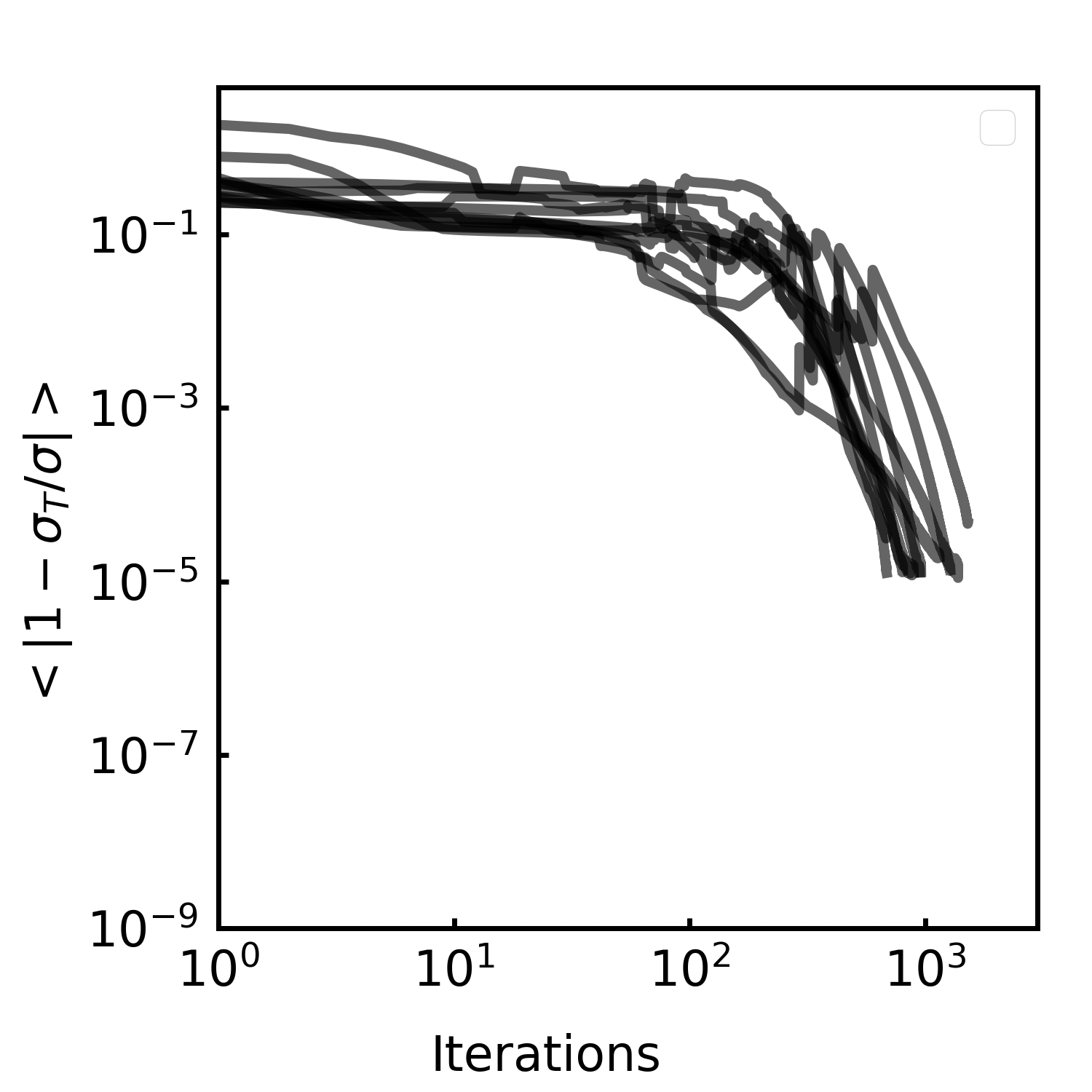}
        \caption{}
        \label{fig:1a}
    \end{subfigure}
    \begin{subfigure}{0.19\linewidth}
        \includegraphics[width=\linewidth]{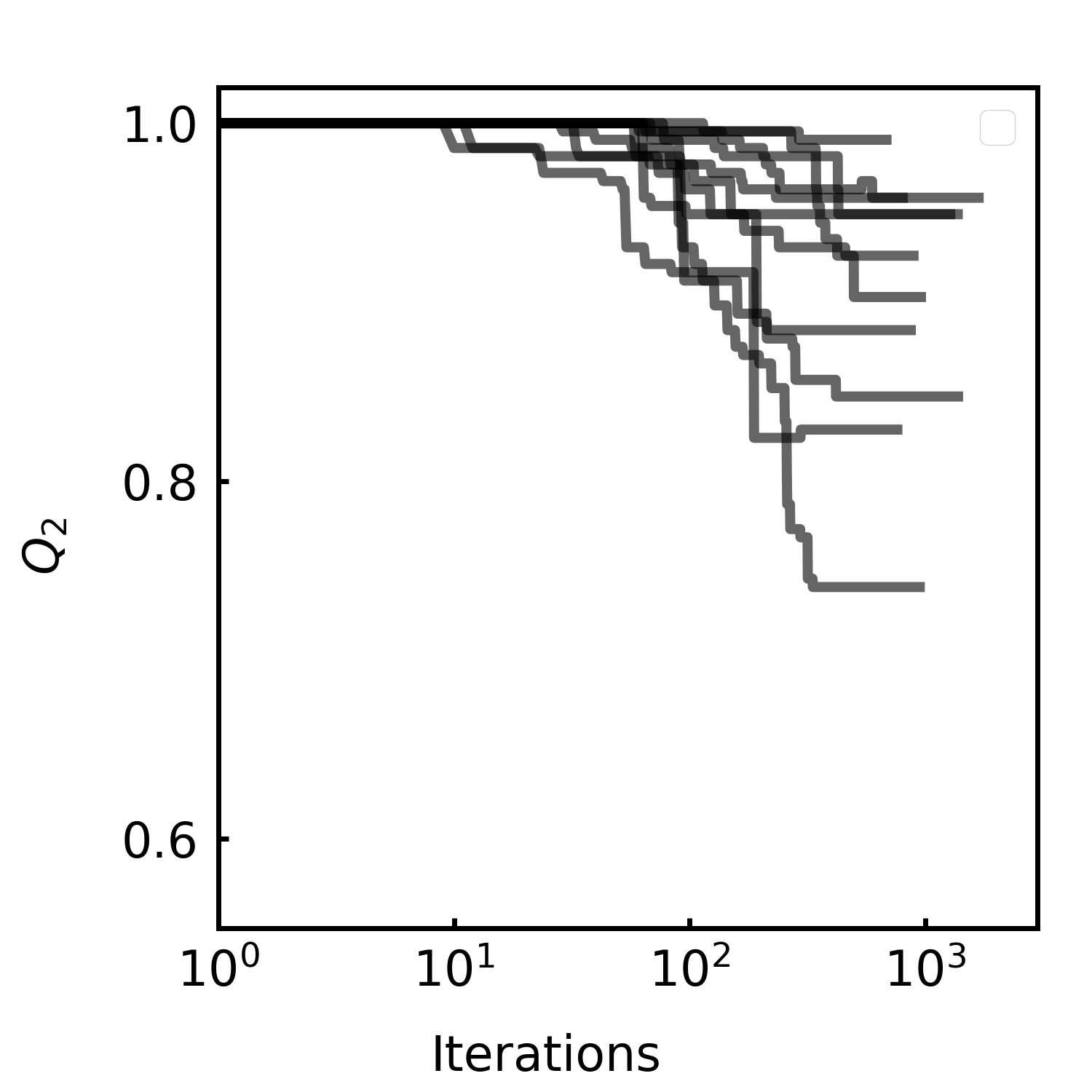}
        \caption{}
        \label{fig:1a}
    \end{subfigure}
    \begin{subfigure}{0.19\linewidth}
        \includegraphics[width=\linewidth]{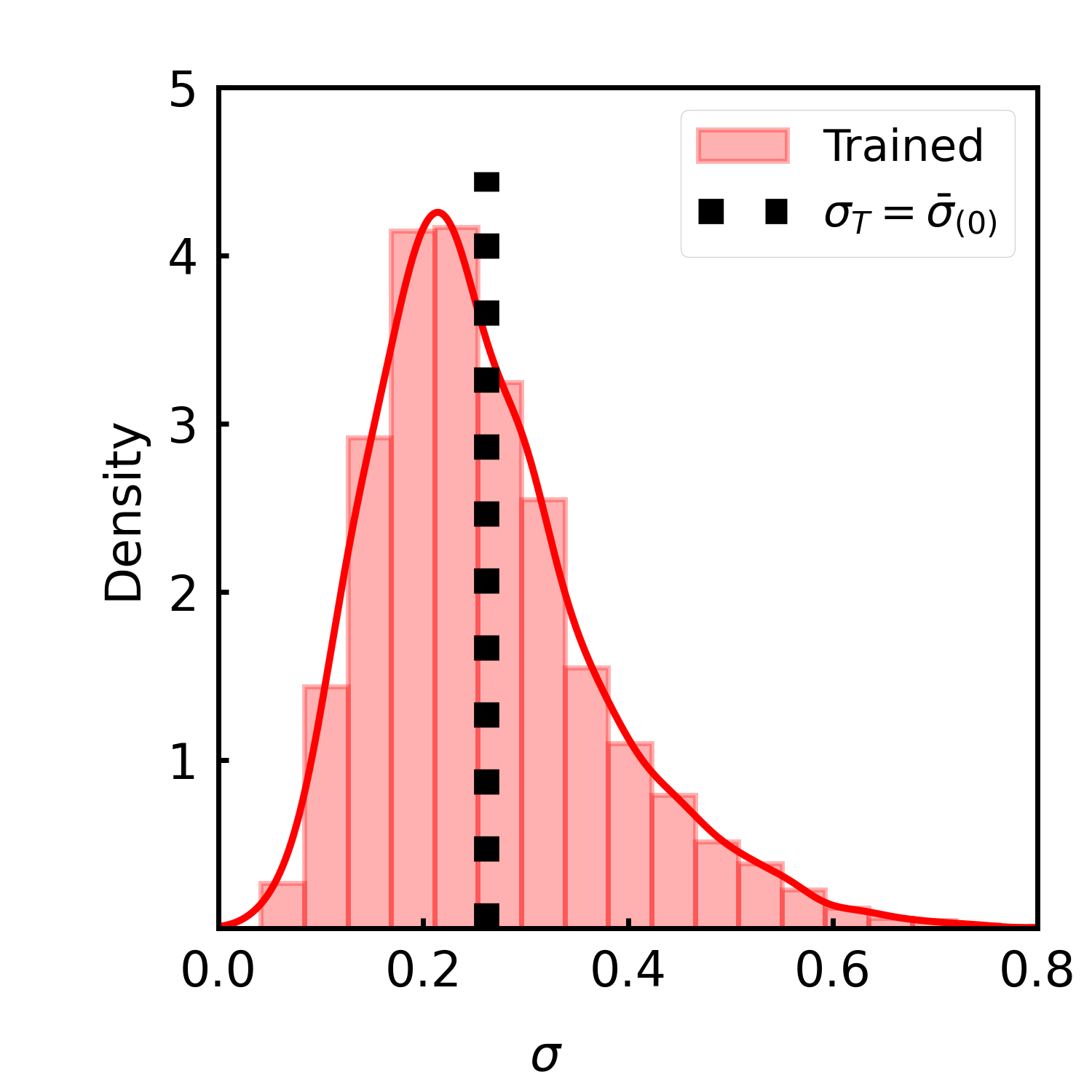}
        \caption{}
        \label{fig:1a}
    \end{subfigure}
    \begin{subfigure}{0.19\linewidth}
        \includegraphics[width=\linewidth]{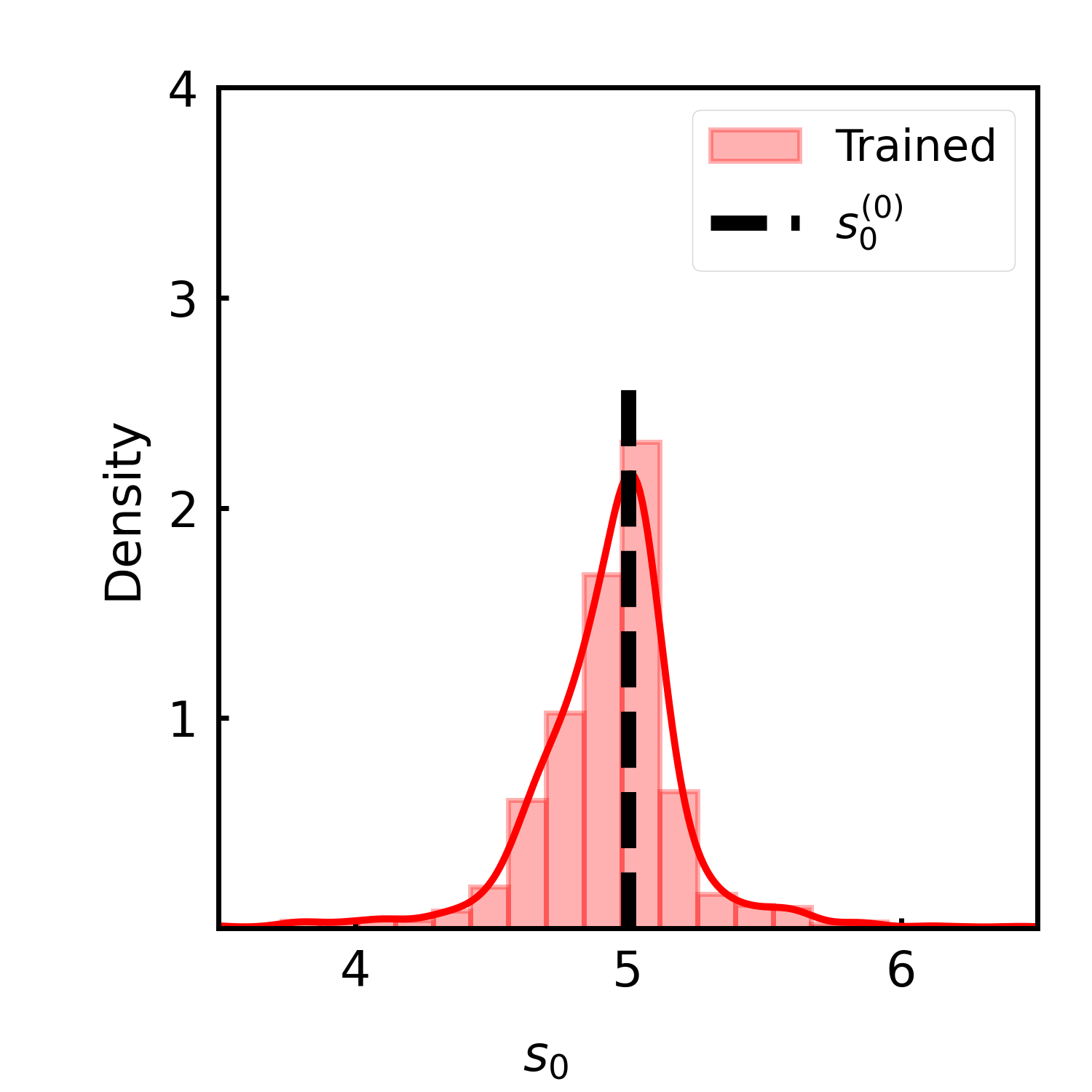}
        \caption{}
        \label{fig:1b}
    \end{subfigure}
    \caption{{\it Reducing geometric frustration enhances learning.} The same parameters are used as in Figure 1,  with the exception of $k_V = 1$. The row and column layout logic, as well as color scheme, are also repeated from Figure 1.}
  \label{fig:S1}

\end{figure*}
\begin{figure*}[t]
\centering
    \begin{subfigure}{0.32\linewidth}
        \includegraphics[width=\linewidth]{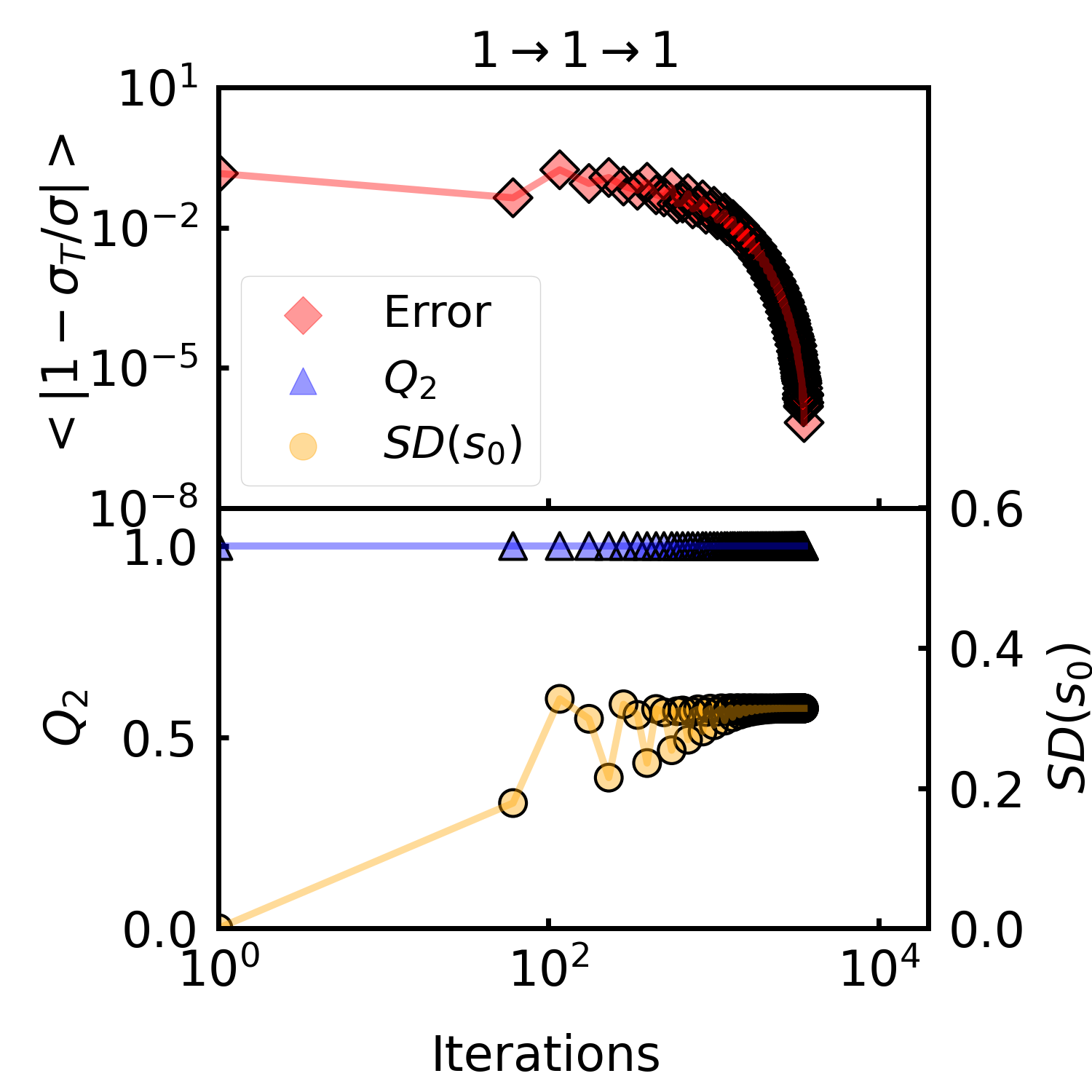}
        \caption{}
        \label{fig:single_cell_snapshots}
    \end{subfigure}
    \begin{subfigure}{0.32\linewidth}
        \includegraphics[width=\linewidth]{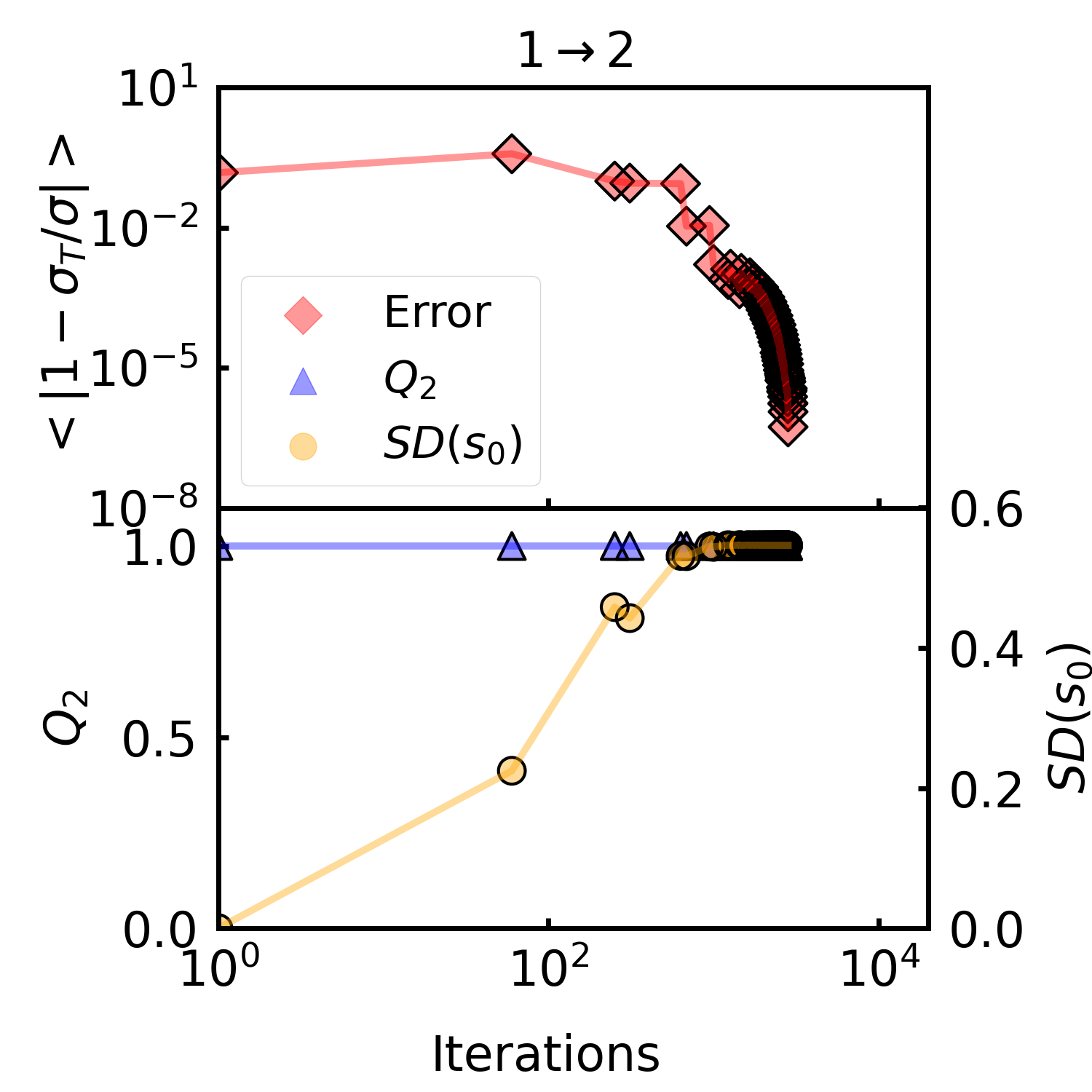}
        \caption{}
        \label{fig:single_cell_errors}
    \end{subfigure}
    \begin{subfigure}{0.32\linewidth}
        \includegraphics[width=\linewidth]{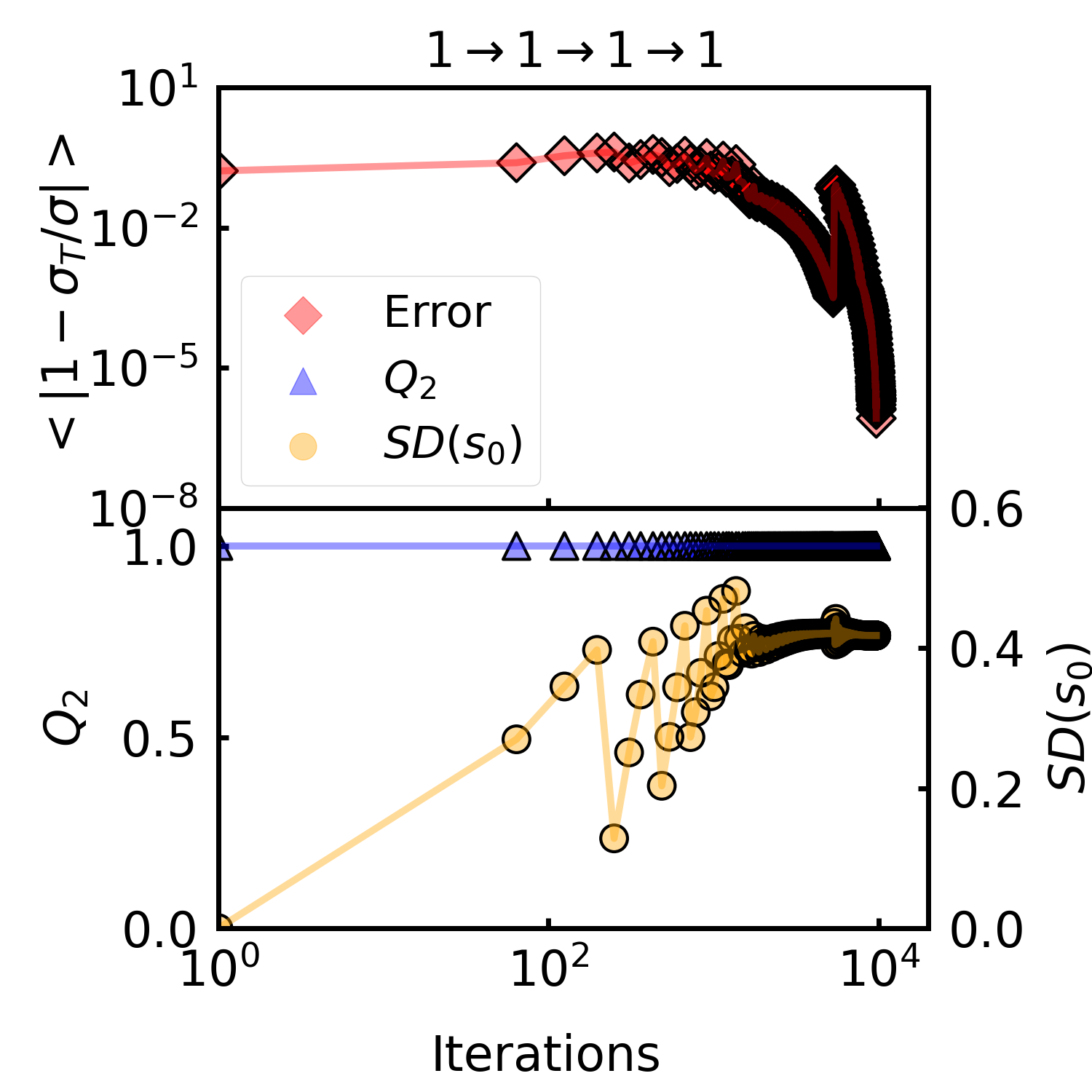}
        \caption{}
        \label{fig:single_cell_overlaps}
    \end{subfigure}
    \vfill
    \begin{subfigure}{0.32\linewidth}
        \includegraphics[width=\linewidth]{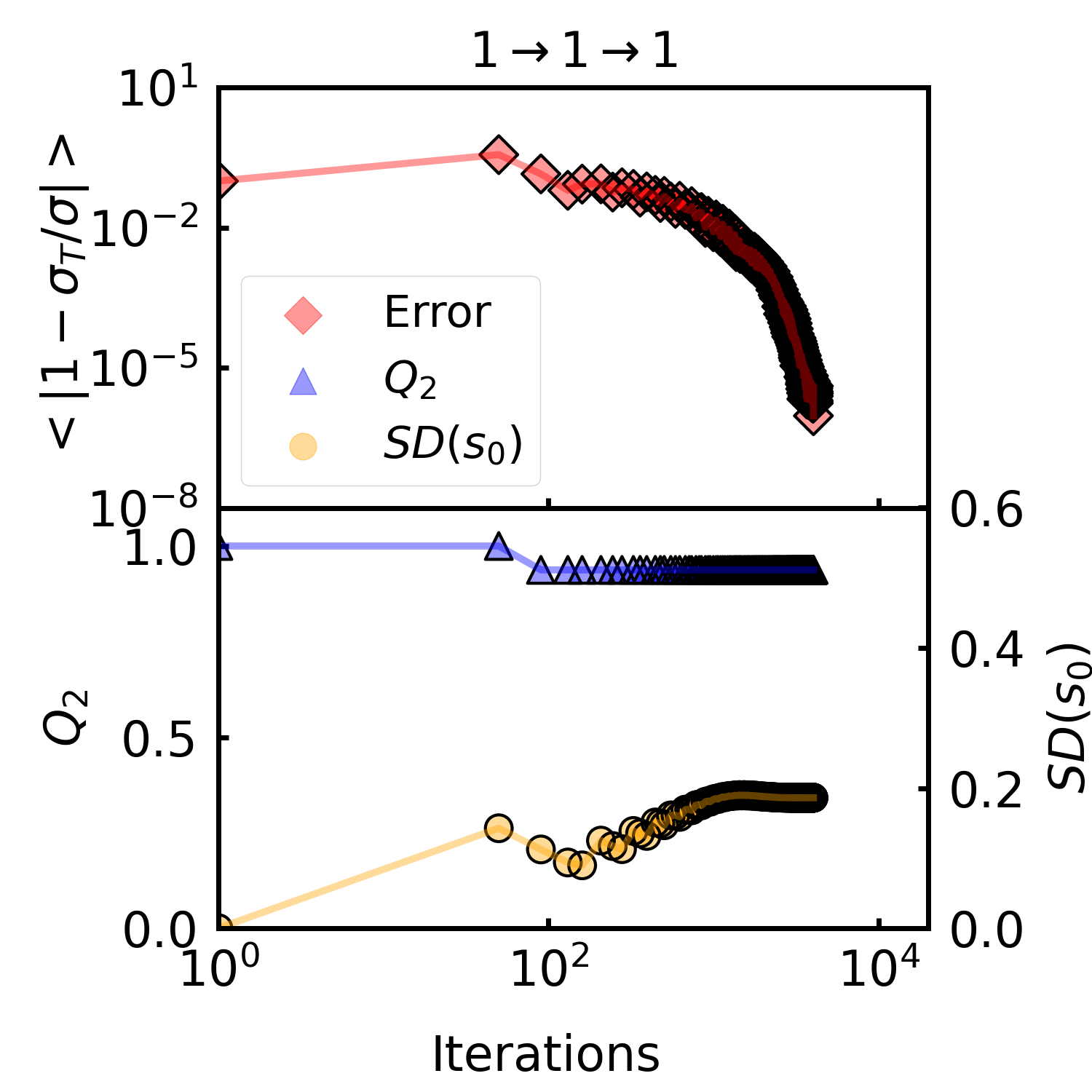}
        \caption{}
        \label{fig:single_cell_sigma_histogram}
    \end{subfigure}
    \begin{subfigure}{.32\linewidth}
        \includegraphics[width=\linewidth]{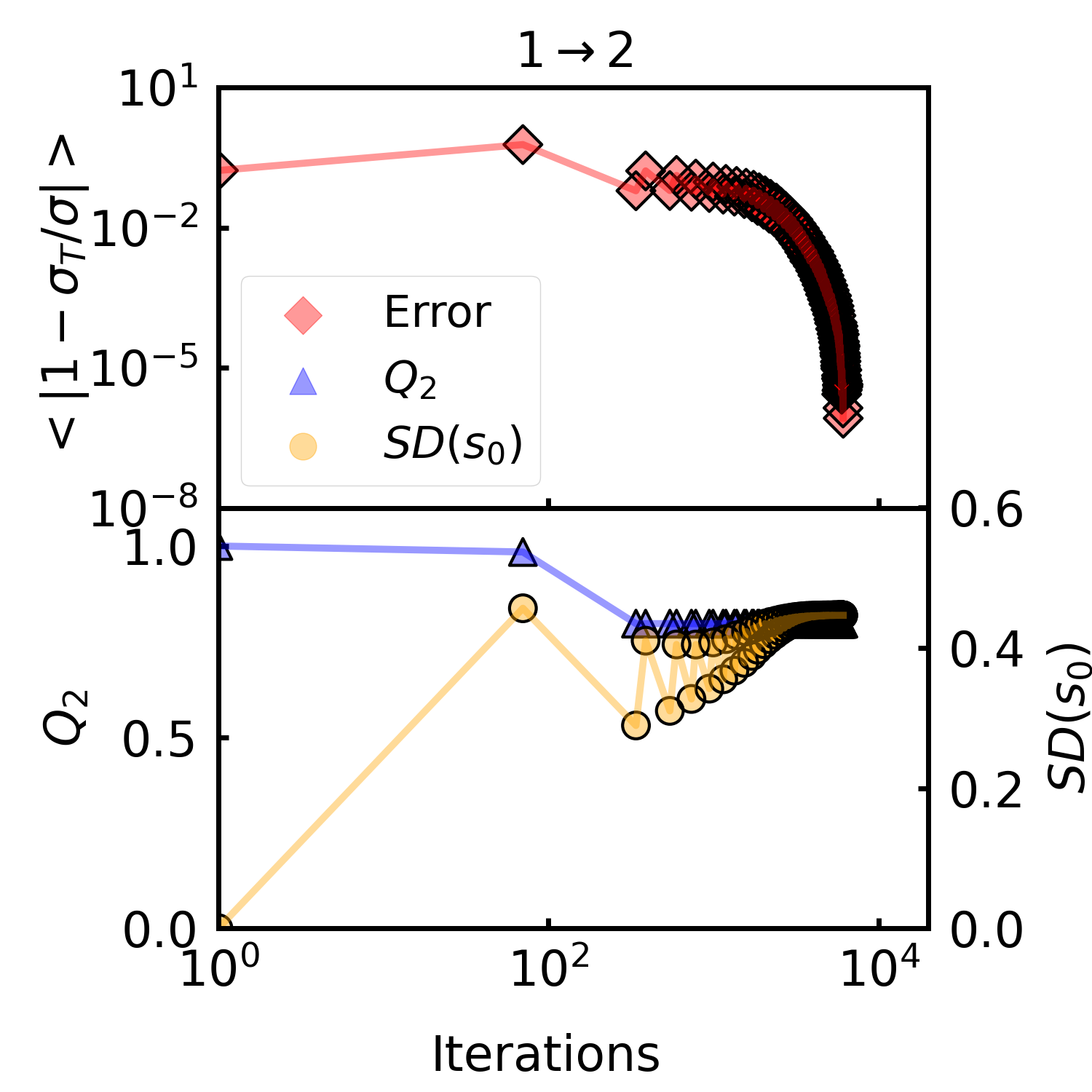}
        \caption{}
        \label{fig:single_cell_s0_histogram}
    \end{subfigure}
    \begin{subfigure}{0.32\linewidth}
        \includegraphics[width=\linewidth]{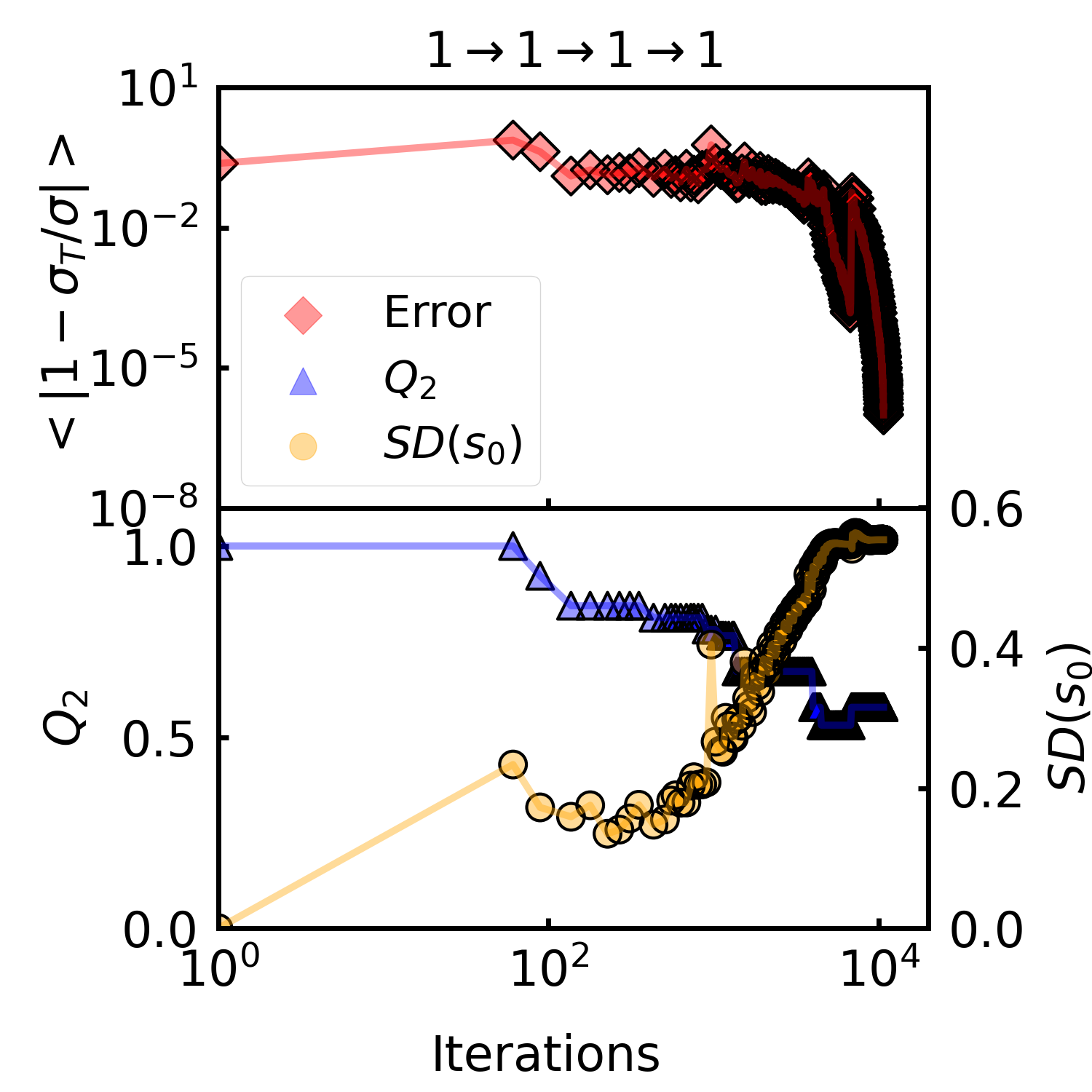}
        \caption{}
        \label{fig:1a}
    \end{subfigure}
    \vfill
    \begin{subfigure}{0.32\linewidth}
        \includegraphics[width=\linewidth]{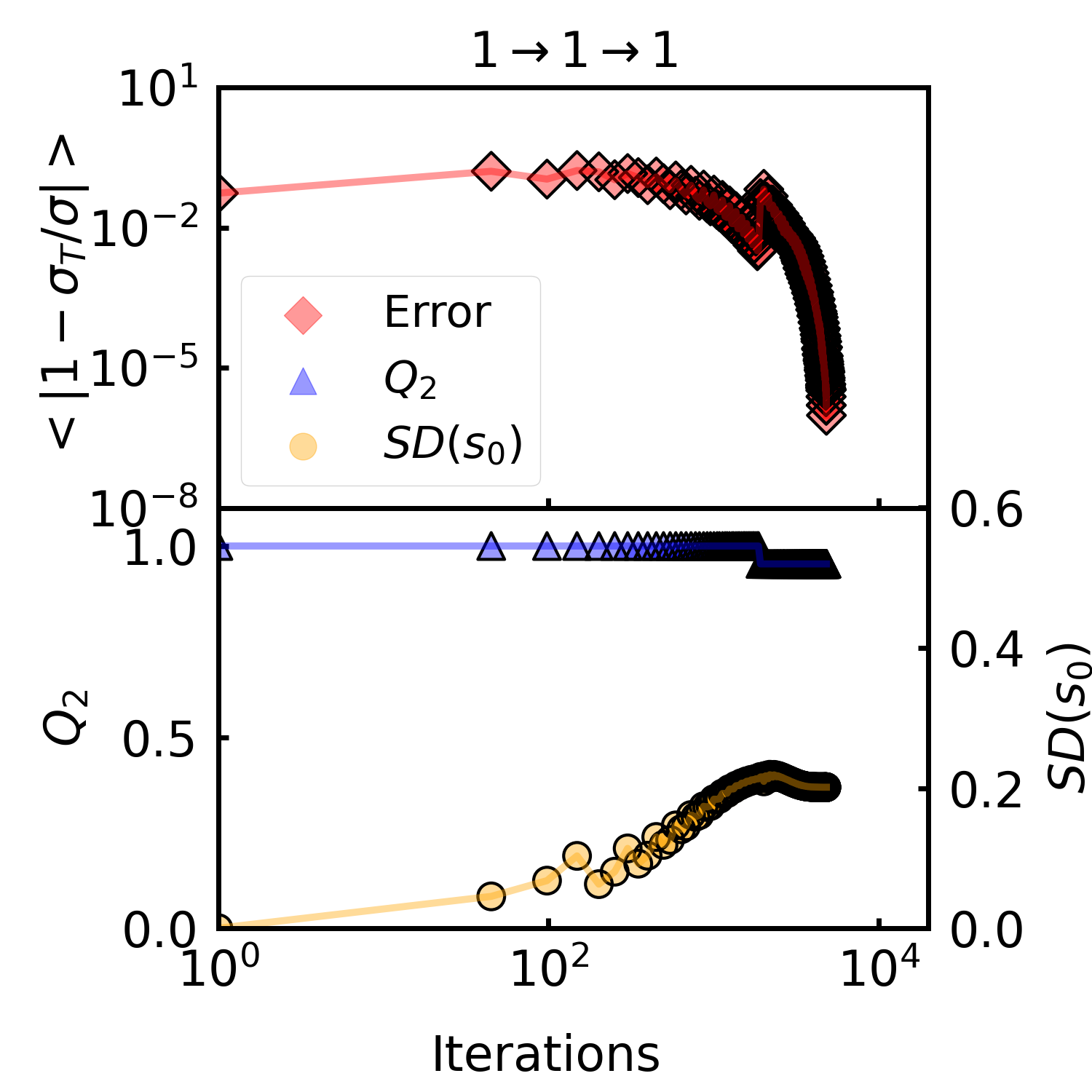}
        \caption{}
        \label{fig:1a}
    \end{subfigure}
    \begin{subfigure}{0.32\linewidth}
        \includegraphics[width=\linewidth]{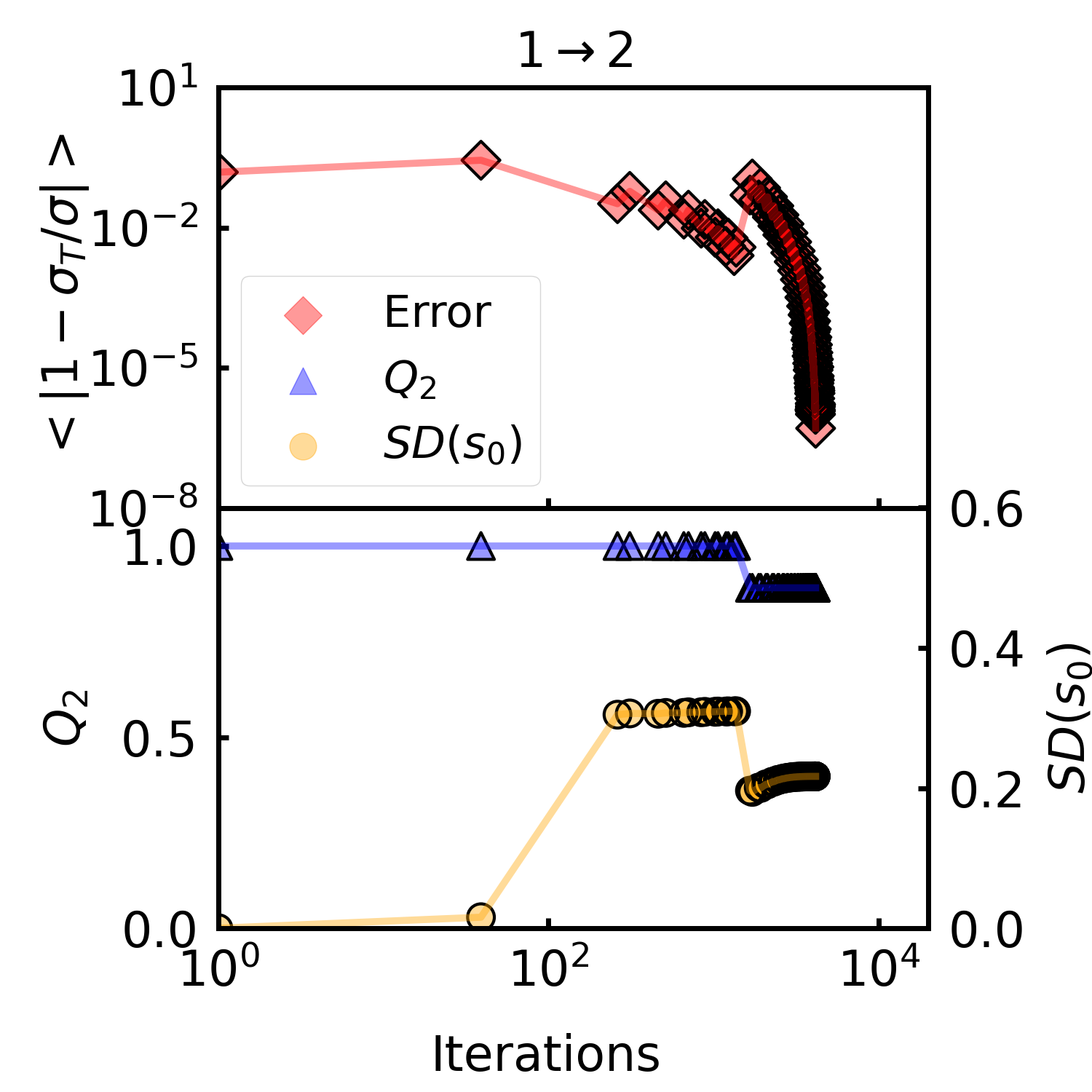}
        \caption{}
        \label{fig:1a}
    \end{subfigure}
    \begin{subfigure}{0.32\linewidth}
        \includegraphics[width=\linewidth]{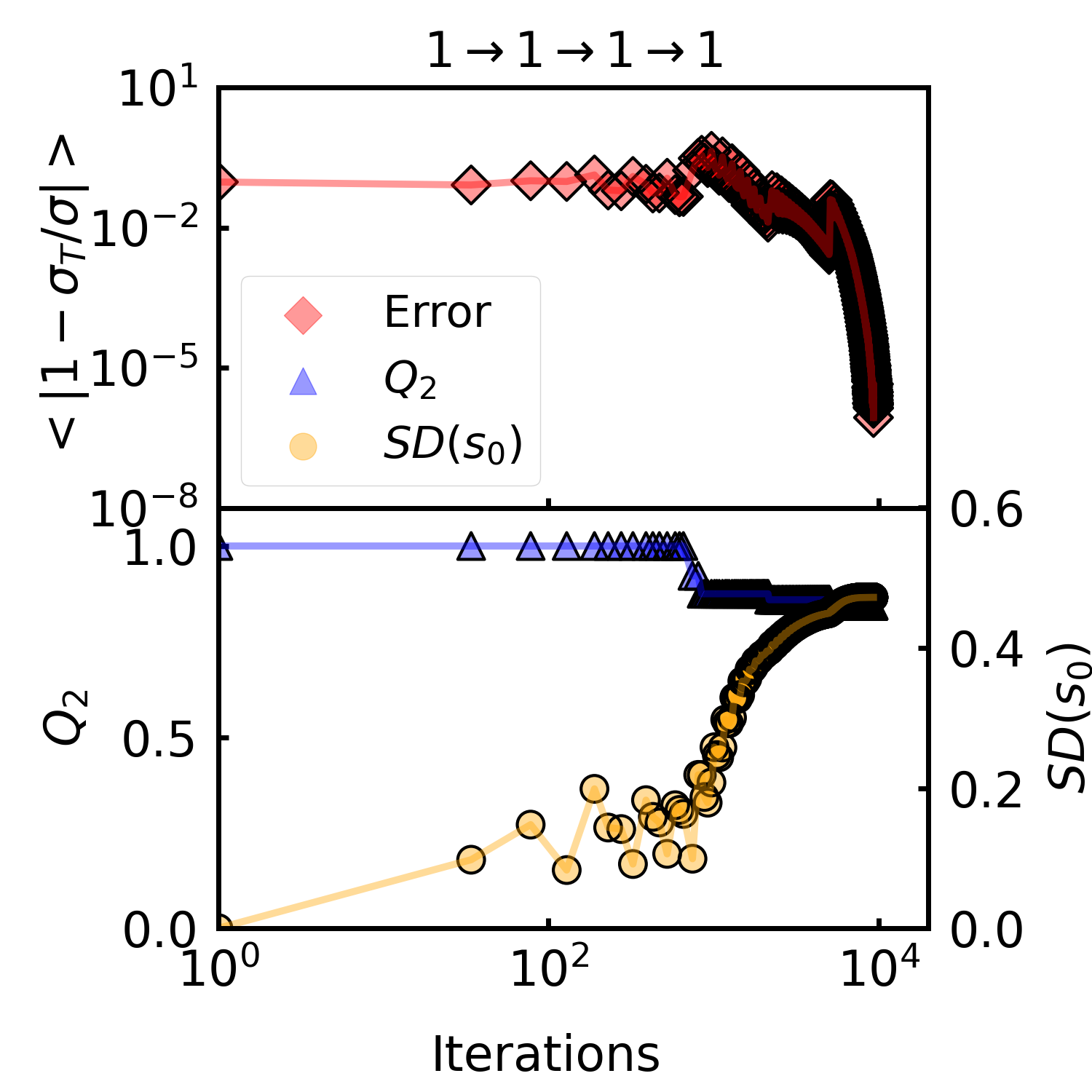}
        \caption{}
        \label{fig:1a}
    \end{subfigure}

    \caption{{\it Dual mechanisms of learning patterns: smooth and topological.} The first row contains instance of training with no reconnections for patterns (a) $1\rightarrow1\rightarrow$1 (b) $1\rightarrow2$ and (c) $1\rightarrow 1\rightarrow 1\rightarrow 1$ sequences. The second row (d)-(f) contains corresponding instances where reconnections lead to a decrease in error. The third row (g)-(i) contains corresponding instances where reconnections lead to an increase in error.}
  \label{fig:1}

\end{figure*}

\begin{figure*}[t]
\centering
    \begin{subfigure}{0.22\linewidth}
        \includegraphics[width=\linewidth]{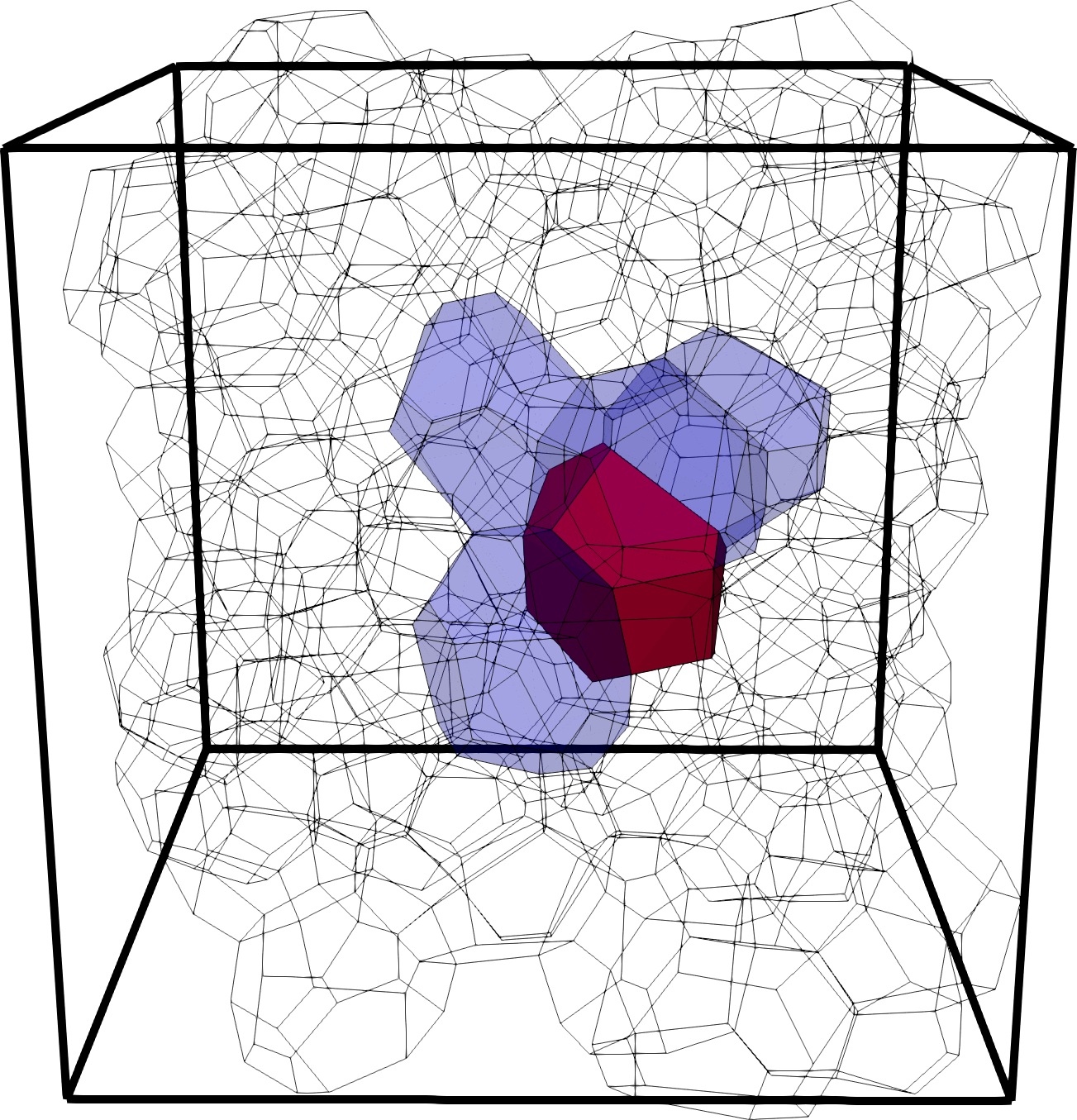}
        \label{}
    \end{subfigure}
    \begin{subfigure}{0.22\linewidth}
        \includegraphics[width=\linewidth]{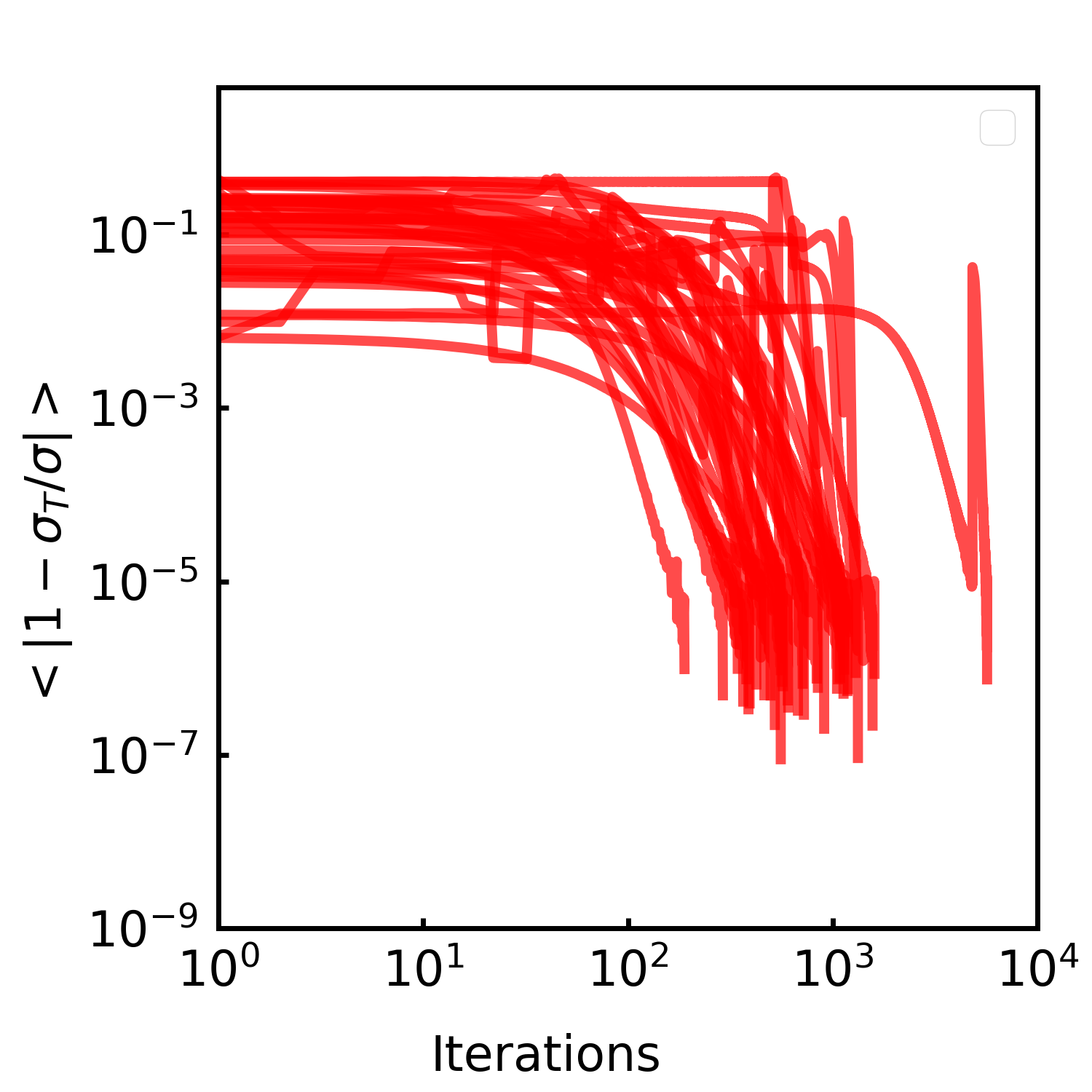}
        \label{}
    \end{subfigure}
    \begin{subfigure}{0.22\linewidth}
        \includegraphics[width=\linewidth]{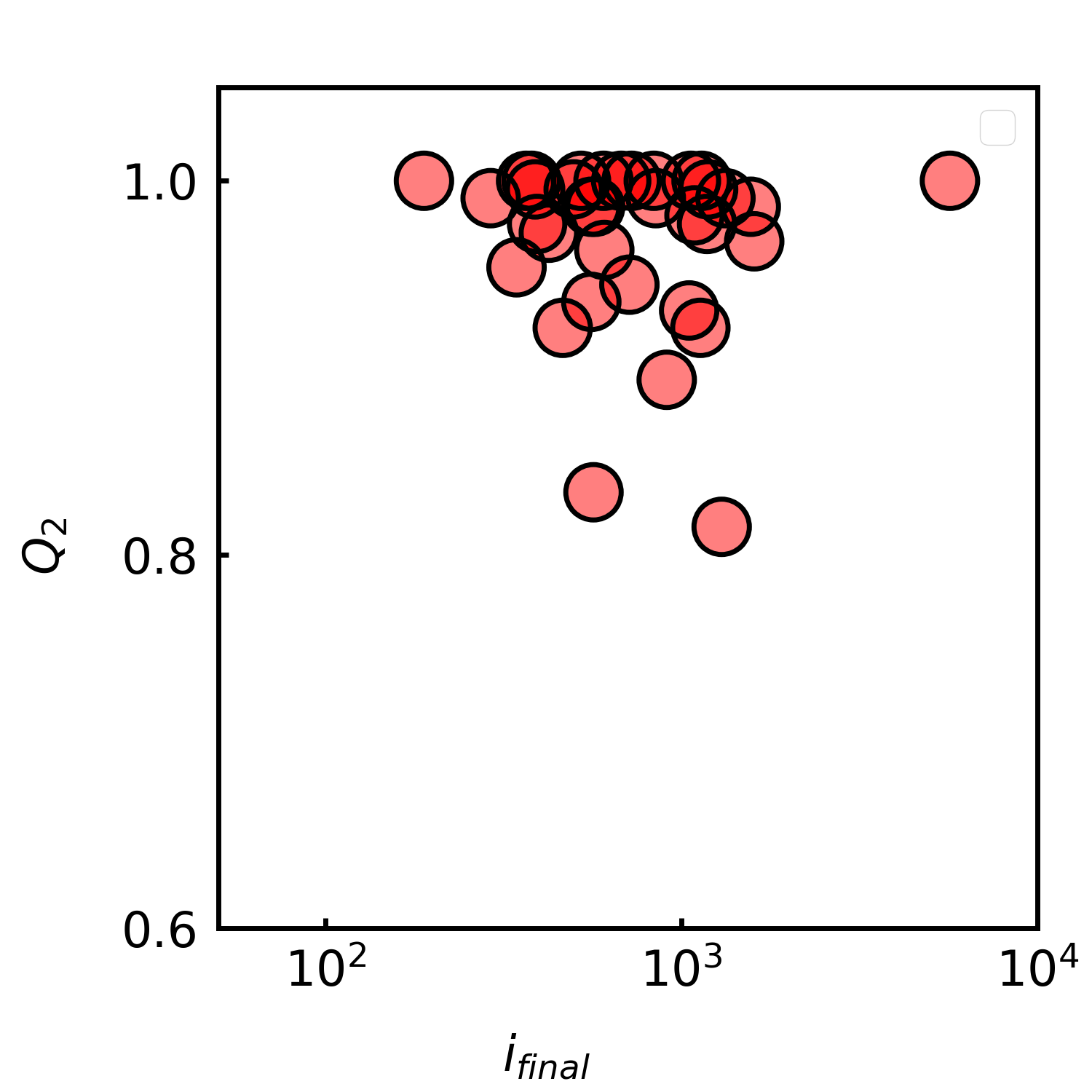}
        \label{}
    \end{subfigure}
    \begin{subfigure}{0.22\linewidth}
        \includegraphics[width=\linewidth]{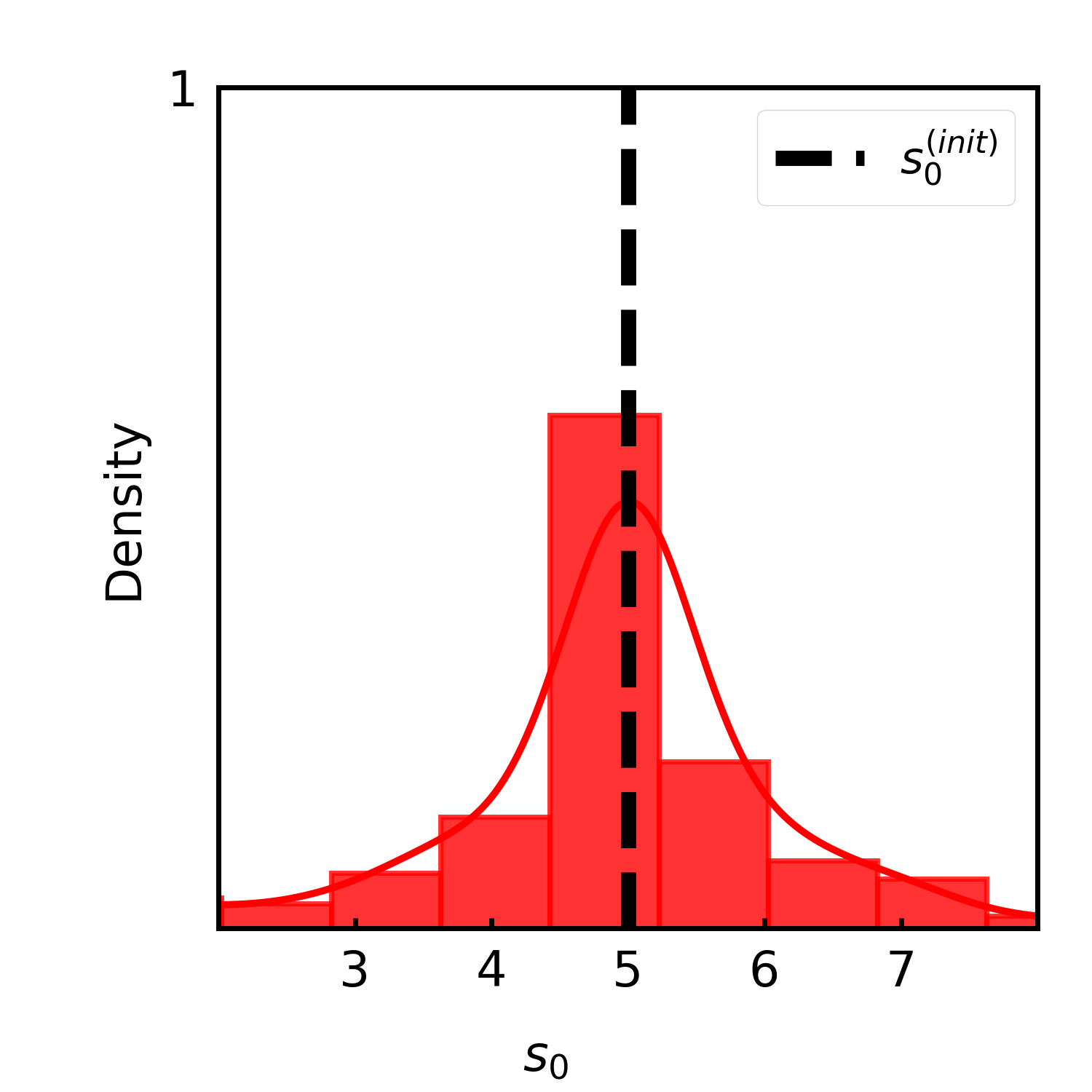}
        \label{}
    \end{subfigure}
    \caption{{\it Minimal system size for learning.} Training cells in a small spheroid 4 hidden cells and one target cell i.e. a spheroid of 5 cells.}
\end{figure*}

\begin{figure*}[t]
\centering
    \begin{subfigure}{0.48\linewidth}
        \includegraphics[width=\linewidth]{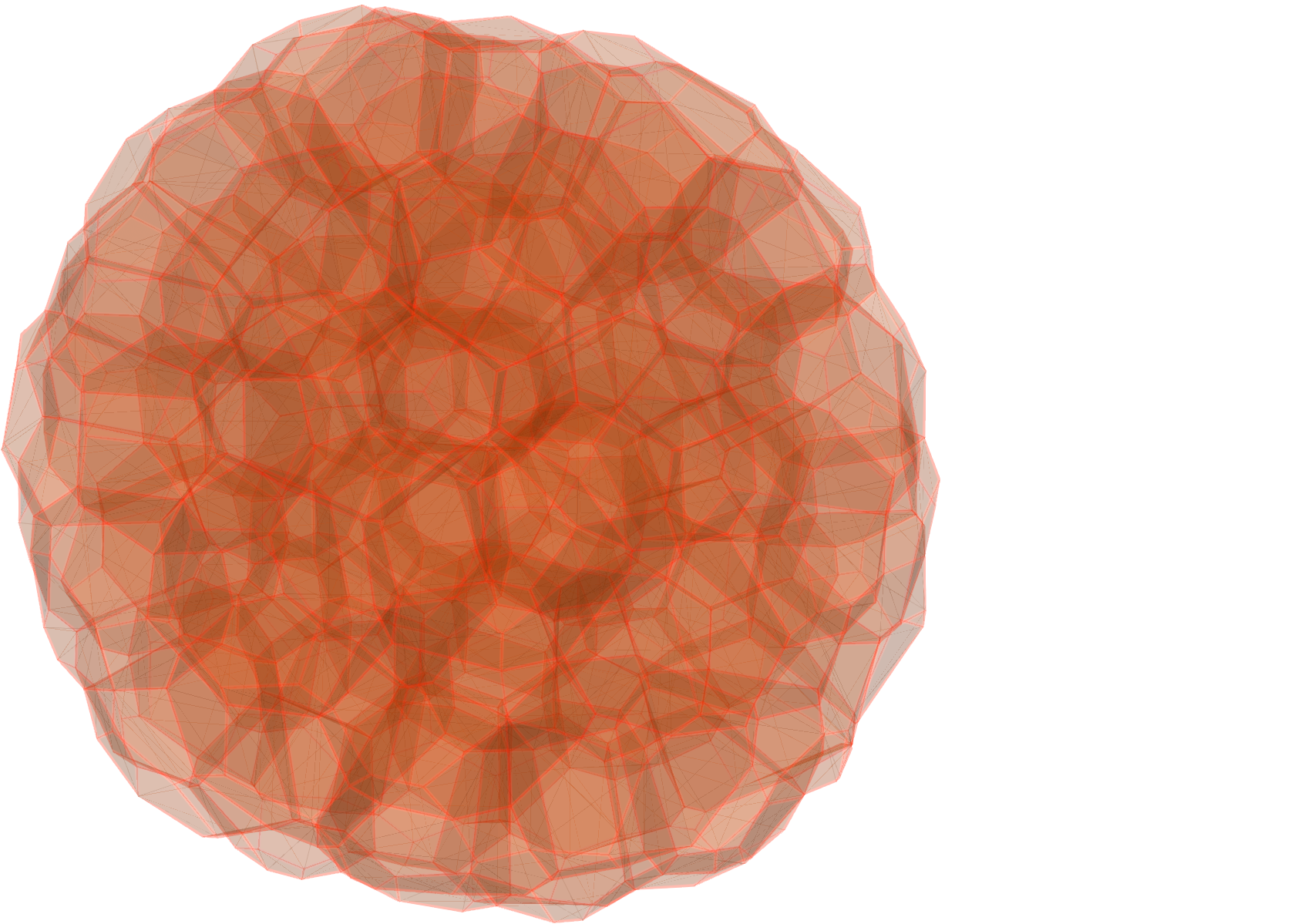}
        \label{}
    \end{subfigure}
    \begin{subfigure}{0.48\linewidth}
        \includegraphics[width=\linewidth]{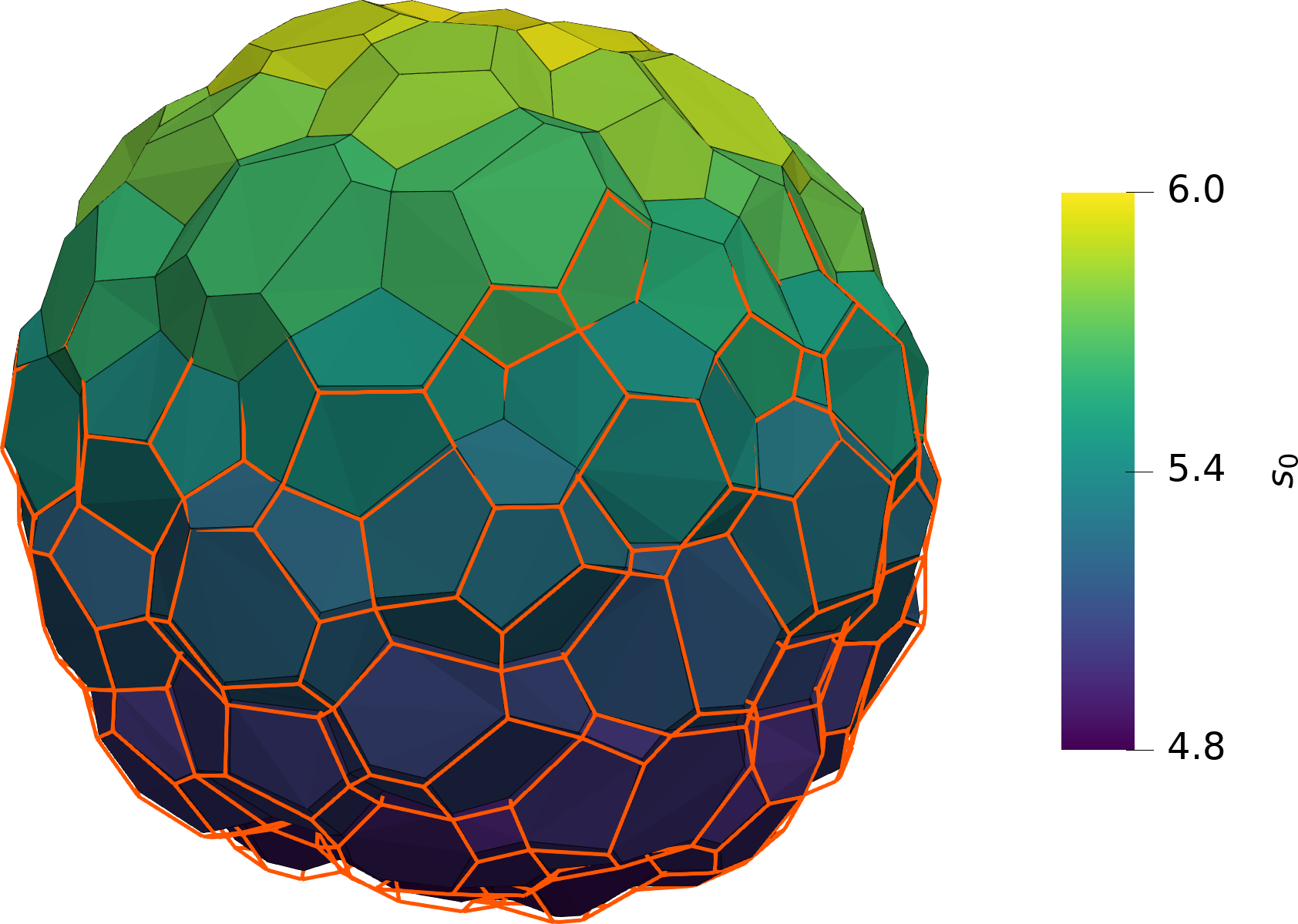}
        \label{}
    \end{subfigure}
    \caption{{\it A biophysical motivation: spheroids with variable target cell shape indices can impact spheroid shape.} The orange wire-frame depicts a spheroid with a homogeneous distribution of $s_0 = 5.4$, stabilized with overdamped motion. The cells are then assigned a $s_0$ value between 4.8 and 6.0 (mapping linearly by the projection of their distance from the spheroid center to an axis pointing up the page). The resulting spheroid, stabilized with overdamped motion, is depicted in by the surface with the assigned color map}
\end{figure*}

\end{document}